\documentclass[twocolumn,twocolappendix]{aastex6}

\usepackage{amsmath}
\usepackage[normalem]{ulem}
\usepackage{footnote}
\usepackage{subfigure}
\usepackage{xspace}
\usepackage[utf8]{inputenc}
\usepackage{amssymb}

\newcommand\msun{\ensuremath{\rm{M}_{\odot}}\xspace}                   
\newcommand\M{\ensuremath{\rm{M}}\xspace}                   
\newcommand\mzams{\ensuremath{\rm{M_{ZAMS}}}\xspace}                   
\newcommand\semi{\ensuremath{\alpha_{\rm{sc}}}\xspace}                 
\newcommand\thermo{\ensuremath{\alpha_{\rm{th}}}\xspace}               
\newcommand\overshoot{\ensuremath{f_{\rm{ov}}}\xspace}            
\newcommand\overshootD{\ensuremath{f_{\rm{ov,D}}}\xspace}         
\newcommand\mesh{\ensuremath{\delta_{\rm{mesh}}}\xspace}          
\newcommand\var{\ensuremath{w_t}\xspace}                          
\newcommand{\Mdot}{\ensuremath{\dot{\rm{M}}}}                          
\newcommand{\alphaMLT}{\ensuremath{\alpha_{\mathrm{MLT}}}}        
\newcommand{\Tc}{\ensuremath{T_{\mathrm{\!c}}}\xspace}                   
\newcommand{\rhoc}{\ensuremath{\rho_{\mathrm{c}}}\xspace}                
\newcommand{\netapprox}{{\tt approx22.net}\xspace}                
\newcommand{\net}[1]{{\tt mesa\_{#1}.net}}                        
\newcommand{\maxdm}{{\ensuremath{\Delta {\mathrm M}_{\rm{max}}}}} 
\newcommand{\Ye}{\ensuremath{Y_{\mathrm{e}}}\xspace}              
\newcommand{\Yec}{\ensuremath{Y_{\mathrm{e,c}}}\xspace}           

\newcommand{\code}[1]{\texttt{#1}}
\newcommand{\mesa}{\code{MESA}\xspace}
\newcommand{\MESA}{\mesa}
\newcommand{\mltpp}{\code{mlt++}\xspace}

\newcommand{\kmpers}{\ensuremath{\rm{km}\ \rm{s}^{-1}}}

\newcommand{\timetillche}{\ensuremath{\log_{10}\left(\left(\rm{\tau_{CHe}-\tau}\right)/\rm{yr}\right)}}
\newcommand{\timetillcc}{\ensuremath{\log_{10}\left(\left(\rm{\tau_{cc}-\tau}\right)/\rm{yr}\right)}}

\newcommand{\hecenter}{\ensuremath{\rm{X}(^4\rm{He_{c}})}}
\newcommand{\ocenter}{\ensuremath{\rm{X}(^{16}\rm{O_{c}})}}

\newcommand{\cpost}{\ensuremath{X\left(^{12}\rm{C}_{\rm{post}}\right)}\xspace}

\begin{document}

\title{On Variations Of Pre-Supernova Model Properties}

\author{
R.~Farmer,\altaffilmark{1}
C.E.~Fields,\altaffilmark{1,2,3}
I.~Petermann,\altaffilmark{1,3}
Luc~Dessart,\altaffilmark{4}
M.~Cantiello,\altaffilmark{5}
B.~Paxton,\altaffilmark{5}
F.X.~Timmes\altaffilmark{1,3}
}

\altaffiltext{1}{School of Earth and Space Exploration, Arizona State University, Tempe, AZ, USA}
\altaffiltext{2}{Department of Physics and Astronomy, Michigan State University, East Lansing, MI 48824, USA}
\altaffiltext{3}{Joint Institute for Nuclear Astrophysics, USA}
\altaffiltext{4}{Universite C\^ote d'Azur, OCA, CNRS, Lagrange, France}
\altaffiltext{5}{Kavli Institute for Theoretical Physics, University of California, Santa Barbara, CA 93106, USA}

\email{rjfarmer@asu.edu}

\shorttitle{Variations Of Pre-Supernova Models} 
\shortauthors{Farmer et al.}

\begin{abstract}

We explore the variation in single star $15\textrm{--}30$\,\msun,
non-rotating, solar metallicity, pre-supernova \MESA models due to changes in the number of isotopes in
a fully-coupled nuclear reaction network and adjustments in the mass resolution.
Within this two-dimensional plane we quantitatively detail the range of core
masses at various stages of evolution, mass locations of the main
nuclear burning shells, electron fraction profiles, mass fraction profiles,
burning lifetimes, stellar lifetimes, and compactness parameter at core-collapse
for models with and without mass loss.
Up to carbon burning we generally find mass resolution has a larger impact on the
variations than the number of isotopes, while the number of isotopes plays
a more significant role in determining the span of the variations 
for neon, oxygen and silicon burning. 
Choice of mass resolution dominates the variations in the structure of the 
intermediate convection zone and secondary convection zone 
during core and shell hydrogen burning respectively, where we find
a minimum mass resolution of $\approx$0.01\,\msun is necessary to achieve convergence
in the helium core mass at the $\approx$5\% level. On the other hand,
at the onset of core-collapse we find $\approx$30\% variations in the 
central electron fraction and mass locations of the main nuclear burning shells,
a minimum of $\approx$127 isotopes is needed to 
attain convergence of these values at the $\approx$10\% level.

\end{abstract}


\section{Introduction}
\label{sec:introduction}

The end evolutionary phases of massive stars remain a rich site of fascinating
challenges that include the interplay between
convection \citep{meakin_2007_ab,viallet_2013_aa}, 
nuclear burning \citep{couch_2015_aa}, 
rotation \citep{heger_2000_aa,rogers_2015_aa,chatzopoulos_2016_aa},
radiation transport \citep{jiang_2015_aa},
instabilities \citep{garaud_2015_aa,wheeler_2015_aa}, 
mixing \citep{maeder_2012_aa}, 
waves \citep{rogers_2013_aa,fuller_2015_aa,aerts_2015_ab},
eruptions \citep{humphreys_1994_aa, kashi_2016_aa}, 
and binary partners \citep{justham_2014_aa, marchant_2016_aa}.
This bonanza of physical puzzles is closely linked with 
compact object formation by core-collapse supernovae (SN) \citep{timmes_1996_ac, eldridge_2004_aa,ozel_2010_aa}
and the diversity of observed massive star transients \citep[e.g.,][]{van-dyk_2000_aa,ofek_2014_aa,smith_2016_aa}. 
Recent observational clues that challenge conventional wisdom 
\citep{zavagno_2010_aa,vreeswijk_2014_aa,boggs_2015_aa,jerkstrand_2015_aa,strotjohann_2015_aa}, 
coupled with the expectation of large quantities of data
from upcoming surveys \citep[e.g.,][]{creevey_2015_aa,papadopoulos_2015_aa,sacco_2015_aa,yuan_2015_aa},
new measurements of key nuclear reaction rates and techniques for assessing 
reaction rate uncertainties \citep{iliadis_2011_aa,wiescher_2012_aa,sallaska_2013_aa},
and advances in 3D pre-SN modeling \citep{couch_2015_aa,muller_2016_aa,jones_2016_aa}, 
offer significant improvements in our quantitative understanding of the end states of massive stars.

One end state, core-collapse SN, is the result of another end
state, massive star progenitors undergoing gravitational collapse
\citep[e.g.,][]{sukhbold_2014_aa,perego_2015_aa,sukhbold_2016_aa}.  
The amount of mass of an isotope 
that can be injected into the interstellar medium
\citep[e.g.,][]{woosley_1995_aa,limongi_2003_aa,nomoto_2013_aa}
depends on the structure of the star at the point of core-collapse.
In turn, the pre-SN structure depends on the evolutionary
pathway taken by the massive star during its lifetime
\citep[e.g.,][]{nomoto_1988_aa,jones_2013_aa}.

This paper is novel in two ways. First, we scrutinize the structure and
evolution of single massive stars from the pre-main sequence (pre-MS)
to the onset of core-collapse with multiple, possibly large, in-situ
nuclear reaction networks. For the first time, we quantify aspects of
the structure and evolution that are robust, or can be made robust, with respect to
variations in the nuclear reaction network used for the entire
evolution.  For example, we explore the diversity of pre-SN model
properties such as the mass locations of the major burning stages,
core masses at various stages of evolution, mass fraction profiles,
and electron fraction profiles.  Second, for each nuclear reaction
network we investigate the impact of methodically and systematically
changing the mass resolution on the structure and evolution of a
set of massive stars. 

In \S\ref{sec:method} we discuss the software instruments, input physics, 
reaction networks, and model choices.
In \S\ref{sec:results} we present the results for different reaction networks,
and mass resolutions with and without mass loss. 
In \S\ref{sec:discussion} we discuss our results and their implications.

\section{Stellar models}
\label{sec:method}

Models of 15, 20, 25 and 30\,\msun
are evolved using the Modules for Experiments in Stellar Astrophysics software instrument
\citep[henceforth \MESA, version 7624,][]{paxton_2011_aa,paxton_2013_aa,paxton_2015_aa}.
All models begin with a metallicity of $Z=0.02$ and a solar abundance distribution from \citet{grevesse_1998_aa}.
The models are evolved without mass loss or with the ``Dutch'' wind
loss scheme \citep{nieuwenhuijzen_1990_aa, nugis_2000_aa,vink_2001_ab,glebbeek_2009_aa}
with an efficiency $\eta$=0.8 for these non-rotating models \citep{maeder_2001_aa}.
Each stellar model is evolved from the pre-MS
until core-collapse, which we take as when any location inside the
stellar model reaches an infall velocity of 1000 $\kmpers$.  
\added{To compute the infall velocity we set \MESA's \texttt{v\_flag=.true.} which adds
a, hydrodynamic, radial velocity term to the model. This additional variable is evolved 
from the  zero-age main sequence (ZAMS) until core-collapse, though it only becomes relevant
to the evolution after core oxygen burning.}

All
the \MESA inlists and many of the stellar models are publicly
available\footnote{\url{http://mesastar.org/results}}.

\subsection{Mass and Temporal Resolution}\label{sec:num_res}

\MESA provides several controls to specify the mass resolution of a
model.  Sufficient mass resolution is required to accurately determine
gradients of stellar structure quantities, but an excessive number of
cells impacts performance.  One parameter for changing the mass
resolution in regions of rapid change is \texttt{mesh\_delta\_coeff}
(\mesh), which acts a global scale factor limiting the change in
stellar structure quantities between two adjacent cells.  Smaller
values of \mesh increases the number of cells. Another parameter
controlling mass resolution is the maximum fraction a star's mass in a
cell, \texttt{max\_dq}. That is, the minimum number of cells in a
stellar model is 1/\texttt{max\_dq}.  We use $\mesh=1.0$ and
$\texttt{max\_dq}=\maxdm/\rm{M}_{\star}(\tau)$, where $\rm{M}_{\star}(\tau)$ is
the mass of the stellar model in solar mass units at time $\tau$ and
\maxdm\ is a parameter we vary between $0.1$\,\msun
and $0.005$\,\msun. \added{We choose to vary \texttt{max\_dq} instead of 
\texttt{mesh\_delta\_coeff} to enable us to set a minimum level of mass 
resolution in the model.}

\MESA also offers a rich set of timestep controls.
The parameter \var broadly controls the temporal resolution by modulating the
magnitude of the allowed changes between time steps. 
At a finer level of granularity, \texttt{dX\_nuc\_drop\_limit} limits the 
maximum allowed change of the mass fractions between timesteps for mass fractions
larger than \texttt{dX\_nuc\_drop\_min\_X\_limit}. 
We use
$\var=5\times10^{-5}$,
$\texttt{dX\_nuc\_drop\_limit}=10^{-3}$, and
$\texttt{dX\_nuc\_drop\_min\_X\_limit}=10^{-3}$ 
for evolution between
the pre-MS to the onset of core Si-burning, where we 
loosen the criteria to allow larger timesteps; 
$\var=5\times10^{-5}$,
$\texttt{dX\_nuc\_drop\_limit}=5\times 10^{-2}$, and 
$\texttt{dX\_nuc\_drop\_min\_X\_limit}=5\times 10^{-2}$.

The timestep control \texttt{delta\_lg\_XH\_cntr\_min}
regulates the timestep as hydrogen is depleted in the core, which aids resolving 
the transition from the ZAMS to the terminal age main sequence (TAMS).
Similarly, the timestep controls 
\texttt{delta\_lg\_XHe\_cntr\_min}, 
\texttt{delta\_lg\_XC\_cntr\_min},
\texttt{delta\_lg\_XNe\_cntr\_min},
\texttt{delta\_lg\_XO\_cntr\_min}, and 
\texttt{delta\_lg\_XSi\_cntr\_min}
control the timestep as one of the major fuels is depleted in the core.
These timestep controls are useful for obtaining, for example, 
convergence of mass shell locations, smoother transitions as a stellar model
exits core H-burning in the HR diagram \citep[i.e., the ``Henyey Hook'',][]{kippenhahn_2012_aa},
and smoother trajectories in the central temperature \Tc~-~central density \rhoc plane.
We use a mass fraction value of 10$^{-6}$ for all these fuel
depletion timestep controls. \added{Additionally, we use \MESA's default timestep controls for controlling 
changes in the hydrodynamics. At the point where hydrodynamics becomes important, during Si burning,
we find we are limited by the \texttt{delta\_lg\_XSi\_cntr\_min} control rather than by changes in the 
hydrodynamics.}

\begin{figure*}[!htb]
\centering{\includegraphics[width=2.0\columnwidth]{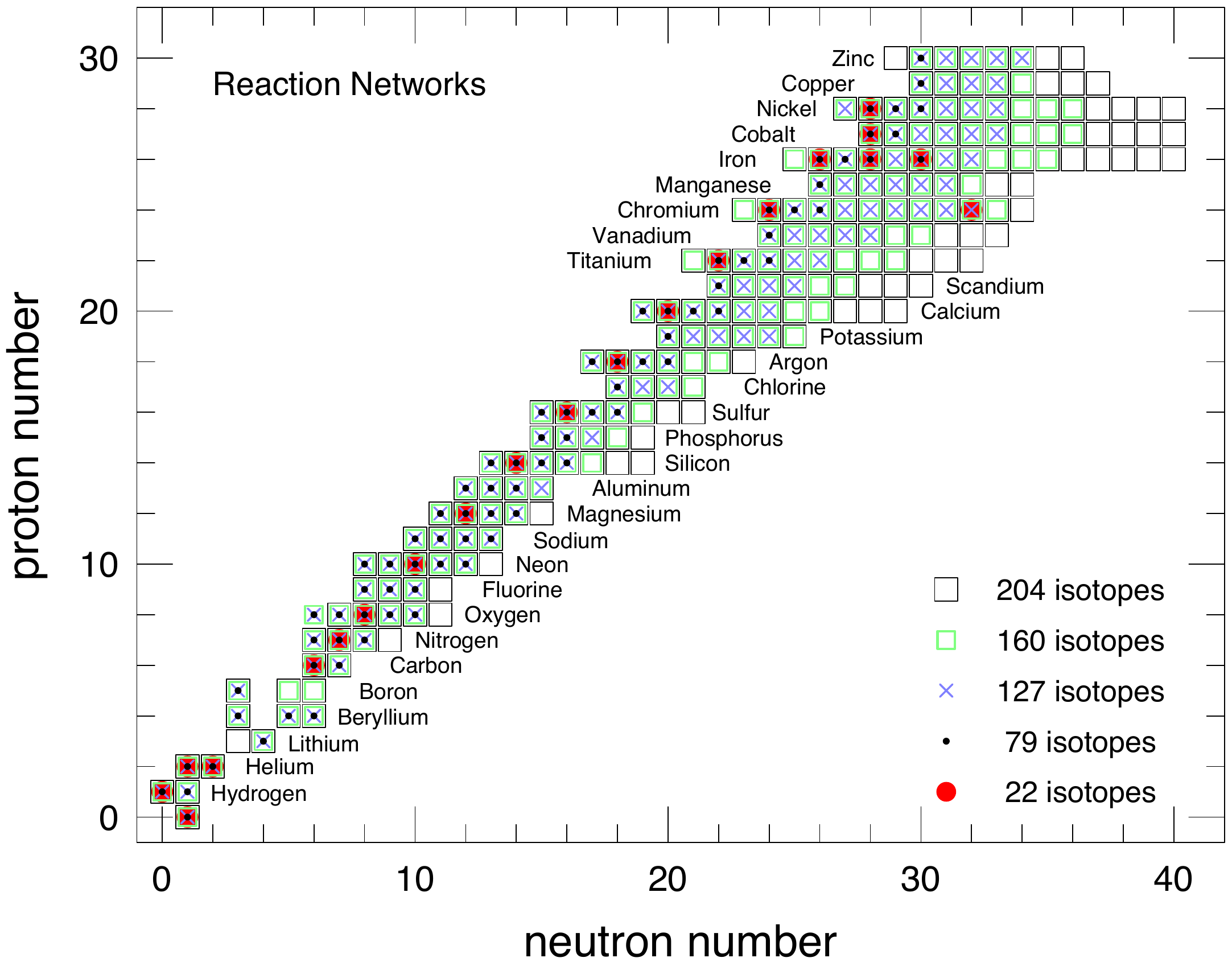}}
\caption{
The five reaction networks used to quantify the variance of the
properties of the pre-SN cores as a function of the number of
isotopes in a network. 
}\label{fig:nzplane}
\end{figure*}

\subsection{Nuclear Reaction Networks}\label{sec:nets}

\MESA evolves models of massive stars from the pre-MS to the onset of
core-collapse with the nuclear burning fully coupled to the
hydrodynamics using a single, possibly large, reaction network (see \citet{paxton_2015_aa}).
This capability avoids the challenges of a) operator splitting errors from
evolving the hydrodynamics and nuclear burning independently; b) stitching together
different solution methods for different phases of evolution, for
example, combining a reaction network with equilibrium methodologies like QSE or NSE, \added{or 
with the use of an adaptive network that modifies itself based on the most populous isotopes present}; 
or finally c) evolving a stellar model with a small reaction network while carrying
along a larger reaction network that does not impact the energy generation
rate or composition of the stellar structure, i.e., passive co-processing.
\MESA's unified approach is not just a solution to issues of accuracy
or self-consistency.  It offers an improvement by providing a single
solution methodology -- in situ reaction networks evolved
simultaneously with the hydrodynamics.

We evolve each stellar model with one of the five nuclear reaction
networks shown in Figure \ref{fig:nzplane}.  We consider a small
network, {\tt approx21\_cr60\_plus\_co56.net} (hereafter \netapprox), 
where each reaction pathway has been predetermined
(i.e., a ``hardwired'' network). Such approximate networks,
an $\alpha$-chain backbone with aspects of H-burning, 
heavy ion reactions, and iron-group photodisintegration, are a traditional
workhorse in massive star models \citep[e.g.,][]{weaver_1978_aa,woosley_1988_aa,heger_2000_aa,heger_2010_aa,chatzopoulos_2012_aa}.
We also use four ``softwired'' networks,
where after specifying the isotopes, all allowed reaction pathways
between the isotopes are linked.  Specifically, we consider \net{79}
which contains isotopes up to $^{60}$Zn (black dots); \net{127}
which adds neutron rich isotopes in
the iron group (purple crosses); \net{160} which adds more neutron
rich isotopes and a few proton rich isotopes (green squares); and
\net{204} which adds isotopes lighter than $^{36}$Cl (black squares), and includes the isotopes identified
in \citet{heger_2001_aa} as important for the electron fraction, \Ye, 
in core-collapse models.

The softwired reaction networks are designed to yield
approximately the same final \Ye as a 3298 isotope reaction
network in one-zone burn calculations of the thermodynamic history 
of 15\,\msun cores \citep{paxton_2015_aa}. For example, in one-zone burn calculations the
\net{204} reaction network gives a final $\Ye=0.4032$, while the 3298
isotope reaction network gives $\Ye=0.4039$.

All forward thermonuclear reaction rates are from the JINA reaclib
version V2.0 2013-04-02 \citep{cyburt_2010_aa}. Inverse rates are
calculated directly from the forward rates (those with positive
$Q$-value) using detailed balance, rather than using fitted
rates. The nuclear partition functions used to calculate the inverse rates are
from \citet{rauscher_2000_aa}.  Electron
screening factors for both weak and strong thermonuclear reactions are
from \citet{alastuey_1978_aa} with plasma parameters from
\citet{itoh_1979_aa}.  All the weak rates are based (in order of
precedence) on the tabulations of \citet{langanke_2000_aa},
\citet{oda_1994_aa}, and \citet{fuller_1985_aa}.
Thermal neutrino energy losses are from \citet{itoh_1996_aa}.

\subsection{Mixing}\label{sec:mix}

We treat convection using the \mltpp approximation 
\citep{paxton_2013_aa}.  This prescription takes the convective
strength as computed by mixing length theory (MLT) and reduces the
superadiabaticity in radiation dominated convection zones (e.g., near
the iron opacity peak).  This decreases the temperature gradient in
these regions and the artificial suppression can be viewed as treating
additional, un-modeled, energy transport mechanisms in these regions
\citep[e.g.,][]{jiang_2015_aa}.  Values of $1.6 \lesssim
\alpha_{\rm{MLT}} \lesssim 2.2$ have been inferred from comparing
observations with stellar evolution models
\citep[][]{noels_1991_aa,miglio_2005_aa,aerts_2010_aa}, and 3D
hydrodynamic simulations of the deep core and surface layers also
suggests a similar range
\citep{viallet_2013_aa,trampedach_2014_aa,jiang_2015_aa}. These
efforts inform our baseline choice of \alphaMLT=1.5, though this is by no
means the only choice \citep{dessart_2013_aa}.

\begin{figure*}[!htb]
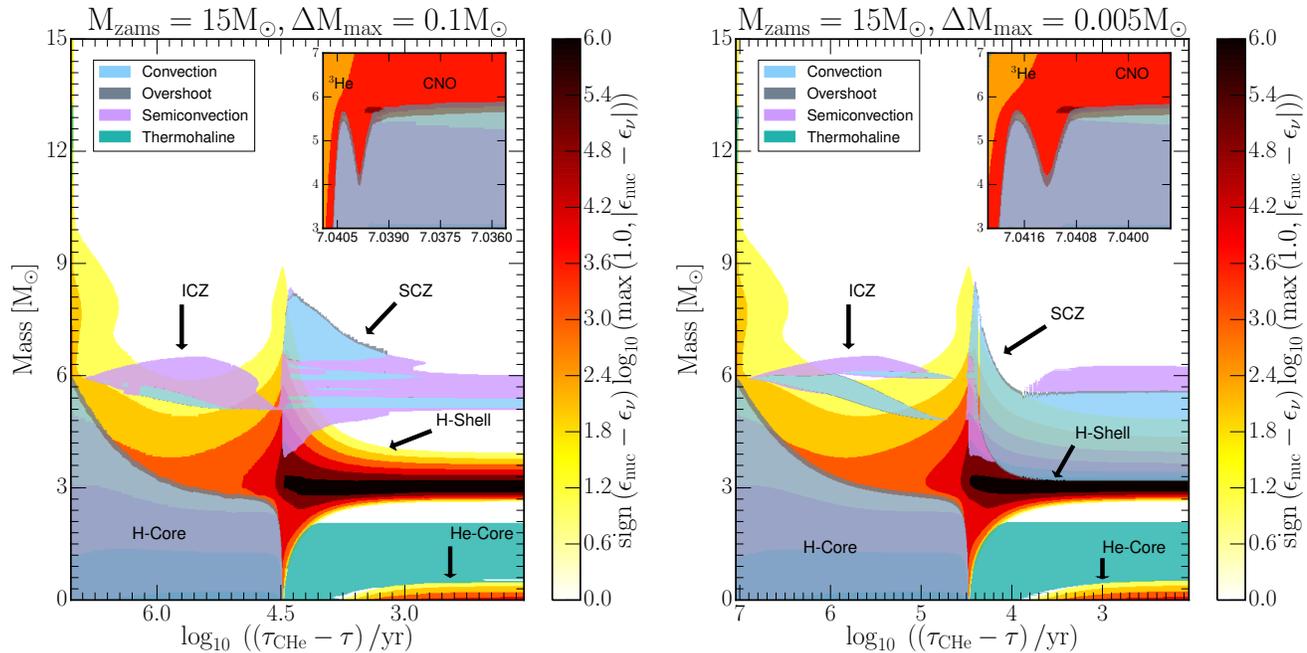

\begin{subfigure}
  \centering{\includegraphics[width=\columnwidth]{{{15msun_nml_res0.1_ch}}}}
\end{subfigure}
\begin{subfigure}
  \centering{\includegraphics[width=\columnwidth]{{{15msun_nml_res0.005_ch}}}}
\end{subfigure}
\caption{ 
Kippenhahn plot for a solar metallicity $M_{\rm{ZAMS}}=15$\,\msun model evolved with the \netapprox\ and
without mass loss. Left panel, mass resolution of $\maxdm=0.1$\,\msun; Right panel, 
mass resolution of $\maxdm=0.005$\,\msun. The age shown on the x-axis is time until 
convection begins in core He-burning, set such that the SCZ can be temporally resolved.
Yellow to red regions denote logarithmic increases in the
net nuclear energy generation rate. Blue marks convective regions, grey identifies
overshoot regions, purple designates semiconvective regions and green labels 
thermohaline mixing regions. 
Labeled are the intermediate convection 
zone (ICZ), secondary convection zone (SCZ), H-core, H-shell, and He-core.
The inset Kippenhahn diagram shows the behavior 
of the convective core during the transition between $^3$He and CNO burning.
}
\label{fig:kip_15_low_high}
\end{figure*}

Convective overshooting is assumed to be an exponential decay beyond
the Schwarzschild boundary of convection \citep{herwig_1997_aa}.
The convective diffusion coefficient, $D_{{\rm conv},0}$, is measured
at a distance \overshootD inside the convection zone, overshooting 
then extends beyond the edge of the convection zone a fraction \overshoot
of the local pressure scale height $\lambda_{\rm{P,0}}$,
\begin{equation}\label{eq:over}
 D_{\rm{OV}}=D_{{\rm conv},0}\exp\left(-\frac{2z}{\overshoot \lambda_{\rm{P,0}}}\right)
\enskip,
\end{equation}
where $D_{\rm{OV}}$ is the overshooting diffusion coefficient at a radial
distance $z$\ from the edge of the convection zone boundary.  Our
baseline choices for the overshoot mixing are $\overshoot = 0.004$ and
$\overshootD = 0.001$. This choice is motivated by recent calibration of a 1D 25\,\msun
pre-SN model to idealized 3D hydrodynamic simulations of turbulent O-burning 
shell convection \citep{jones_2016_aa}. 
\MESA offers the flexibility to have different
values for convection driven by different types of nuclear burning (H,
He, C and others) and whether the burning occurs near the core or in a
shell, but for simplicity we take these values to be the same for all
regions.

Thermohaline mixing is included in the models when \mbox{$\nabla_{\rm{T}}-\nabla_{\rm{ad}}\leq B \leq
0$}, where B is the Br\"{u}nt composition gradient. The thermohaline mixing 
diffusion coefficient $D_{\rm{th}}$ 
is parameterized as
\begin{equation}
 D_{\rm{th}}=\thermo \frac{3K}{2\rho C_{p}}
\frac{B}{\left(\nabla_{\rm{T}} - \nabla_{\rm{ad}}\right)}
\enskip ,
\end{equation}
where $\thermo$\ is a dimensionless parameter, $K$ is the radiative
conductivity, and $C_{\rm{p}}$ is the specific heat at constant pressure
\citep{ulrich_1972_aa,kippenhahn_1980_aa}.
Continuing attempts to calibrate $D_{\rm{th}}$ include multidimensional
simulations of fingering convection under stellar conditions
\citep{traxler_2011_aa,brown_2013_aa,garaud_2015_aa}.  We set
$\thermo$=2.0 except during core silicon burning where we set
$\thermo$=0.0.

Finally, we include semiconvection
when $\nabla_{\rm{ad}} < \nabla_{\rm{T}} < \nabla_{\rm{L}}$, where
$\nabla_{\rm{L}} = \nabla_{\rm{ad}} + B $. We take the strength of 
semiconvective mixing $D_{\rm{semi}}$ as
\citet{langer_1983_aa,langer_1985_aa},
\begin{equation}\label{eq:semi}
 D_{\rm{sc}}=\semi \left(\frac{K}{6C_p\rho}
\right)\frac{\nabla_{\rm{T}}-\nabla_{\rm{ad}}}{\nabla_{\rm{L}}
-\nabla_{\rm{T}}},
\end{equation}
where \semi is a dimensionless parameter.  Ongoing efforts to
calibrate 1D semiconvection models include multidimensional simulations of
double diffusive convection \citep{zaussinger_2013_aa,spruit_2013_aa}
and comparing massive star models with observations
\citep{yoon_2006_aa}. While recent work suggests that to a first order
approximation, the effect of semiconvection can be neglected 
\citep{moll_2016_aa}, we adopt a baseline choice of $\semi=0.01$.

\section{Results}\label{sec:results}

We present the variations in the \MESA version 7624 pre-SN model properties due to 
changes in the nuclear reaction network and mass resolution in the order 
of the major burning stages; 
H-burning in \S\ref{sec:burn_h}, stellar lifetimes in \S\ref{sec:lifetimes}, 
He-burning in  \S\ref{sec:burn_he}, 
C-burning in  \S\ref{sec:burn_c}, 
Ne-burning in  \S\ref{sec:burn_ne}, 
O-burning in  \S\ref{sec:burn_o}, 
Si-burning in  \S\ref{sec:burn_si}, 
the onset of core-collapse in \S\ref{sec:cc},
and representative pre-SN nucleosynthesis yields in \S\ref{sec:nucleo}.

\subsection{Core/Shell H burning}\label{sec:burn_h}

Figure \ref{fig:kip_15_low_high} shows the MS and giant branch (GB) 
evolution of a 15\,\msun model at two resolutions $\maxdm=0.1$\,\msun and $\maxdm=0.005$\,\msun,
evolved without mass loss and with the \netapprox network.
As stars evolve off of the pre-MS and onto the ZAMS, they begin
burning hydrogen in a convective core, predominantly via the CNO
cycle, but also via primordial $^3$He in the $pp$ chain
\citep{hansen_2004_aa,iben_2013_aa,iben_2013_ab}.
This convective region expands outwards to 
consume approximately half the mass of the star.
Once the primordial $^3$He is exhausted, at $\timetillche \approx 7.04$ in 
Figure \ref{fig:kip_15_low_high}, 
the convection region recedes and the core temperature and density
increase. 
At this point in the evolution most of the primordial $^{12}$C is
already piled up in $^{14}$N due to the proton capture onto $^{14}$N
reaction rate \citep{arnett_1996_aa,iliadis_2007_aa}.
Due to the strong temperature dependence of the CNO cycle, the energy
production increases and the convective region expands slightly
further out in mass coordinate than the first peak at $\timetillche
\approx 7.039$. Due to the increased energy generation from the CNO
cycle the star itself also begins expanding at this point.

\begin{figure*}[!htb]
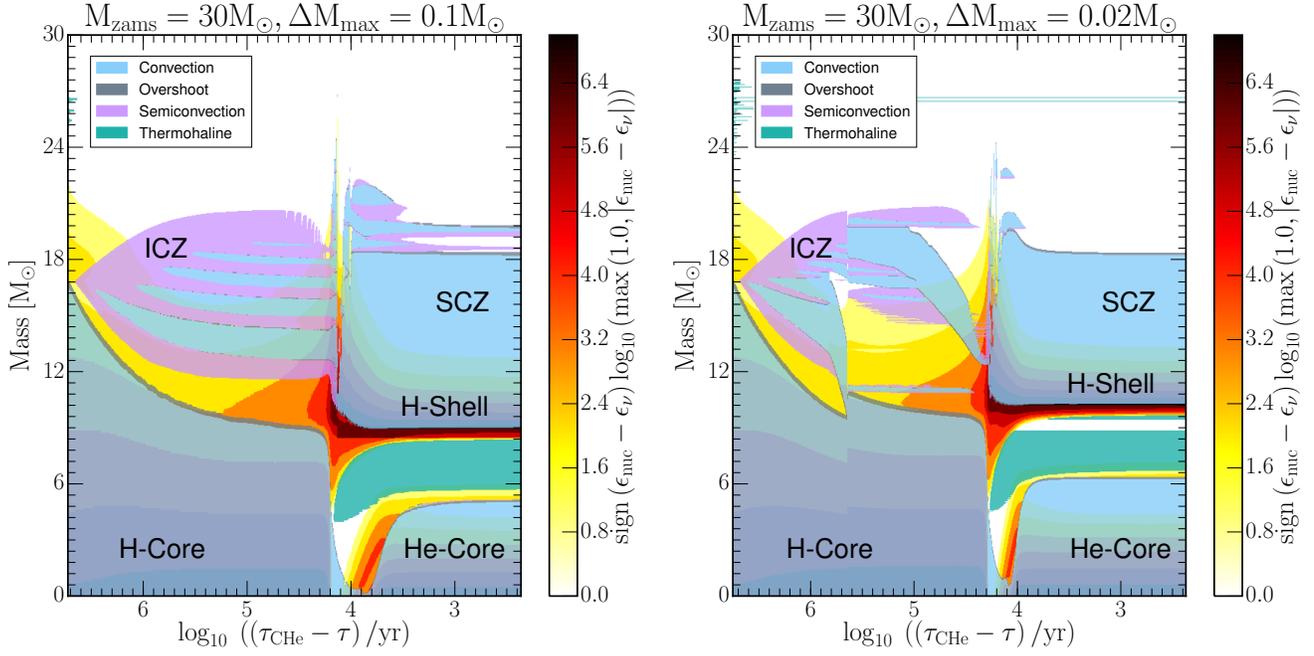

\begin{subfigure}
  \centering{\includegraphics[width=\columnwidth]{{{30msun_nml_res0.1_model_17}}}}
\end{subfigure}
\begin{subfigure}
  \centering{\includegraphics[width=\columnwidth]{{{30msun_nml_res0.02_model_97}}}}
\end{subfigure}
\caption{
Same as Figure \ref{fig:kip_15_low_high} but for a solar metallicity
$M_{\rm{ZAMS}}=30$\,\msun model evolved with the \net{79}\ and without mass loss.
The left panel is for a mass resolution of $\maxdm=0.1$\,\msun,
and the right panel is for a mass resolution of $\maxdm=0.02$\,\msun. As convection never
ceases in the core, the x-axis is set such that it has a similar scale as 
Figure \ref{fig:kip_15_low_high}, and is measured until an arbitrary point during
core helium burning.
}
\label{fig:kip_30_low_high}
\end{figure*}

\begin{figure*}[!htb]
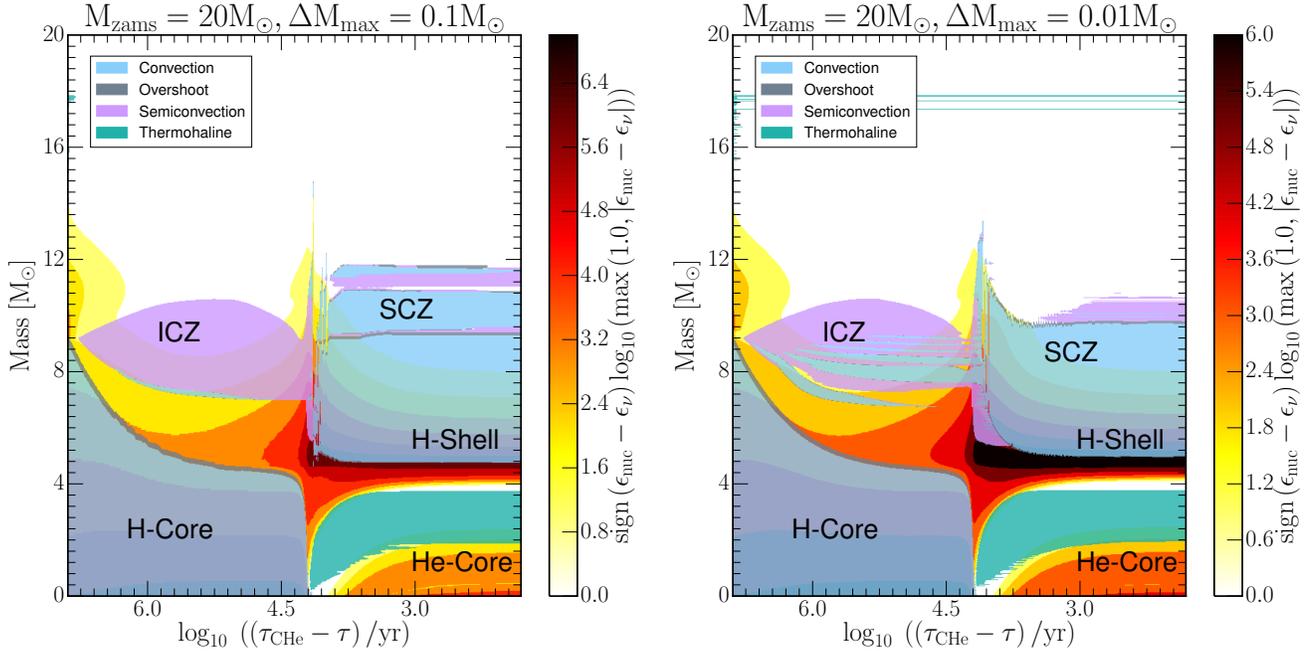

\begin{subfigure}
  \centering{\includegraphics[width=\columnwidth]{{{20msun_ml_res0.1_model_28}}}}
\end{subfigure}
\begin{subfigure}
  \centering{\includegraphics[width=\columnwidth]{{{20msun_ml_res0.01_model_148}}}}
\end{subfigure}
\caption{
Same as Figure \ref{fig:kip_15_low_high} but for a solar metallicity
$M_{\rm{ZAMS}}=20$\,\msun model evolved with the \net{127}\ and mass loss.  
The left panel corresponds to mass resolution of $\maxdm=0.1$\,\msun, while
the Right panel corresponds to a mass resolution of $\maxdm=0.01$\,\msun. The age shown 
on the x-axis is
the time until convection begins in the He core. Note the change in color bar scale
between the left and right panels; lower mass resolution models
release more energy during H-shell burning.
}
\label{fig:kip_20_low_high}
\end{figure*}

The convective core shrinks during core H-burning.
This recession leaves behind a composition gradient ($\mu-$gradient, $\nabla_{\mu}$) and 
lower levels of nuclear burning. Our models form a layered structure 
of semiconvective and convective regions, labeled an intermediate convection 
zone (ICZ) in Figure \ref{fig:kip_15_low_high}, at $\timetillche \approx 5.75$
at the approximate mass coordinate of the maximal extent of the core 
convection zone ($\approx 6$\,\msun) \citep{langer_1985_aa, heger_2000_aa,Hirschi_2004_aa}.
As the mass resolution or the initial mass of the stellar model increases the 
structure and behavior the ICZ changes.
For the $M_{\rm{ZAMS}}=15$\,\msun models, in Figure \ref{fig:kip_15_low_high}, when the 
mass resolution
is increased from $\maxdm=0.1$\,\msun to $\maxdm=0.005$\,\msun, the 
fraction (in mass coordinates) of semiconvection in the ICZ decreases 
while the amount of convection increases. 

Figure \ref{fig:kip_15_low_high} shows that as the mass resolution 
increases the fine structure of the H-core convection zone,
the maximal extent of the H-core extent and the edge of the overshoot region 
are better determined. All of which determine how much fresh H fuel is 
pulled into the core during the MS and how much fuel is available just outside 
the convective core to form the ICZ. At higher resolution less hydrogen is burned
when the ICZ forms, which makes the star more compact due to the decreased burning. 
As less hydrogen has been burned, $\nabla_{\mu}$ is smaller.

Semiconvection occurs when $\nabla_{\rm{ad}} < \nabla_{\rm{T}} < \nabla_{\rm{L}}$ where
$\nabla_{\rm{L}}$ is a function of $\nabla_{\mu}$. If $\nabla_{\mu}$ is large enough
it can maintain a semiconvective region over the entire ICZ. 
If $\nabla_{\mu}$ is small in a local region, then the semiconvective region 
may transition into a convective region \citep{langer_1985_aa}.
This occurs in Figure \ref{fig:kip_15_low_high} in the higher resolution
$\maxdm=0.005$\,\msun panel where $\nabla_{\mu}$ is
smaller, allowing more convection to develop inside the semiconvective
region \citep{mowlavi_1994_aa}. This is most clearly seen as the
bifurcation of the ICZ in the lower resolution $\maxdm=0.1$\,\msun
(left panel), while in the right panel the convection zone is embedded
inside the semiconvective region. Higher mass resolutions also
capture the substructure in the $\nabla_{\rm{L}}$, which become the seeds
for the convective regions to form.

\citet{langer_1985_aa} showed similar variations in the ICZ structure 
by varying the strength of the \semi parameter. As the
efficiency of semiconvection increases, the fine structure
increases. Our \MESA models achieve a similar result by varying the 
mass resolution for a fixed \semi.

A secondary convective zone (SCZ) then forms, at $\timetillche \approx$
4.5, when H-shell burning begins, in both models. Figure \ref{fig:kip_15_low_high}
shows how the structure and behavior of the SCZ changes with mass
resolution. The low resolution model (left panel) has layered
semiconvection/convection that stays outside of the H-shell. \replaced{The
higher resolution  model (right panel) is convection dominated and
propagates into the H-shell burning region}{ In the
higher resolution  model (right panel), the SCZ is dominated by convection which
propagates into the H-shell burning region}. This merger of the SCZ and
the H-shell leads to fresh fuel being injected into the H-shell and
allows the star to live longer. If the star has mass loss, 
then the longer H-burning lifetimes create larger He-cores which will last longer and thus 
increase the amount of mass lost (\S\ref{sec:lifetimes}).

This difference in behavior is due to the SCZ forming within the mass
coordinates previously occupied by the ICZ semiconvective/convective
mixing region. The larger convectively mixed region during the ICZ in
the higher resolution model implies that the $\mu-$gradient left
behind by the receding H-core is reduced or even zero. In turn, this
reduction in the $\mu-$gradient allows convection to form over a
larger mass range in the SCZ. That is, a preceding semiconvective
phase \replaced{to reduce the $\mu-$gradient is not necessary}{is not necessary to reduce the $\mu-$gradient}. 
This is shown in
Figure \ref{fig:kip_15_low_high} by the relative difference in mass
covered by semiconvection and convection at $\timetillche\approx$4.5.

\begin{figure*}[!htb]
  \centering{\includegraphics[width=1.0\textwidth]{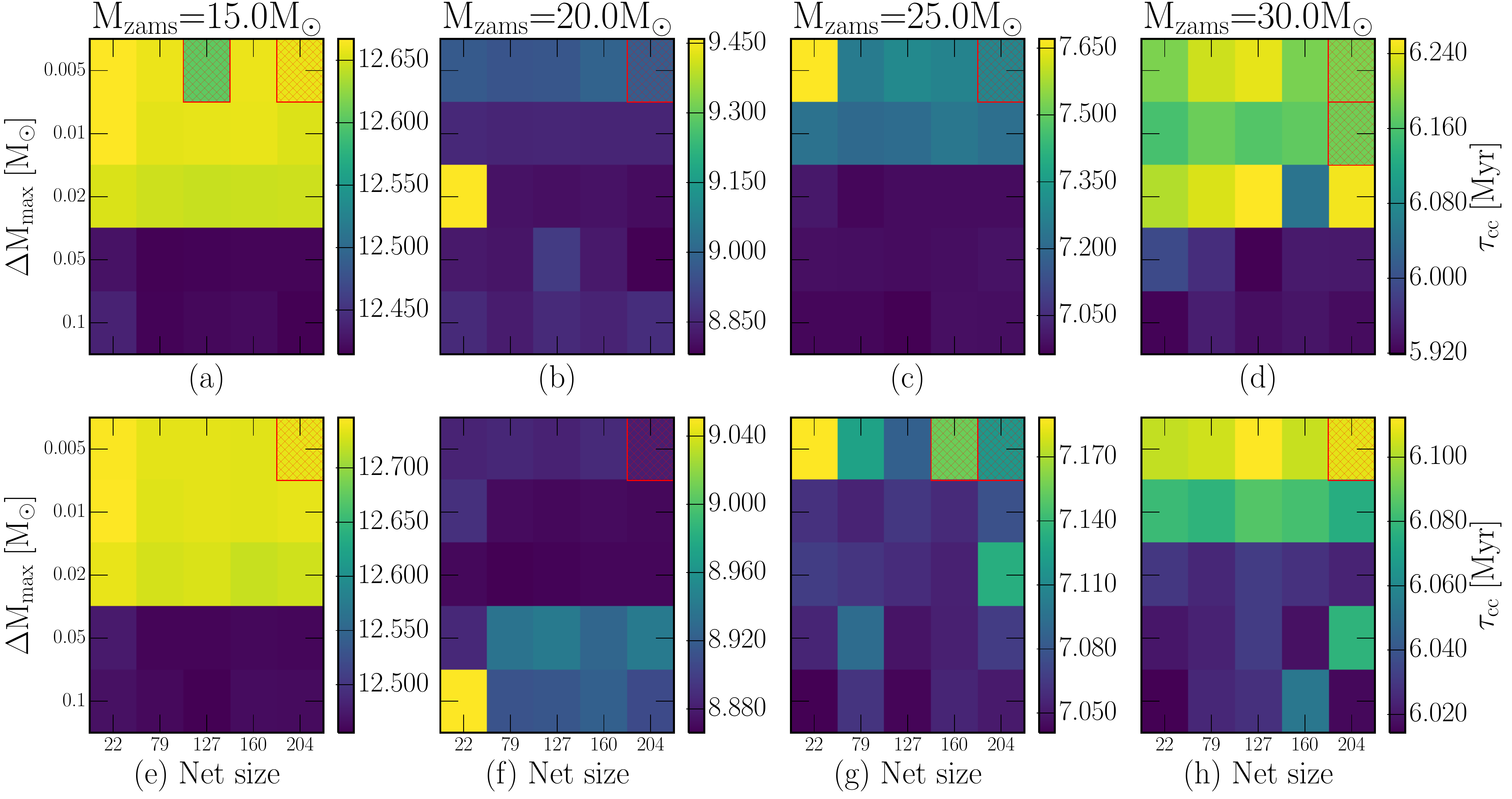}}
\caption{ 
Lifetimes in mega years till core collapse as a function of mass resolution and number of isotopes 
in the nuclear reaction 
network. Top panel: models without mass loss,
Bottom panel: models with mass loss.
Hatching indicate models that did not reach core-collapse, but do reach silicon burning.
}
\label{fig:age_nml}
\end{figure*}

The $\mzams=30$\,\msun models without mass loss are qualitatively similar, and
Figure \ref{fig:kip_30_low_high} shows a representative set of \MESA models.
When the ICZ forms, multiple layers of alternating semiconvection and
convection form \citep{langer_1985_aa}.  At lower mass resolutions
(left panel) this structure begins at $\M \approx 17$\,\msun and propagates
inward to $\M \approx 13$\,\msun.  As the mass resolution increases (right panel),
the convective fingers start at a similar mass coordinate but will
propagate deeper into the star.  Figure
\ref{fig:kip_30_low_high} shows an example of the fingers moving
sufficiently inwards to penetrate the H-burning core. There is a brief
increase in the nuclear burning, as shown by the near step function in
burning at $\timetillche \approx 5.5$ due to the ICZ injecting unburnt
hydrogen into the core. This leads to a longer MS lifetime for these
models. This penetration of the core is a binary process, either the
ICZ penetrates the core or it does not.  Thus, one may expect step
functions in the MS lifetime (\S\ref{sec:lifetimes}).

For the $\mzams=20$\,\msun models with mass loss, Figure
\ref{fig:kip_20_low_high} shows a third possible state. As the mass
resolution increases the SCZ, which formed at the TAMS, decreases in extent. 
The left panel of Figure \ref{fig:kip_20_low_high}
shows that for a relatively coarse mass resolution of $\maxdm=0.1$\,\msun
the SCZ forms multiple convection/semiconvection zones out to $\M
\approx 12$\,\msun, while the right panel shows that for a finer mass
resolution of $\maxdm=0.01$\,\msun the SCZ forms a single convection
zone out to $\M \approx 10$\,\msun. The increased extent of the SCZ
region allows more hydrogen fuel to be mixed inwards towards the core, and
increases the hydrogen mass fraction by a few percent at the H-burning
shell. This allows the H-shell to burn more energetically and for
longer. Thus we would expect the models with lower spatial resolution
to evolve slower to core collapse, but this effect is limited due to the small
change in hydrogen mass fraction.

\subsection{Stellar Lifetimes}\label{sec:lifetimes}

Figure \ref{fig:age_nml} shows the lifetime for the non mass losing
models (top panel) and for models with mass loss (bottom panel). The most striking feature is the sharp
bifurcation of the lifetimes as a function of mass resolution,
most prominently seen in the $\mzams=15$\,\msun and $30$\,\msun models without
mass loss at a spatial resolution of $\maxdm=0.05$\,\msun.  This is a direct consequence
of the differences in behavior of the ICZ and SCZ during the MS and GB
phases. Models where the ICZ penetrate the MS H-core, or where the SCZ
penetrates the H-shell as in the $15$\,\msun models, live up to $\approx
5\%$ longer due to the injection of fresh hydrogen fuel. \added{Models with hatching
indicate those that do not reach core-collapse, though all reach at least Si burning. Thus we include
them in our comparisons up to Si burning.}

Stellar models with and without mass loss have similar ages, with a
trend for stars with mass loss to have shorter lifetimes
\citep[e.g.,][]{eid_2004_aa}.  The spread in age is larger for models
without mass loss than those with mass loss. This is chiefly due to
differences in the size of the timesteps, thus this is a numerical
artifact.  While all models have the same temporal control \var, those
with mass loss require a smaller timestep to keep the variation per
step within the limits specified by \var.  Smaller timesteps allow the
mass lost to be treated more accurately, as the amount of mass loss
per step depends on the stellar parameters at the start of the step,
thus smaller timesteps will better capture fast changes in the star's structure.

Compared to the pre-SN lifetimes from 
\citet{paxton_2011_aa}, \citet{limongi_2000_aa},
\citet{woosley_2002_aa}, and \citet{Hirschi_2004_aa}, we find in general,
the total stellar lifetimes considered agree to within $\approx \pm 5 \%$. 
The models of \citet{woosley_2002_aa} tend to predict longer lifetimes. 
Specifically, for the $\mzams=15$\,\msun model, they predict an $\approx 3 \%$ longer 
lifetime until core-collapse. This value increases to $\approx 7 \%$ for the 
$\mzams=25$\,\msun model. \citet{limongi_2000_aa} predict shorter lifetimes with
$\approx 7 \%$ difference for their $\mzams=25$\,\msun model. The models 
of \citet{paxton_2011_aa} and \citet{Hirschi_2004_aa} agree with our values 
to within $\lesssim 1\%$ for $\mzams=15$ and $20$\,\msun models,  and spread to 
$\pm 2\%$ for $\mzams=25$\,\msun. Each set of pre-SN models considered 
here for comparison use a slightly differing set of input physics. For example, 
our models use a more efficient value for convective overshoot than in the 
\citet{paxton_2011_aa} models and consider larger nuclear reaction networks. 
In addition, each set of pre-SN models considered use different 
(and often unspecified) mass and temporal resolutions.
We find considerable agreement among the total 
stellar lifetimes, with the largest discrepancies occurring at 
$\mzams=25$\,\msun.

Models with the same mass resolution show little variation in the
final age with respect to changes in the number of isotopes in the reaction network. 
For example, the $\mzams=15$\,\msun 
models with and without mass loss in Figure \ref{fig:age_nml}
show uniformity in the stellar age as a function of network size on
either side of the bifurcation point. This is due to efficient mixing
of the composition by convection; it does not make much difference
where or when the SCZ penetrates the H-shell, for once it does, the
H-shell is injected with similar amounts of fresh fuel (see Figure
\ref{fig:kip_15_low_high}).  As the initial mass of the stellar model
increases, the magnitude of the spread in the stellar lifetime
decreases. This is due to smaller variations in when and where the ICZ
penetrates the H-core.

There are outlier models for the smallest network, \netapprox:
the 
$\mzams=20$\,\msun model without mass loss and $\maxdm=0.02$\,\msun,
the $\mzams=25$\,\msun model without mass loss and $\maxdm=0.005$\,\msun, and  
the $\mzams=20$\,\msun model with mass loss and $\maxdm=0.1$\,\msun.
These are caused by the SCZ penetrating deep into the star, from the edge of the H-burning shell
to the radiative region at the center of the star. This injects enough unburnt hydrogen 
to prolong the lifetimes.

The $\mzams=20$\,\msun models without mass loss in Figure
\ref{fig:age_nml} show a lifetime bifurcation at a finer spatial
resolution of $\maxdm=0.005$\,\msun than the
$\mzams=15$\,\msun. In addition, the
$\mzams=20$\,\msun models with mass loss show the bifurcation is
inverted -- increasing resolution decreases the stars lifetime -- due
to the relative extent of the SCZ region discussed for Figure
\ref{fig:kip_20_low_high}.

For the $\mzams=25$\,\msun and 30\,\msun models, Figure \ref{fig:age_nml} shows the
lifetime bifurcation point varies depending on whether there is mass
loss or not. The stars without mass loss bifurcate at $\maxdm=0.01$\,\msun and
$\maxdm=0.02$\,\msun respectively, while the mass losing models bifurcate
at $\maxdm=0.005$\,\msun and $\maxdm=0.01$\,\msun, effectively one level of
resolution higher.  This variation may be due to our grid of
mass resolutions being insufficient to resolve the
bifurcation.

\begin{figure*}[!htb]
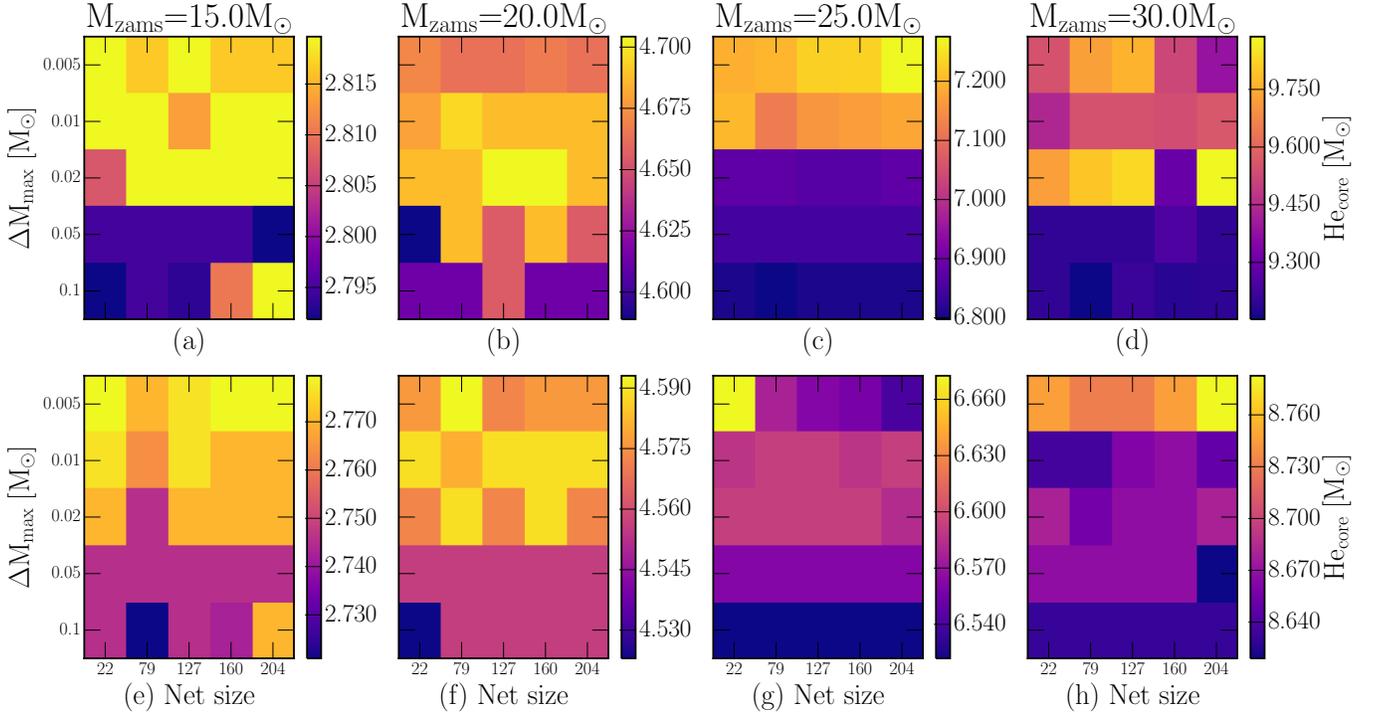

  \centering{\includegraphics[width=1.0\textwidth]{{{core_he_start_all}}}}
\caption{ 
The $^4$He core mass at the formation of the He core, defined as where 
$X(^{1}\rm{H})<0.01$ and $X(^{4}\rm{He})>0.1$.
Top panel: \MESA models without mass loss, Bottom panel: models with mass loss.
}
\label{fig:he_core_startche_nml}
\end{figure*}

\begin{figure*}[!htb]
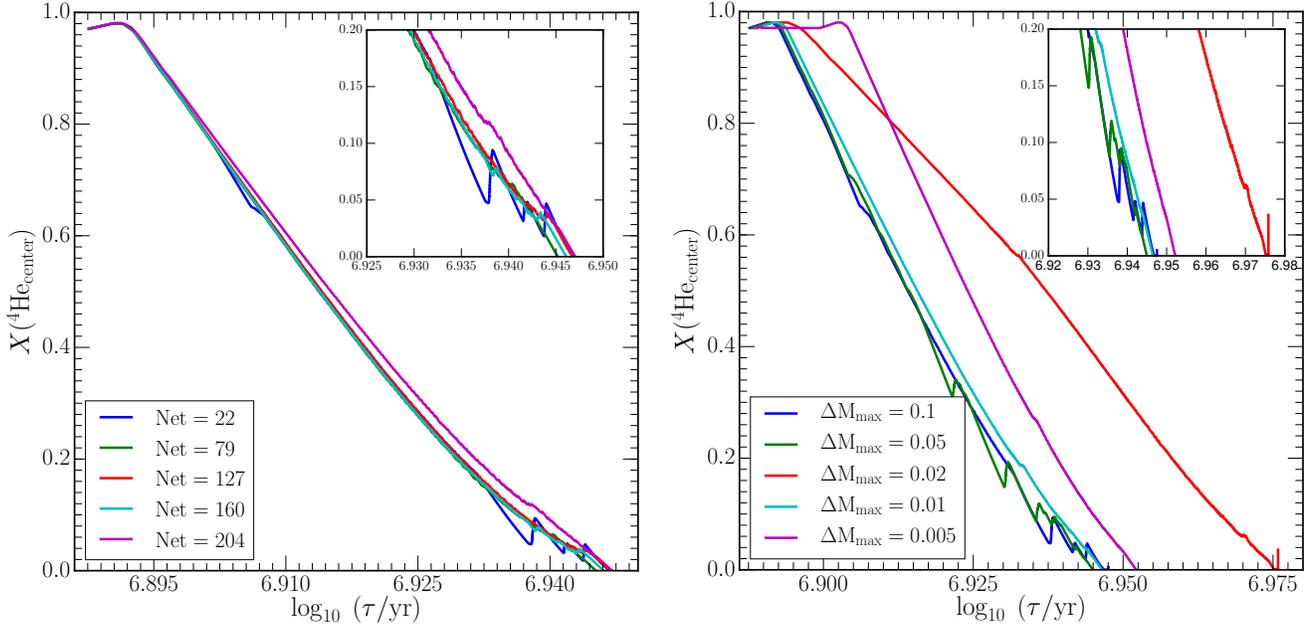

\begin{subfigure}
  \centering{\includegraphics[width=\columnwidth]{{{core_he4_o16_net}}}}
\end{subfigure}
\begin{subfigure}
  \centering{\includegraphics[width=\columnwidth]{{{core_he4_o16_res}}}}
\end{subfigure}
\caption{
Mass fraction of $^4$He as a function of time during core He-burning 
for the $\mzams=20$\,\msun models without mass loss. Left panel: models as a function of number of 
isotopes for a fixed resolution of $\maxdm=0.1$\,\msun. Right panel: models as a function of 
mass resolution for a fixed \netapprox network. The inserts show a zoom in when $X(^4\rm{He})<0.2$.
}
\label{fig:che_centerhe}
\end{figure*}

We encourage careful consideration of the mass resolution when
modeling massive stars with \MESA. Not only to the choice \MESA's \mesh parameter,
which only increases the number of zones if \MESA detects a large
enough spatial gradient, but also the number of zones used throughout
the model. Much of the variation seen in the MS and GB phases is set
by the stellar structure at the start of the MS, which is early enough
that \mesh may not have sufficient time to act. From Figures
\ref{fig:age_nml} and for our choice of \semi, we recommend a
mass resolution of at least $\maxdm=0.01$\,\msun, which equates to
at least $\approx 2000$ mass cells during the pre-MS and MS evolution in order to 
reasonably resolve the ICZ and SCZ features. In addition, we 
recommend temporal controls that limit the timestep as major fuels 
are depleted in the core (see \S\ref{sec:num_res}).

\subsection{Core He-burning}\label{sec:burn_he}

Figure \ref{fig:he_core_startche_nml} shows the He-core mass at the
onset of convective core He-burning as a function of mass resolution
\maxdm \ and the number of isotopes in the nuclear reaction network.
We define the He-core mass
as the location where $X(^4\rm{He})>0.1$ and $X(^1\rm{H})<0.01$ and
measured when the star reaches the base of the GB.
They share many of the same features -- bifurcations and outliers --
discussed for the stellar lifetimes in Figure \ref{fig:age_nml}.  For
the $\mzams=15$\,\msun models without mass loss the initial He-core
mass ranges from $\approx$~2.79 to 2.81\,\msun, while models with mass
loss span \replaced{$\approx 2.72\text{--}2.77$\,\msun}{$\approx$~2.72 to 2.77\,\msun}.  For the
$\mzams=20$\,\msun models without mass loss the initial He-core
mass range between $\approx$~4.6 to 4.7\,\msun, while models with mass
loss span between $\approx$~4.5 to 4.6\,\msun.  Our
$\mzams=25$\,\msun models without mass loss show
initial He-core mass ranges of $\approx$~6.8 to 7.2\,\msun, while models with
mass loss span $\approx$~6.54 to 6.66\,\msun.  The
$\mzams=30$\,\msun models without mass loss show the
initial He-core mass ranges between $\approx$ 9.2 to 9.8\,\msun, while models
with mass loss spans $\approx$~8.63 to 8.78\,\msun.  In general, a longer
core H-burning phase produces a larger He-core mass.
The initial
He-core masses in Figure \ref{fig:he_core_startche_nml} and their range
as a function of mass resolution and number of isotopes in the reaction network
are commensurate with the He-core masses found in numerous studies
\citep[e.g.,][]{nomoto_1988_aa,langer_1991_aa,wellstein_1999_aa,woosley_2002_aa,petermann_2015_aa,choi_2016_aa}.

\begin{figure*}[!htb]
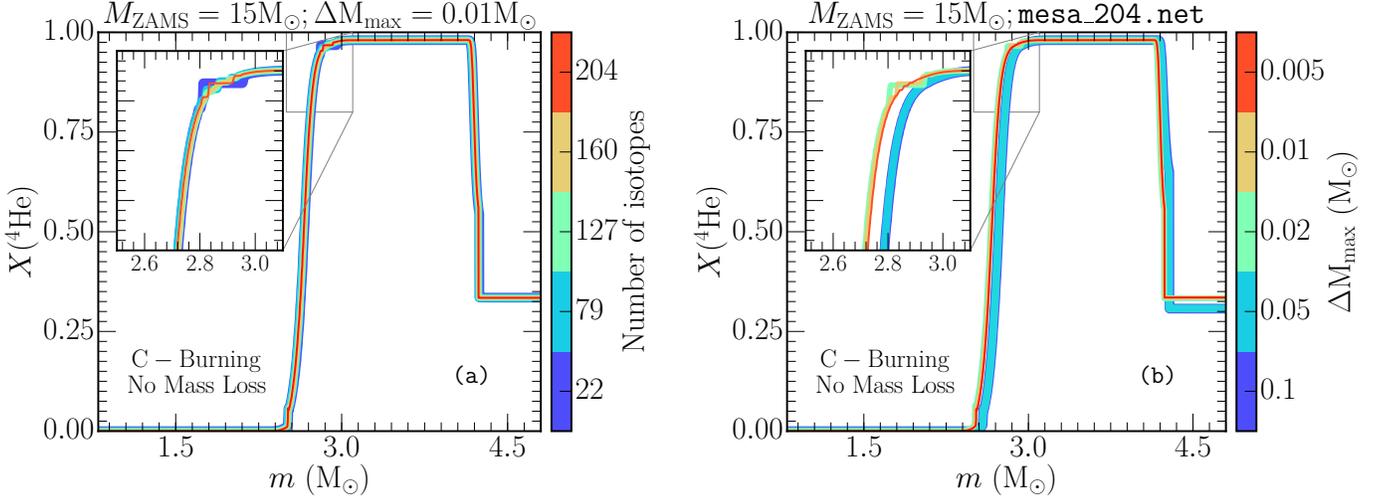

\begin{subfigure}
  \centering{\includegraphics[width=1.05\columnwidth,trim={0.5cm 1.25cm 0.5cm 1cm},clip]{{{15m_m_v_he_abund_v_net_nML_w_inset}}}}
\end{subfigure}
\begin{subfigure}
  \centering{\includegraphics[width=1.05\columnwidth,trim={0.5cm 1.25cm 0.5cm 1cm},clip]{{{15m_m_v_he_abund_v_dq_nML_w_inset}}}}
\end{subfigure}
\caption{
$^{4}$He mass fraction profile at the onset of C-burning for the
$\mzams=15$\,\msun model without mass loss as a function of 
\texttt{(a)} the number of isotopes in the nuclear reaction network and 
\texttt{(b)} the mass resolution \maxdm.
}
\label{fig:m_he_at_cburn}
\end{figure*}

Figure \ref{fig:che_centerhe} shows the evolution of the central
helium mass fraction, \hecenter, during the core He-burning phase of the 20\,\msun models
without mass loss. These central abundances, due to \replaced{convection}{convective mixing},
also represent the mass fraction values over the entire He-burning core.
The core of these stars begin with $\hecenter\approx~1$ and $\ocenter\approx~0$
and, after $\approx$~10$^6$ yr, and end with $\hecenter\approx~0$ and
$\ocenter\approx0.8$, with the remainder being mostly $^{12}$C.

The left panel of Figure \ref{fig:che_centerhe} shows the evolution of
\hecenter \ as a function of the number of isotopes in the nuclear
reaction network, at a mass resolution of $\maxdm=0.1$\,\msun for the
$\mzams=20$\,\msun models without mass loss.
The slopes
of the lines shows the models are evolving at similar rates. The
positive offset in the \net{204} is caused by the core being slightly cooler
at the start of core helium convection, for the \netapprox the temperature is 
$177\times10^6$~K and for \net{204} it is $175\times10^6$~K.  Nevertheless, the \net{204} model
reaches \hecenter$\approx$0 at the same time as the other models. The
spikes in \hecenter \ for the small \netapprox, expanded \replaced{by}{in} the inset
plot, are due to core breathing pulses
\citep[henceforth CBP, see][]{castellani_1985_aa}. These occur when the $\hecenter$ 
value is small such that if the core convection expands slightly outward in
mass, then a small entrainment of unburnt $^4$He leads to a large
increase in the nuclear energy production \citep{straniero_2003_aa}.
For the \netapprox models, the edge of the convection core can move
inwards and outwards $\approx$~0.5\,\msun on timescales of $\approx$~200 yr.  
The models with larger nuclear networks have significantly
smaller CBPs as their convection cores move inwards and outwards
$\approx$~0.05\,\msun, entraining a significantly smaller amount of
additional $^4$He fuel.

The left panel of Figure \ref{fig:che_centerhe}, shows that the \netapprox
almost doubles \hecenter\ during a CBP. The nuclear energy generated
from a CBP is sufficient to expand the stellar envelope as can be seen
as a blue loop in theoretical HR diagrams
\citep[e.g.,][]{constantino_2016_aa}.  The impact of a CBP is also
mirrored in the evolution of \ocenter, and leads to a larger \ocenter
\ at the end of core He-burning and thus a smaller $^{12}$C mass
fraction. All these models at constant \maxdm \ have between 1600-1700
zones and a similar timestep distribution at this evolutionary stage,
thus the differences between CBPs are not due to changes in
resolution.  The presence of CBPs has been suspected as being a
numerical artifact
\citep{caputo_1989_aa,boothroyd_1993_aa,constantino_2016_aa}, and we
show that it can also be due to the choice of the nuclear network.

The right panel of Figure \ref{fig:che_centerhe} shows the evolution of
\hecenter\ as a function of mass resolution for \netapprox. There
is a wide variation in \hecenter. Both the $\maxdm=0.1$\,\msun and $\maxdm=0.05$\,\msun
models have CBPs, with the $\maxdm=0.05$\,\msun models starting their CBPs
at earlier times with a larger \hecenter.  This reinforces the notion
that these \netapprox CBPs are purely numerical artifacts.  Increasing
the resolution above $\maxdm=0.05$\,\msun (which is also necessary for the
ICZ to penetrate the H-core) is sufficient to prevent the CBP from
forming. Models with CBPs have between 1500-2000 zones, while models
without CBP have between 2000-7000 zones.

During core He-burning, the convective core grows, largely because the
mass of the He-core itself grows from the overlying H-burning
shell. This growth of the He-core has two effects. First, as the mass
of the He-core grows so does the core luminosity, but the radius of the
convective core stays nearly the same. This causes the density of the
core to steadily increase as core He-burning proceeds. Second, as the mass of the He-core 
rises the ratio of the gas pressure to the total pressure decreases, 
which favors efficient convection in the core.

\begin{figure*}[!htb]
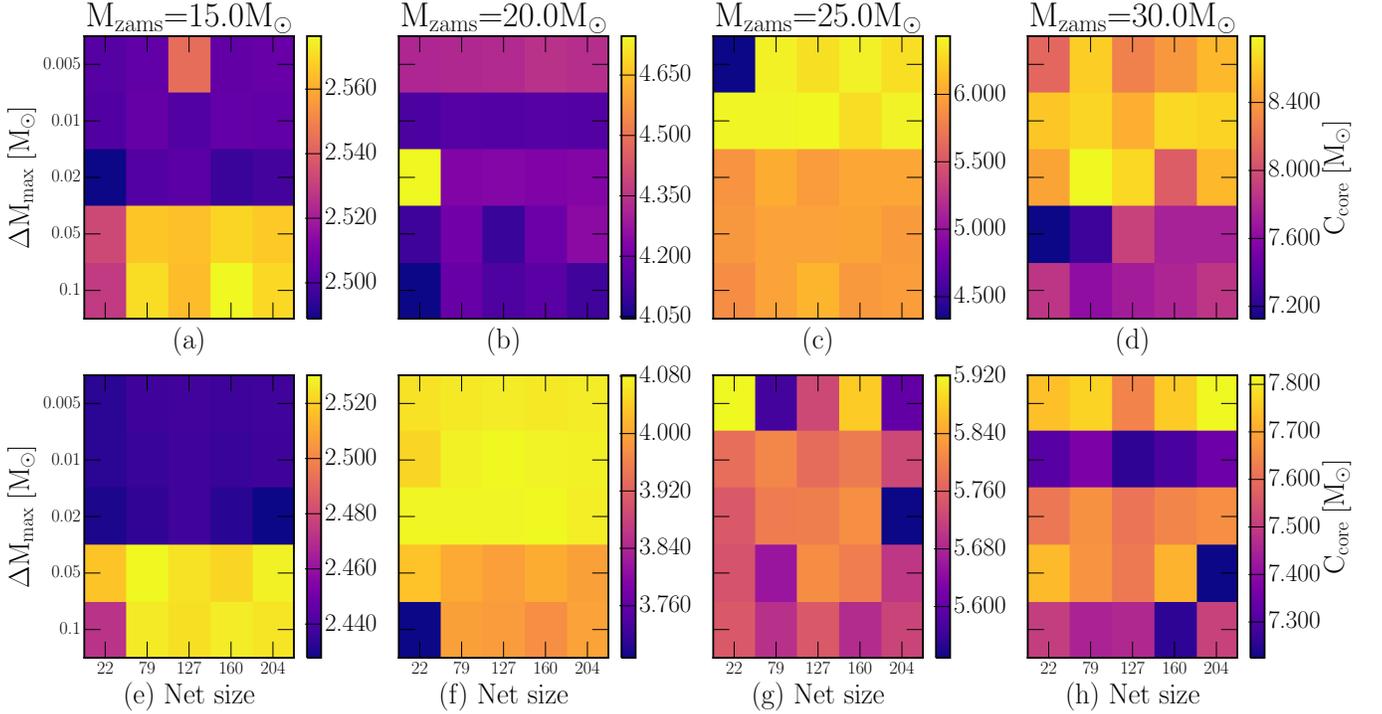

  \centering{\includegraphics[width=1.0\textwidth]{{{c_core_start_all}}}}
\caption{ 
Carbon core mass at carbon ignition as a function of the number of isotopes in the 
network and mass resolution.
Top: for \MESA models without mass loss. Bottom: for models with mass loss.
}
\label{fig:c_core_start}
\end{figure*}

The core prior to C-ignition is composed of the ashes of core
He-burning, mainly $^{12}$C and $^{16}$O in a $\approx$1:4 ratio for the choice of
$^{12}$C($\alpha,\gamma$)$^{16}$O reaction rate from the JINA reaclib
version V2.0 2013-04-02 \citep[see Figure \ref{fig:che_centerhe}, and][]{west_2013_aa}.  
There are also other
isotopes present in trace amounts such as the neutron-rich
$^{21,22}$Ne, $^{25,26}$Mg from processing the ashes of the CNO
elements during He-burning; about 1\% of X($^{20}$Ne) from
$^{16}$O($\alpha,\gamma$)$^{20}$Ne during He-burning; the light
s-process (which we do not follow); and other heavy elements present
since the pre-MS from the initial composition.

Figure~\ref{fig:m_he_at_cburn} shows the $^4$He mass fraction profile
during C-burning for a subset of the
$\mzams=15$\,\msun models. The left panel plots the X($^4$He)
profile as a function of the number of isotopes in the network for the
models without mass loss and a mass resolution of
$\maxdm=0.01$\,\msun.  The inset plot is a zoom in of the inner edge of
the He-shell.  We find the inner and outer mass location of the
He-shell agree within $\approx \pm$ 0.05\,\msun across all network
sizes.  The right panel plots the X($^4$He) profile as a function of
mass resolution for the models without mass loss and that use the largest
reaction network considered, \net{204}. The inset plot is a zoom of
the inner edge of the He-shell.  We find that mass resolutions of
$\maxdm=0.1$ and $0.05$\,\msun do not sufficiently resolve the He-burning
shell locations.  For mass resolutions $\maxdm \lesssim 0.02$\,\msun we
find convergence of the He-shell inner and outer boundaries.

\subsection{C-burning}\label{sec:burn_c}

Figure \ref{fig:c_core_start} shows the C-core mass at the onset of
C-burning as a function of mass resolution and the number of isotopes
in the nuclear reaction network.
The core mass is defined as the mass location where 
$X(^{12}\rm{C})>0.1$ and $X(^4\rm{He})<0.01$,
measured when the core reaches $\log_{10}\,\left(\rm{T_c/K}\right)=8.95$.
These maps share many of the same variations, chiefly
the bifurcations and outliers, that are inherited from core H-burning 
(see Figure \ref{fig:age_nml}).  
For
the $\mzams=15$\,\msun models without mass loss the initial C-core
mass is in the range $\approx2.49\textrm{--}2.57$\,\msun, while models with mass
loss span $\approx 2.43 \textrm{--} 2.52$\,\msun.  For the
$\mzams=20$\,\msun models without mass loss have initial C-core
masses spanning $\approx 4.0 \textrm{--} 4.3$\,\msun, while models with mass
loss span $\approx 3.9 \textrm{--} 4.1$\,\msun.  Our
$\mzams=25$\,\msun models without mass loss show
initial C-core masses of $\approx 5.7\textrm{--}6.3$\,\msun, while models with
mass loss span $\approx 5.6 \textrm{--} 5.8$\,\msun.  The
$\mzams=30$\,\msun models without mass loss have an
initial C-core mass range of $\approx 7.8 \textrm{--} 8.7$\,\msun, while models
with mass loss span $\approx 7.3 \textrm{--} 7.8$\,\msun.

Comparing with Figure \ref{fig:he_core_startche_nml}, the
$\mzams=15$\,\msun models with initially more massive He-core
have less massive C-cores at carbon ignition.  This is due to less
massive ZAMS models having a larger electron degeneracy in the C-core,
a larger \rhoc, and thus an enhanced screening factor for
the $^{12}$C+$^{12}$C reaction rate.  For models more massive than the
$\mzams=15$\,\msun, those with larger He-cores have the largest C-core at carbon
ignition.

\begin{figure*}[!htb]
  \centering{\includegraphics[width=0.475\textwidth]{{{20msun_nml_res0.1_net160_cburn}}}
             \includegraphics[width=0.475\textwidth]{{{20msun_nml_res0.05_net160_cburn}}}}
  \\
  \centering{\includegraphics[width=0.475\textwidth]{{{20msun_nml_res0.02_net160_cburn}}}}
\caption{ 
Kippenhahn plots for the $\mzams=20$\,\msun models without mass loss and the
\net{160} reaction network.  Top left, $\maxdm=0.1$\,\msun; Top right,
$\maxdm=0.05$\,\msun; Bottom, $\maxdm=0.02$\,\msun.  Time is measured in
years until core-collapse. Red/orange regions denote vigorous burning,
while purple denotes significant cooling and electron degeneracy.
Light blue are regions with convection, grey regions with convective
overshoot and purple regions with semiconvection, thermohaline mixing
is not shown for clarity. At the coarsest resolution (top left) carbon burns
under radiative condition with one convective episode. At finer resolution
(top right) carbon ignites under radiative/convective conditions followed
by two convective flashes, while at the higher resolutions (bottom), C-ignition
occurs under convective conditions with three episodes
of convective C-burning.
}
\label{fig:kips_cburn}
\end{figure*}

All the $\mzams = 15$\,\msun models undergo convective C-ignition at the center of
the star followed by a series of 3 convective flashes, where each
additional flash ignites at the approximate maximum mass location of
the previous convective C-flash.  All the $\mzams=25$\,\msun and 30\,\msun models
show C-ignition under radiative conditions that propagates outwards in
mass, followed by a single convective flash. The $\mzams=20$\,\msun models have
a more complex C-ignition that straddles the boundary between
radiative and convective carbon burning \citep{timmes_1996_ac,heger_2000_aa,Hirschi_2004_aa}.

Figure \ref{fig:kips_cburn} shows the differences in core C-ignition
for the $\mzams=20$\,\msun model without mass loss and with the \net{160}
network.  As the mass resolution increases, C-ignition
transitions from radiative conditions at the coarsest resolution
($\maxdm=0.1$\,\msun) to a mixture of convective and radiative conditions at
intermediate resolutions ($\maxdm=0.05$\,\msun), and finally to purely
convective conditions at higher resolutions ($\maxdm=0.02$\,\msun). 
Models with yet finer mass resolution are similar to the
$\maxdm=0.02$\,\msun case. 

Carbon burning in all cases begins at the center.  In the case of the
$\mzams=20$\,\msun model with the coarsest resolution shown in Figure
\ref{fig:kips_cburn}, the central temperature is large enough 
($T_{\mathrm{9,c}} \approx$ 0.8) 
to burn carbon despite the central regions still being
dominated by thermal neutrino cooling (light purple). Only near
$\M \approx 0.3$\,\msun at \timetillcc~$\approx$~1.7 does the nuclear energy
generation rate from burning overtake the thermal neutrino cooling
rate, and thus the color in the Kippenhahn transitions to red.

As the C-ignition conditions transition from
radiative to convective, the time spent in the core C-burning phase
increases from $\approx 45$ yr to $\approx 250$yr.  After carbon
burning ceases, the star has $\approx 1$ year until core-collapse.  As
the mass resolution increases the mass location of the innermost
boundary of the $^4$He shell (see Figure \ref{fig:m_he_at_cburn})
decreases, while the \added{mass location of the} outer boundary of the $^{12}$C shell
increases, by $\approx 0.03$\,\msun. In addition, as the resolution
increases $\log_{10}\,\left(\rm{T_c/K}\right)$ decreases from 8.90 to 8.88 and
$\log_{10}$($\rho_c$/(g cm$^{-3}$)) decreases from 5.36 to 5.20.
The combustion at the top of the final \replaced{convection zone}{carbon flash} is predominantly 
neutron captures, with the neutrons provided by $^{22}$Ne,
onto the ashes of C-burning to form $^{23}$Na,$^{25,26}$Mg and $^{27}$Al. 

Figure~\ref{fig:20m_abund_w_c12} shows the mass fraction profiles 
at core C-depletion for the 
$\mzams=20$\,\msun model with mass loss, $\maxdm=0.01$\,\msun, 
and the \net{204} reaction network.
The composition of the core prior to Ne-ignition is the ashes of core
C-burning, mainly $^{16}$O, $^{20,21}$Ne, $^{23}$Na, $^{24,25,26}$Mg,
$^{26,27}$Al, to a smaller extent $^{29,30}$Si, $^{31}$P, and other
heavy elements present since the pre-MS from the initial composition
and from any s-processing during helium burning
\citep{arnett_1969_ab,arnett_1972_aa,endal_1975_aa,lamb_1976_aa,
arnett_1985_aa,woosley_1995_aa,woosley_2002_aa,Hirschi_2004_aa,bennett_2012_aa}.


\begin{figure}[!htb]
  \centering{\includegraphics[width=\columnwidth,trim={1cm 1.25cm 0cm 1cm},clip]{{{20m_abund_w_c12_v2}}}}
\caption{Mass fraction profiles at core C-depletion for the $\mzams=20$\,\msun 
model with $\maxdm=0.01$\,\msun, \net{204}, and mass loss.
The corresponding evolutionary time is $\timetillcc \approx 0.48$.
}
\label{fig:20m_abund_w_c12}
\end{figure}


\begin{figure*}[!htb]
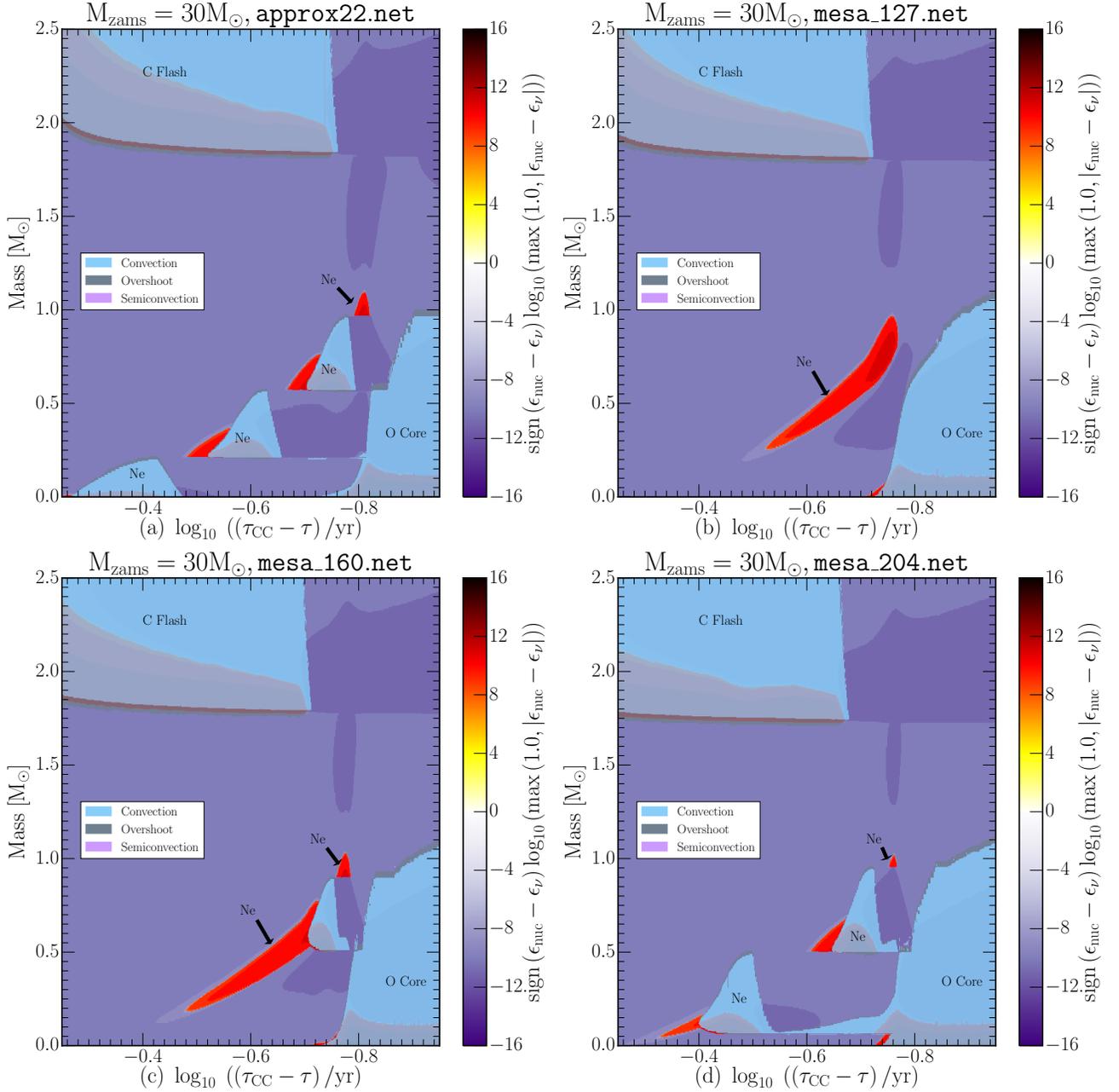

  \centering{\includegraphics[width=0.475\textwidth]{{{30msun_ml_approx_res0.01_neon}}}
             \includegraphics[width=0.475\textwidth]{{{30msun_ml_127_res0.01_neon}}}}
  \\
  \centering{\includegraphics[width=0.475\textwidth]{{{30msun_ml_160_res0.01_neon}}}
	     \includegraphics[width=0.475\textwidth]{{{30msun_ml_205_res0.01_neon}}}}
\caption{
Core Ne-burning Kippenhahn plots of the $\mzams=30$\,\msun models with mass loss and 
$\maxdm=0.01$\,\msun for different nuclear reaction networks. Top left (a), \netapprox; Top 
right (b),
\net{127}; Bottom left (c), \net{160}; Bottom right (d), \net{204}.  Time is measured in
years until core-collapse.
}
\label{fig:kips_neon}
\end{figure*}

\replaced{The \cpost values are not zero}{Carbon is not completely destroyed}, as would be expected for complete
combustion, due to the short amount of time the final convective
flash lasts. The initial flashes in Figure \ref{fig:kips_cburn} last
tens or hundreds of years depending on resolution and number of isotopes
in the reaction network. The final flash lasts only a few
years, which is insufficient to burn all the carbon over the
approximately few solar masses the final convection zone covers. The
flash does not survive longer due to the convection being driven by
the burning at the base of the convection zone. This burning front
attempts to propagate towards the central regions, but encounters 
the ashes from the previous flash. This ash has
insufficient $^{12}$C to sustain the proto-flame and the convection
dies. The situation has similarities to C-burning flames in 
Super Asymptotic Giant Branch models
\citep{denissenkov_2013_ab,chen_2014_aa,farmer_2015_aa}.
The carbon left behind is thus predominantly concentrated at the
top of the CO core, and may ignite again as a carbon shell (see
Figure \ref{fig:kips_si}).

\subsection{Ne-burning}\label{sec:burn_ne}

Neon is the next abundant nucleus to burn in balanced power via the
$^{20}{\rm Ne}(\gamma,\alpha)^{16}{\rm O}$ photodisintegration reaction 
\citep{arnett_1974_aa} at a core temperature of
$T_9$~$\approx$~1.5 and core density of $\log_{10}\rho\approx6.6$ g cm$^{-3}$.  
The net result is that 2($^{20}$Ne)
$\rightarrow$ $^{16}$O + $^{24}$Mg at a rate determined by how fast
$^{20}$Ne captures $\alpha$-particles from the equilibrium set up
between $^{16}$O and $^{20}$Ne
\citep[e.g.,][]{busso_1985_aa,thielemann_1985_aa,chieffi_1998_aa,woosley_2002_aa}.

For the $\mzams=15$\,\msun models, Ne-ignition occurs
predominately at the center which drives a $\approx 0.5$\,\msun
convection zone for approximately one month. As $\mzams$ increases to 20
and 25\,\msun, this initial Ne-flash can be followed by a subsequent
Ne-flash which may or may not drive a convective region once O-burning
has become vigorous.  For the $\mzams=30$\,\msun models the
initial Ne-flash is followed by one or more additional Ne-flashes which
propagate outwards in mass to $\approx 1$\,\msun.

Figure \ref{fig:kips_neon} shows a variety of Ne-burning behaviors for
the $\mzams=30$\,\msun models with mass loss.  The model using
the \netapprox reaction network (upper left) shows a series of outward
moving convective flashes starting at $\timetillcc=-0.2$ and ending at
$\timetillcc=-0.8$.  The model using the \net{127} reaction network
(upper right) shows neon ignites in a weak radiative flash, that lasts
$\approx 1$ month.  For the \net{160} model (lower left) there is an
extensive off-center radiative burning region which transitions into a convective
flash. Finally, the \net{204} model contains a series of off-center
flashes, where a pocket of convection persists from the first ignition 
of neon to the ignition of oxygen.

\begin{figure}[!htb]
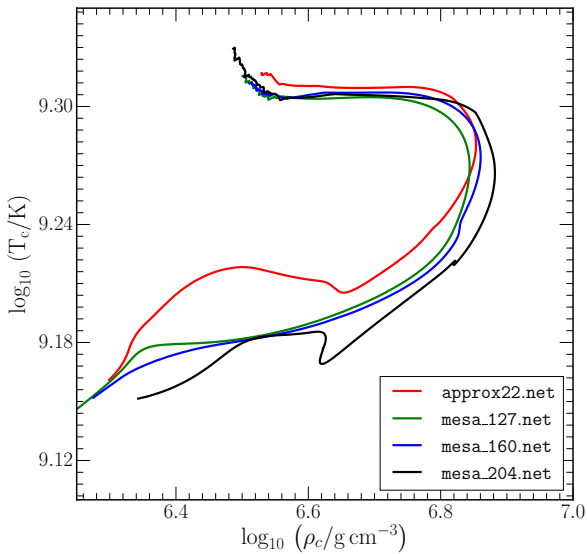

  \centering{\includegraphics[width=\columnwidth]{{{core_rho_temp}}}}
\caption{
Evolution of the central temperature and density for the 
$\mzams=30$\,\msun models shown in Figure \ref{fig:kips_neon}.
}
\label{fig:neon_core_temp_rho}
\end{figure}

Figure \ref{fig:neon_core_temp_rho} shows the evolution of \Tc
and \rhoc for the models shown in Figure
\ref{fig:kips_neon}. The $\mzams=30$\,\msun models start
Ne-burning at $\log_{10}\,\left(\rm{T/K}\right)\approx 9.17$ and 
$\log_{10}$($\rho_c$/(g cm$^{-3}$))~$\approx$~6.6. As
Ne-burning progresses the density and temperature increase until core
O-burning begins with an accompanying creation of a central convection
zone which lowers \rhoc.  The tracks are well converged.
The maximum temperature difference is from $T_{\mathrm{9,c}}$~$\approx$~1.65 
to $\approx$~1.50, 
a $\approx$~8\% difference, with the largest offset in the
\netapprox.  As the number of isotopes in the reaction network
increases, the core is denser for a given temperature.  The larger
networks thus undergo increased neutrino cooling, which is density
dependent but not dependent on the isotopes in the network. This
increased cooling rate is what prevents the neon from vigorously
igniting at the center, and prevents the ignition from driving a
central convection zone.

\begin{figure*}[!htb]
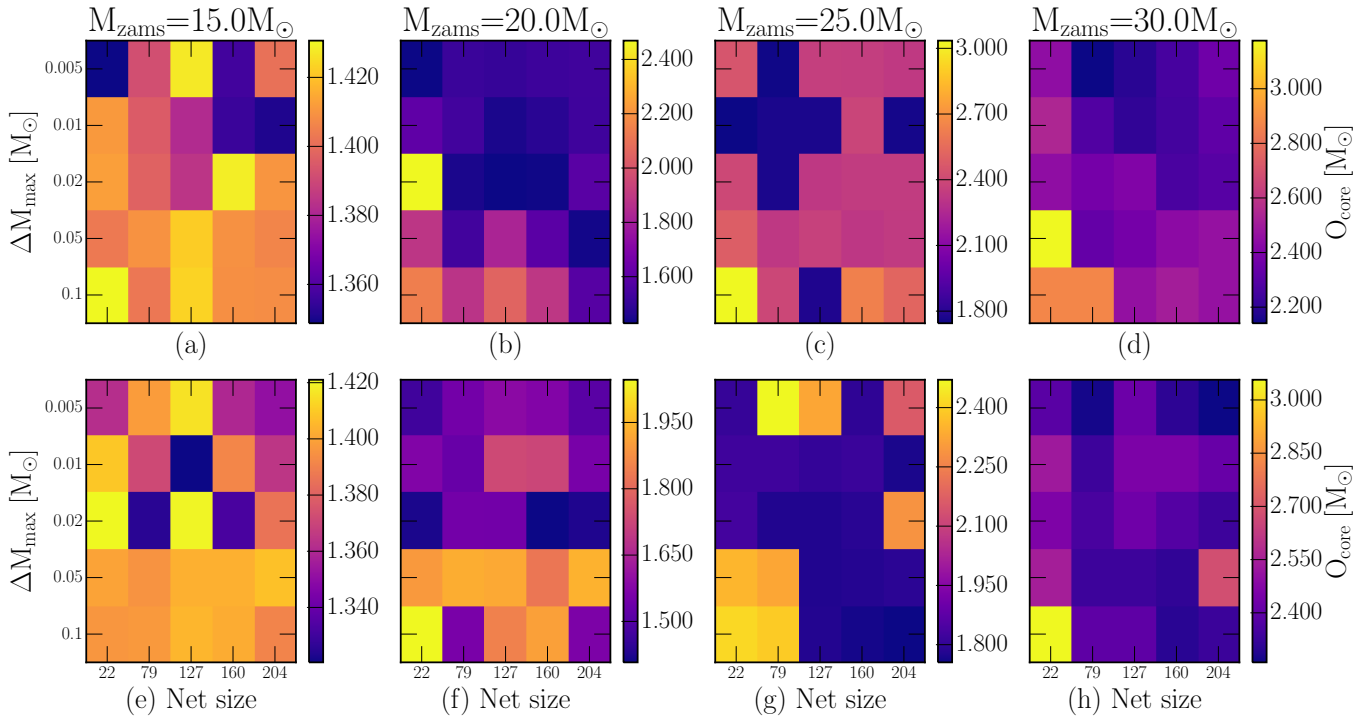

  \centering{\includegraphics[width=1.0\textwidth]{{{o_core_start_all}}}}
\caption{ 
Oxygen core mass at oxygen ignition, defined when X($^{16}\rm{O_c})=0.7$,
as a function of the number of isotopes in the reaction network and mass resolution.
Top: for \MESA models without mass loss. Bottom: with mass loss.
}
\label{fig:o_core_start}
\end{figure*}

These variations in burning structure due to changes in the number of
isotopes in the nuclear reaction network lead to variations in the
post Ne-burning abundance profiles.  For the
$\mzams=30$\,\msun models in Figure \ref{fig:kips_neon}, the
\netapprox model burns neon the longest amount of time and over the
most mass, and thus shows the largest abundance changes. We find
X($^{24}$Mg) is enhanced by a factor $\approx$~6 compared to the
pre-Ne burning mass fraction. Models with the softwired \net{127},
\net{160}, \net{204} networks, show an X($^{24}$Mg) enhancement factor
that increases as the number of isotopes increases; from $\approx$~1.5
for the \net{127} model to $\approx$~2.0 for the \net{204} model, with
the a total of $\approx 0.05\%$ X($^{24}$Mg) left.
The other main product of
Ne-burning, $^{16}$O, increases from $\approx$~0.7 to $\approx$~0.8
in the inner 1.5\,\msun, with the relative change increasing as the
number of isotopes increases similar to the $^{24}$Mg.


\subsection{O-burning}\label{sec:burn_o}

\begin{figure*}[!htb]
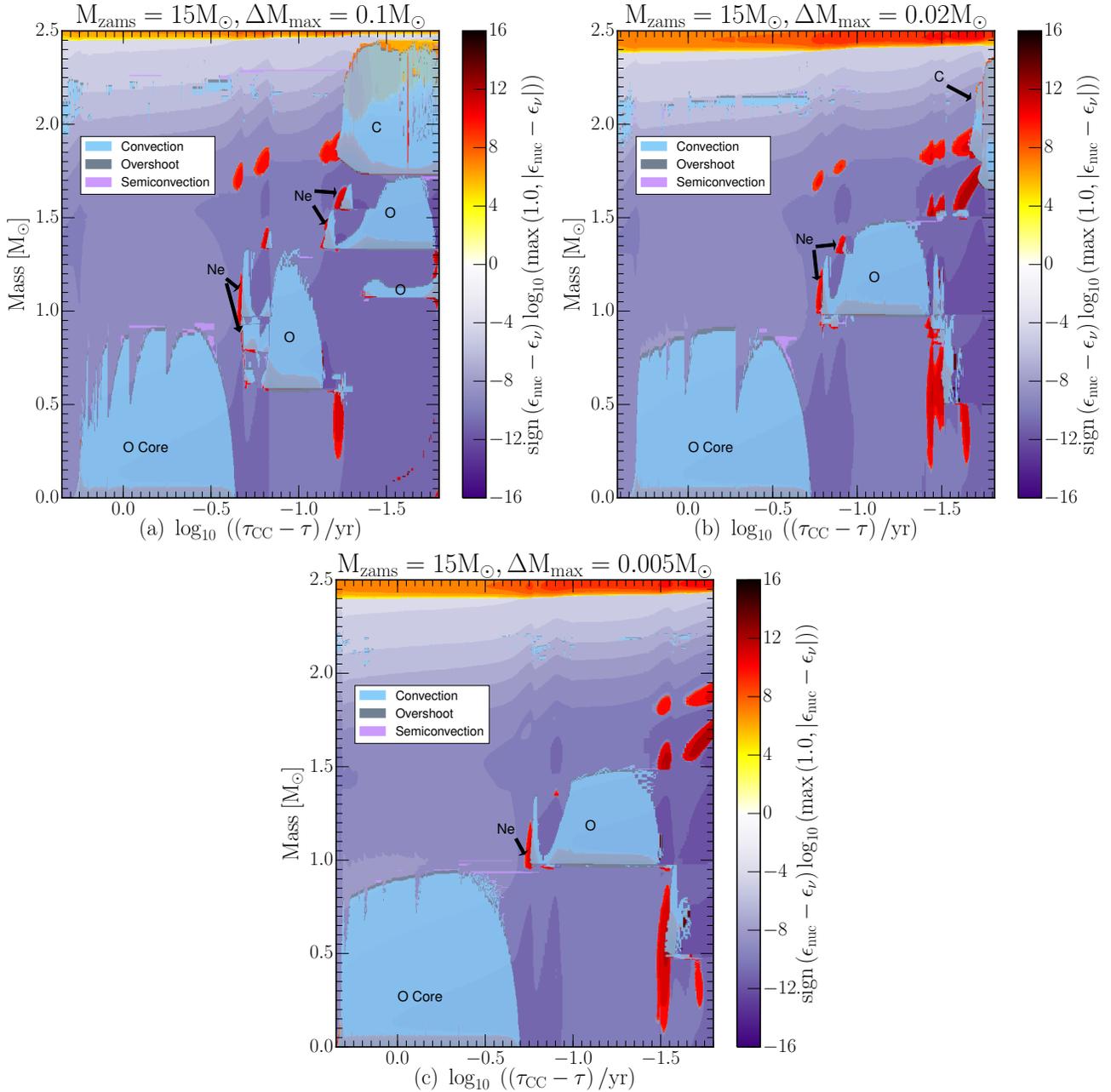

  \centering{\includegraphics[width=0.475\textwidth]{{{15msun_nml_res0.1_oxy}}}
             \includegraphics[width=0.475\textwidth]{{{15msun_nml_res0.02_oxy}}}}
  \\
  \centering{\includegraphics[width=0.475\textwidth]{{{15msun_nml_res0.005_oxy}}}
	     }
\caption{
Core O-burning Kippenhan plots of the $\mzams=15$\,\msun models without mass loss and
\net{204} with various \maxdm. Top left (a), \maxdm=0.1; Top right (b),
\maxdm=0.02; Bottom (c), \maxdm=0.005. Time is measured in
years until core-collapse. Fuels driving selected radiative or convective regions are labeled.
}
\label{fig:kips_oxy}
\end{figure*}

The large $^{16}$O mass fraction, coupled with $^{16}$O+$^{16}$O being
a true fusion reaction and not a photodisintegration driven event like
Ne-burning or Si-burning, ensures that O-burning is a key energetic
and nucleosynthesis stage in the late phases of massive star evolution
\citep{rakavy_1967_aa,arnett_1972_ab,woosley_1972_aa,woosley_1995_aa,eid_2004_aa,sukhbold_2016_aa}.

In 1D stellar evolution instruments such as \MESA, convective mixing
and energy transport is modeled using MLT \added{in a time dependent manner}
\citep{vitense_1953_aa,bohm-vitense_1958_aa,cox_1968_aa}, which is
usually tuned to reproduce solar properties
\citep{asplund_2009_aa}.  However, thermal neutrino cooling speeds up
O-burning to such an extent that the evolutionary timescales are close
to the sound crossing time, so that direct, multi-dimensional
compressible numerical hydrodynamics must be applied for maximum
fidelity to the underlying physics.  Such studies show nuclear burning
tightly couples to turbulent convection so that fuel is consumed in
chaotic episodes \citep{bazan_1998_aa, asida_2000_aa,meakin_2007_aa}.
Core O-burning and shell O-burning are dominated by large scale modes
of fluid flow, which are of such low order that they do not cancel to
a smooth spherical behavior
\citep{couch_2015_aa,chatzopoulos_2016_aa,jones_2016_aa,muller_2016_aa}.
Moreover, 3D simulations of O-burning suggest that MLT gives an
incomplete representation of stellar convection
\citep{arnett_2015_aa}.  Nevertheless, 3D simulations of core
O-burning and beyond are resource intensive and to date have been run
primarily to address hydrodynamic and transport aspects. Such 3D
simulations have not yet been run to assess the detailed
nucleosynthesis.

Figure \ref{fig:o_core_start} shows the O-core masses at the start of
core oxygen convection. 
We define this as the location where $X(^{16}\rm{O})>0.1$ and $X(^{12}\rm{C})<0.01$ and
measured when $X(^{16}\rm{O_c})=0.7$.
In general, as the mass resolution increases the
O-core mass decreases, with the $\mzams=15$\,\msun models having
the smallest spread. This is due to the 15\,\msun models having a more
consistent behavior between the start of carbon and the end of neon burning. 
The O-core masses
range from $1.35-1.4$\,\msun for the 15\,\msun models, $1.5-2.5$\,\msun for the
 $\mzams=20$\,\msun models, $1.8-3.0$\,\msun for the 25\,\msun, and 
$2.25-3.15$\,\msun for the $\mzams=30$\,\msun models. 
The spread in the O-core mass is smaller for the mass losing models 
compared to models without mass loss. Much of the
fine structure seen in Figures \ref{fig:he_core_startche_nml} and 
\ref{fig:c_core_start} has been erased due to the variations in carbon
and neon burning. Panels \texttt{(e)} and \texttt{(f}) still show a clear bifurcation
however, with the lowest resolution models having the largest O-core masses.

\begin{figure}[!htb]
  \centering{\includegraphics[width=\columnwidth,trim={1cm 1.25cm 0cm 1cm},clip]{{{15m_abund_post_o_burn}}}}
\caption{
Mass fraction profiles of abundant isotopes at core O-depletion for the $\mzams=15$\,\msun 
model with $\maxdm=0.01$\,\msun, \net{204}, and mass loss.
The corresponding evolutionary time is $\timetillcc \approx -1.95$.
}
\label{fig:15m_abund_o_depl}
\end{figure}

\begin{figure*}[!htb]
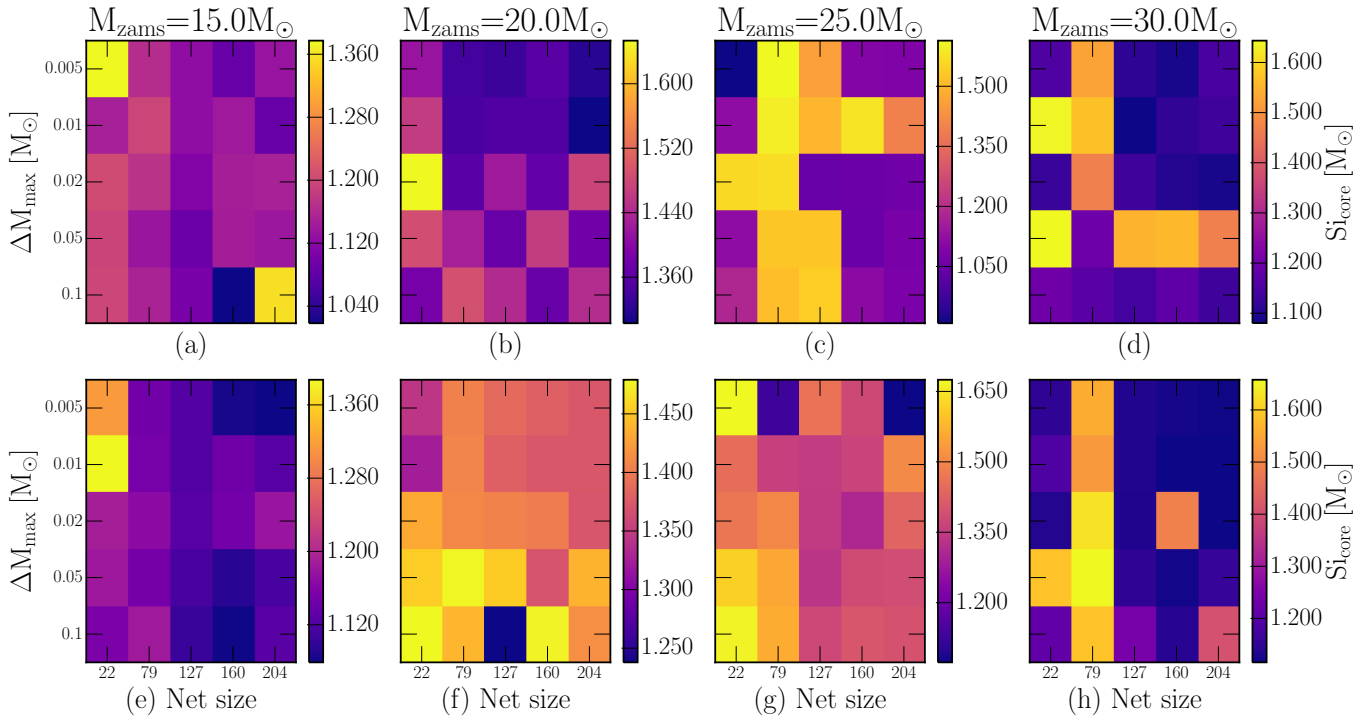

  \centering{\includegraphics[width=1.0\textwidth]{{{core_si_start_all}}}}
\caption{ 
Silicon core mass at silicon ignition, defined when 
$\log_{10}\,\left(\rm{T_c/K}\right)=9.5$,
as a function of the number of isotopes in the reaction network and mass resolution.
Top: for \MESA models without mass loss. Bottom: with mass loss.
}
\label{fig:si_core_start}
\end{figure*}

Figure \ref{fig:kips_oxy} shows the evolution during core O-burning
for the $\mzams=15$\,\msun models without mass loss, the
\net{204} reaction network, and mass resolutions of $\maxdm=0.1$\,\msun,
$0.02$\,\msun, and $0.005$\,\msun.  In each model O-ignition occurs at the
center at $\timetillcc\approx0.4$ and O-burning continues
until $\timetillcc\approx -0.5$. O-burning is concentrated over
$\approx0.1$\,\msun of the core, while a convective region is generated
out to $\approx1.0$\,\msun.  For $\maxdm=0.1$\,\msun, the convective region
undergoes a series of contractions and expansions (in mass
coordinates), analogous to the CBP seen during core He-burning.
During a core oxygen breathing pulse (OBP), when the convective
region reaches its maximum extent (in mass coordinate) and ingests
$^{16}$O, there is a brief increase in the extent of the
central burning region and a corresponding increase in the energy
released in the core. These OBPs have a smaller effect on the $^{16}$O
abundance compared to CBPs, due to the weaker density dependency of
O-burning compared to He-burning. These OBPs are present for all four
masses considered in this work and unlike the helium counterparts
these OBPs are independent of the nuclear network used.  As the mass
resolution increases the OBP behavior changes,
the number of OBPs decreases, the maximum mass the
OBPs extend out to decreases, and they have a shorter lifetime. This suggests
careful treatment of the mass resolution is needed during this stage of a
star's evolution.

Once core O-burning ceases the core contracts until Si-ignition.
During this contraction phase the $\mzams=15$\,\msun models
undergo a series of off-center flashes. These mimic the ignition at
the center by having a series of small neon flashes which lead to a
larger oxygen flash.  The residual carbon left behind from carbon
burning (Figure \ref{fig:20m_abund_w_c12}) can also now ignite as an
additional flash near the edge of the CO core.  For the
$\mzams=15$\,\msun models in Figure \ref{fig:kips_oxy}, as the
numerical resolution increases the number of these flashes decreases,
the $\maxdm=0.1$\,\msun has three oxygen flashes and a number of
smaller neon flashes, while the $\maxdm=0.02$\,\msun and
$\maxdm=0.005$\,\msun has only one oxygen flash and a variable number
of neon flashes.  As the initial mass of the star increases the number
of neon/oxygen flashes decreases and the carbon flash ignites later,
once silicon burning has commenced.

Dominating the ashes of O-burning are Si, S, Ar, Ca in
roughly solar proportions.  Figure~\ref{fig:15m_abund_o_depl} shows
the isotopes $^{28}$Si, $^{32,33,34}$S, $^{35,37}$Cl, $^{36,38}$Ar,
$^{39,41}$K, and $^{40,42}$Ca are present in significant quantities
\citep[e.g.,][]{woosley_1995_aa,limongi_2000_aa,limongi_2003_aa,sukhbold_2016_aa}
for the $\mzams=15$\,\msun model with mass loss and $\maxdm=0.01$\,\msun.
During O-burning several isotopes begin to be made as radioactive
progenitors of stable isotopes, for example, $^{37}$Cl as $^{36}$Ar and $^{41}$K as
$^{41}$Ca. The heaviest isotopes ($A \gg 50$) that were
present in the initial composition at birth, along those heavy
isotopes made from the s-process during He-burning and C-burning
(which we do not follow), begin to be destroyed by photodisintegration
reactions; essentially melting them into the Fe-group.  O-burning also
features the first appearance of quasi-equilibrium clusters
\citep{bodansky_1968_aa, woosley_1973_aa,woosley_1992_aa,hix_1996_aa,meyer_1998_aa}.
These
clusters grow to encompass more isotopes as core O-burning
proceeds. Weak interactions such as $^{33}$S(e$^{-}$,$\nu_e$)$^{33}$P,
$^{35}$35(e$^{-}$,$\nu_e$)$^{35}$S, and
$^{37}$Ar(e$^{-}$,$\nu_e$)$^{37}$Cl decrease the \Ye significantly during
O-burning, especially at O-depletion \citep[e.g.,][]{heger_2001_aa}. For example, 
we find \Yec~$\approx$~0.4778  at core O-depletion 
for the model shown in Figure~\ref{fig:15m_abund_o_depl}, which has decreased from
$\Yec=0.499$ prior to O-burning.

\subsection{Si-burning}\label{sec:burn_si}

Silicon burning is the last exothermic burning stage and produces the
Fe-peak nuclei. Due to Coulomb repulsion, it is improbable that two
$^{28}$Si nuclei will fuse to $^{56}$Ni.  Instead, a
photodisintegration driven rearrangement of the abundances takes
place, originating from equilibria established among individual
reactions with their reverse reactions
\citep{bodansky_1968_aa}.  When such equilibria happen among many
reactions, the material reaches an equilibrium state where nuclei merge
into clusters. Units of interaction are no longer nuclei, but the
clusters themselves, which adapt their properties according to the
local thermodynamic conditions
\citep{woosley_1973_aa,woosley_1992_aa,hix_1996_aa,meyer_1998_aa,the_1998_aa,hix_1999_aa,magkotsios_2010_aa}.

In general, not all reactions are in equilibrium. Consequently, this
state is named quasi-static equilibrium (QSE).  The special case where
all strong and electromagnetic reactions are balanced by their reverse
reactions is called nuclear statistical equilibrium (NSE), because all
mass fractions may be described in terms of statistical properties of
excited nuclear states (e.g., temperature dependent partition
functions) and nuclear structure variables
\citep[masses and $Q$ values; ][]{clifford_1965_aa,hartmann_1985_aa,nadyozhin_2004_aa,jordan_2004_aa,seitenzahl_2008_aa}.

Weak interactions are excluded from these definitions since
for conditions relevant to hadronic physics, they never attain
equilibrium \citep[e.g.,][]{heger_2001_aa,arcones_2010_aa}.  Hence,
equilibrium notions are related only with strong and electromagnetic
interactions. In practice, there is either one cluster in NSE or QSE,
or two QSE clusters, one for the Si-group and one for the Fe-group
nuclei. Even the main products of Si-burning depend quite sensitively on
small changes of the electron fraction, temperature and density.
Although the physics of QSE/NSE is relevant for our models, we reiterate 
that no QSE or NSE approximations are made in our models.

During Si-burning, as for O-burning, the energetics of nuclear burning
tightly couples to turbulent convection, and must be modeled with 3D
simulations to assess the fidelity of the approximations made by 1D
stellar evolution instruments
\citep[e.g.,][]{arnett_2011_ab,couch_2015_aa,muller_2016_aa,jones_2016_aa}.
For example, \citet{couch_2015_aa} found that during the final minutes
of Si-burning in a massive star that the fluctuating, peak convective
speeds of $\approx$~200 km s$^{-1}$ were common and that the speed of
the convection increases to $\approx$~500 km s$^{-1}$ as collapse
approaches and the core contracts. These speeds are not negligible
relative to nominal infall speeds of 1000 km s$^{-1}$ for our
core-collapse initial models. Distilling the essential features of 3D
simulations into models suitable for 1D stellar evolution instruments
remains a challenge for future investigations.

Figure \ref{fig:si_core_start} shows the Si-core mass at Si-ignition.
This is defined as the location where $X(^{28}\rm{Si})>0.1$ and $X(^{16}\rm{O})<0.01$,
measured when $\log_{10}\,\left(\rm{T_c/K}\right)=9.5$.
As the mass resolution increases there is a general trend, with
substantial scatter, for the Si-core \added{mass} to decrease, with the
$\mzams=15$\,\msun and 20\,\msun models having the smallest
spread.  The Si-core masses range from $1.05-1.35$\,\msun for the
15\,\msun models, $1.3-1.6$\,\msun for the $\mzams=20$\,\msun
models, $1.0-1.7$\,\msun for the 25\,\msun \added{models}, and $1.1-1.6$\,\msun for the
$\mzams=30$\,\msun models.  The spread in the Si-core mass is
about the same for the mass losing models compared to models without
mass loss. Nearly all of the fine structure seen in Figures
\ref{fig:he_core_startche_nml} and \ref{fig:c_core_start}, and even the
coarser structure in Figure \ref{fig:o_core_start}, has been largely 
erased before the onset of Si-burning.
The bulk of the silicon core is built during the core oxygen burning phase,
but a number of short flashes, which can be seen in 
Figure \ref{fig:kips_oxy} at \timetillcc$\approx -1.5$, can occur before core Si-burning
commences. These flashes introduce an additional fine structure into the 
core composition and hence the location of the silicon core boundary.

\begin{figure*}[!htb]
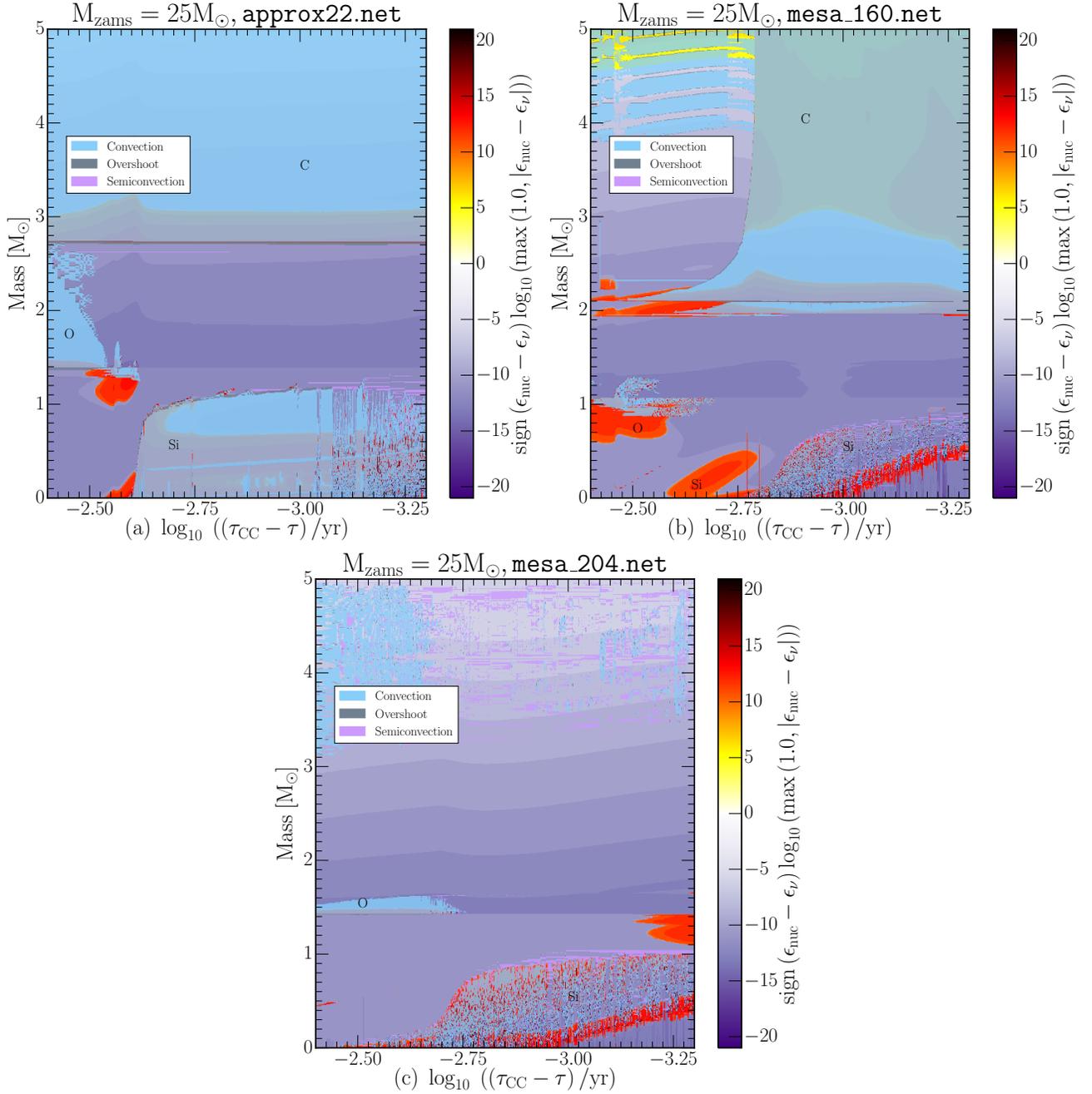

  \centering{\includegraphics[width=0.475\textwidth]{{{25msun_ml_res0.005_approx_si}}}
             \includegraphics[width=0.475\textwidth]{{{25msun_ml_res0.005_160_si}}}}
  \\
  \centering{\includegraphics[width=0.475\textwidth]{{{25msun_ml_res0.005_205_si}}}
	     }
\caption{
Core Si-burning Kippenhan plots of the $\mzams=25$\,\msun models with mass loss and
$\maxdm=0.005$\,\msun for various reaction networks. Top left (a), \netapprox; Top right (b),
\net{160}; Bottom (c), \net{204}. Time is measured in
years until core-collapse. Fuels driving selected radiative or convective regions are labeled.
}
\label{fig:kips_si}
\end{figure*}

\begin{figure*}[!htb]
\begin{subfigure}
  \centering{\includegraphics[width=\columnwidth]{{{si_abun_approx_191}}}}
\end{subfigure}
\begin{subfigure}
  \centering{\includegraphics[width=\columnwidth]{{{si_abun_205_195}}}}
\end{subfigure}
\caption{
Mass Fraction profiles for the two $\mzams=25$\,\msun models shown in 
Figure \ref{fig:kips_si} at $\timetillcc=-3.0$ for (left panel) the 
\netapprox reaction network and (right panel) the \net{204} reaction network.
}
\label{fig:abun_si}
\end{figure*}

Figure \ref{fig:kips_si} shows the evolution during core Si-burning
for $\mzams=25$\,\msun models with mass loss and
$\maxdm=0.005$\,\msun for the \netapprox, \net{160} and \net{204} reaction
networks.  Silicon ignites at the center \replaced{with $\approx 1$}{within one} day 
\replaced{until}{of}
core-collapse.  The initial phase of transforming $^{28}$Si is
sensitive to the thermodynamic conditions and electron captures change the
\Ye continuously during the buildup of the QSE clusters
\citep[e.g.,][]{thielemann_1985_aa}.  

For the \netapprox model in Figure \ref{fig:kips_si},
Si-burning propagates outwards to $\approx 1$\,\msun driving a
convection region over this mass range. The Si-burning region then
recedes to $\approx 0.5$\,\msun while leaving the convection zone
behind.  For the models using the \net{160} and \net{204} reaction networks 
there is a central
burning region occupying $\approx 0.2$\,\msun.  Outside this core
region is a mixed region comprised of many small pockets of
radiative/convective burning.  The larger and more simply-connected
convective region for the \netapprox allows a larger fraction of the
$^{28}$Si to be converted into iron-group elements and to homogenize
the spatial distribution of those iron-group isotopes.  The final \Ye
in the \net{160} and \net{204} models is determined to within  
$\approx$~2\% before the iron-core begins to collapse,
making shell Si-burning a key evolutionary state for determining the
\Yec and the iron-core structure \citep[e.g.,][]{heger_2001_aa}.


Figure \ref{fig:kips_si} also shows a change in the qualitative
behavior of the carbon shell. Carbon burns at the base of a convective
region in the $\mzams=25$\,\msun models using the \netapprox
reaction network, at $\approx 3.0$\,\msun and $\timetillcc\ge -2.25$,
and has been since core O-burning.  In contrast, the models using the
\net{160} reaction network show C-burning in two distinct layers, one
layer at $\approx 2.0$\,\msun and a second layer at
$\approx 2.4$\,\msun, once Si-burning becomes vigorous.
The $\mzams=25$\,\msun models using the \net{204} reaction
network do not ignite a carbon shell, chiefly due to an oxygen flash
at $\approx 1.5$\,\msun and $\timetillcc=-2.25$ where a previous
convection region extended out to $\approx 5$\,\msun.  This region
partially burnt C, Ne, and O leaving behind the mixed
convection/semiconvection due to small scale composition gradients,
as seen in Figure \ref{fig:kips_si} for \net{204}.

Figure \ref{fig:abun_si} shows the composition of the cores at
$\timetillcc=-3.0$ for the $\mzams=25$\,\msun models in Figure
\ref{fig:kips_si}. The left panel corresponds to the stellar model
using the \netapprox reaction network and the right panel corresponds
to the models using the \net{204} reaction network. First, we note that
the \netapprox model has a significantly smoother composition profiles
in the inner $\approx 1.5$\,\msun due to the more simply connected
convective region it exhibits (see Figure \ref{fig:kips_si}) relative
to the multiple-connected convective regions exhibited by the
\net{204} model. Second, the core in the \netapprox model is
predominately $^{54}$Fe, while at the same time until core-collapse
the \net{204} model is a mix of $^{54,56}$Fe and $^{52}$Cr. At the
center of the \net{204} model the composition is $\approx$ 68\%
$^{56}$Fe, $\approx$ 18\% $^{52}$Cr and $\approx$ 2\% $^{54}$Fe, with
the remainder in various iron group elements.  Third, there is a
significant impact of the different shell O-burning behavior in the
2-5\,\msun range, before silicon burning commences (see Figure \ref{fig:kips_si}). 
The model using the
\net{204} reaction network has a smaller $^{16}$O mass fraction that
extends deeper into the core than the model using the \netapprox
reaction network. The O shell has also been polluted by the products
of oxygen burning as well as being depleted in $^{12}$C.

\subsection{Core-Collapse}\label{sec:cc}

When the Fe-core reaches its finite-temperature Chandrasekhar mass 
\citep{baron_1990_aa}, 
\begin{equation}
\begin{split}
& M_{\rm{Ch, eff}} \simeq 5.76 \ \Ye^2 \ \times \\ 
& \left [  1 - 0.057 + \left ( \frac{s_e}{\pi \Ye} \right )^2 + 1.21 \left ( \frac{s_e}{A} \right ) \right ]
\enskip \,\msun ,
\label{eq:mch}
\end{split}
\end{equation}
electron capture and photodisintegration drive the collapse of the Fe-core
with the largest infall speeds being reached near the outer edge of the Fe-core.
Here, $s_e = S_e/(N_A k)$ is a dimensionless average electron
entropy and $A$ is an average atomic weight.
We terminate our \MESA models when any mass coordinate within the Fe-core
exceeds an inward velocity of 1000 km s$^{-1}$. The structural and nucleosynthesis 
properties at this key evolutionary point 
\citep[e.g.,][]{woosley_1995_aa,Hirschi_2004_aa,dessart_2010_aa,chieffi_2013_aa}
can have significant effects on the 
subsequent explosion (if achieved) and nucleosynthesis
\citep[e.g.][]{janka_2012_aa,dolence_2013_aa,ott_2013_aa,couch_2014_aa,
pejcha_2015_ab,couch_2015_aa,muller_2016_aa,jones_2016_aa,bruenn_2016_aa}.


Figure~\ref{fig:fe_core_cc} shows the final Fe-core mass (which we define below) at
core-collapse,\replaced{which we define}{taken} when the radial infall speed reaches $v>1000$
\kmpers, as a function of the mass resolution \maxdm\ and the number of
isotopes in the reaction network.  For a fixed \maxdm, in most cases,
the final Fe-core mass \added{for models} using the \net{127}, \net{160}, and \net{204}
reaction networks agree to within $\approx 0.05$\,\msun.  In some
cases, the \netapprox and \net{79} reaction networks \added{models} produce a more
massive Fe-core ($\approx$~0.07--0.13\,\msun).  This is most noticeable
for the 15\,\msun models without mass loss, where \net{160} and
\net{204} \added{models} yield an Fe core of $\approx$~1.33\,\msun while \added{the} \netapprox \added{model}
gives $M_{\rm{Fe}}$~$\approx$~1.51\,\msun, an $\approx$~14\%
difference.

\begin{figure*}[!htb]
  \centering{\includegraphics[width=1.0\textwidth]{{{fe_core_mass}}}}
\caption{ 
Iron core mass at core-collapse, defined when $v>1000$ \kmpers,
as a function of the number of isotopes in the reaction network and mass resolution.
Top row: for \MESA models without mass loss. Bottom row: for models with
mass loss. White squares
denote models that did not reach core-collapse.
}
\label{fig:fe_core_cc}
\end{figure*}

\begin{figure*}[!htb]
\includegraphics[width=\columnwidth]{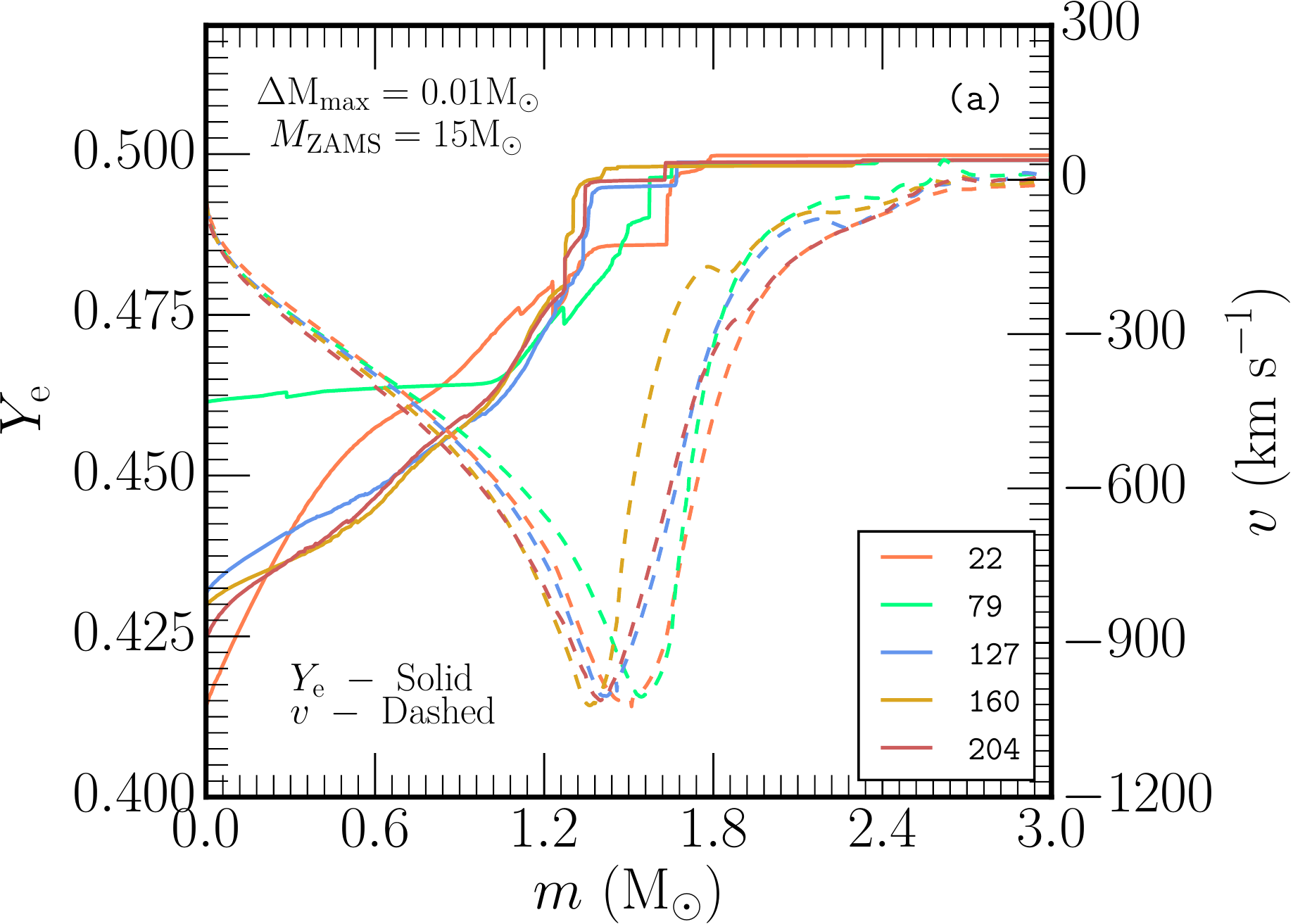}
\includegraphics[width=\columnwidth]{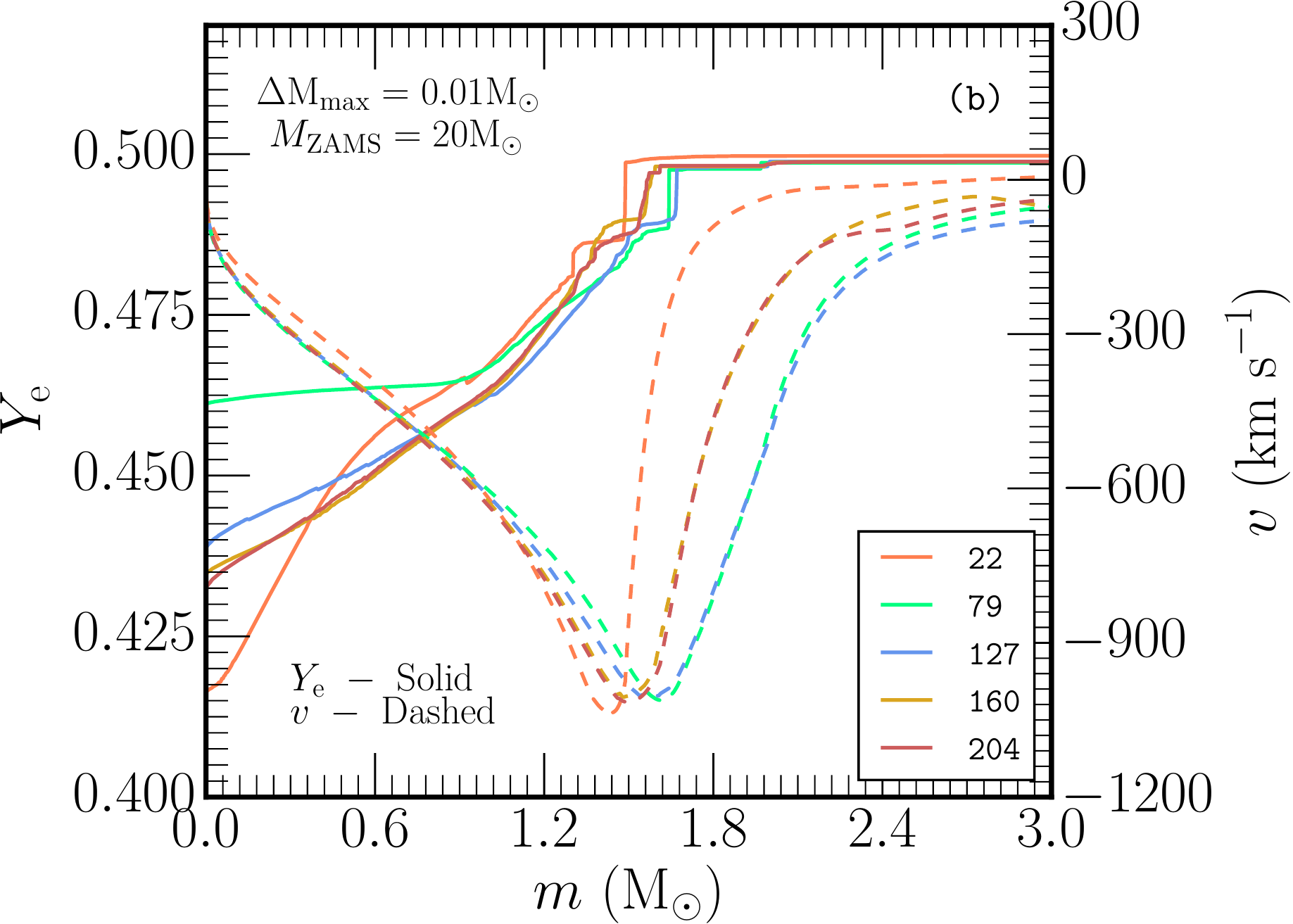} \\
\includegraphics[width=\columnwidth]{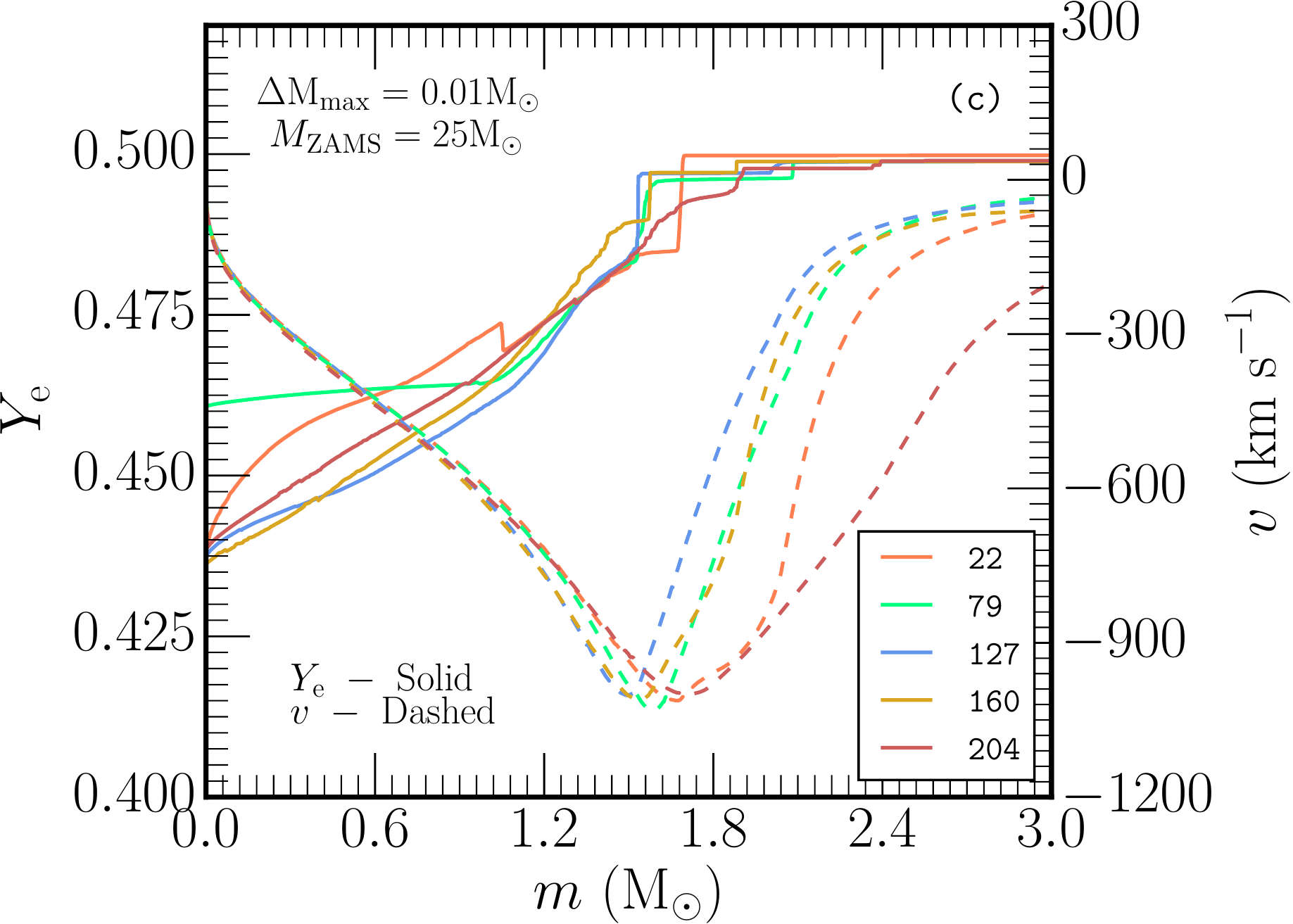}
\includegraphics[width=\columnwidth]{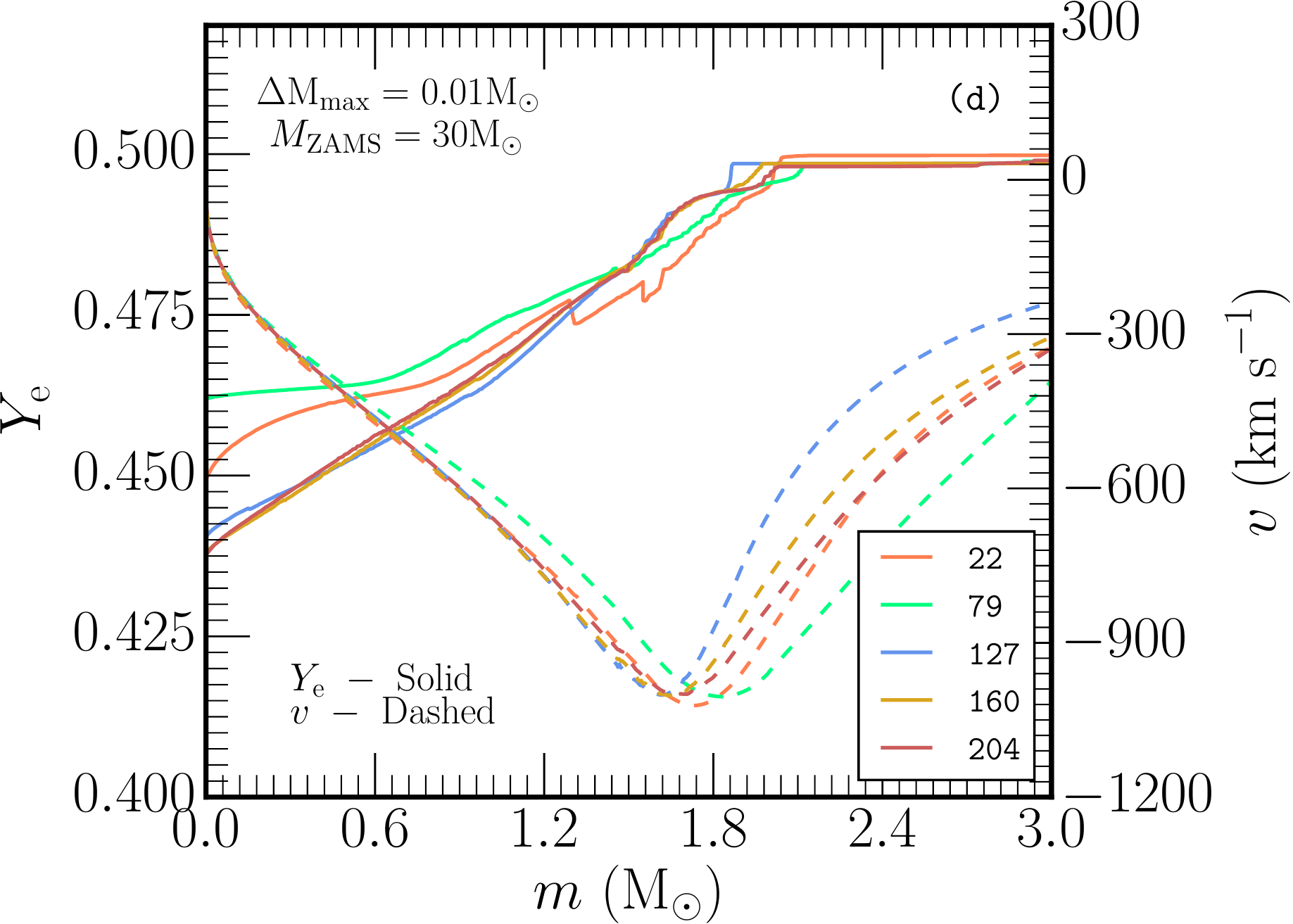}
\caption{
The electron fraction \Ye (solid curves) and radial velocity (dashed curves) at 
core-collapse for 
$\mzams$ = 15, 20, 25, and 30\,\msun models with mass loss and 
$\maxdm = 0.01$\,\msun
Color denotes the number of isotopes used in the reaction network. 
}
\label{fig:ye_vel_ML}
\end{figure*}

The definition of the final Fe-core mass is worthy of mention. We use
\mesa's definition, which is the maximum mass location where
$X\left(^{56}\rm{Fe}\right)>0.1$ and
$X\left(^{28}\rm{Si}\right)<0.01$.  Other options for defining the
Fe-core mass include the location of the maximum infall velocity, a
fixed \Ye value, or the \Ye jump \citep{woosley_1995_aa,heger_2001_aa}. 
Comparing the different
measures, we find the median Fe-core mass using \mesa's abundance
definition is $\approx 0.01$\,\msun less than that when using the peak
velocity location. When using the \Ye jump definition the core mass
is $\approx 0.03$\,\msun greater than the peak velocity location. A
fixed \Ye definition was found to be unsuitable due to variations
between the models in the \Ye value outside the Fe-core, which could
lead to changes as large as $\approx0.5$\,\msun in the core mass location.  
Consideration needs to be
given to finding a consistent and reliable definition for the
final core mass, to enable comparisons between different models. We do
not claim that \mesa's definition is superior.

For a fixed reaction network, say \net{160},
Figure~\ref{fig:fe_core_cc} shows that mass resolution can
have a significant impact on the final Fe-core mass.  The largest
spread occurs for the $\mzams=25$\,\msun where the Fe-core
mass ranges from $\approx$~1.38 - 1.8\,\msun. \replaced{For the 25\,\msun model the
core splits into two groups, the low and high resolutions which is the
similar trend as that seen in its He-core mass (Figure
\ref{fig:he_core_startche_nml}), those models with lower He-core
masses have grown smaller iron core masses.}{For the 25\,\msun model the
core masses split into two groups based on the resolution, a
similar trend to that seen in its He-core mass (Figure
\ref{fig:he_core_startche_nml}), those models with lower He-core
masses have grown smaller iron core masses.} 
However, for mass
resolutions of $\maxdm \leq 0.01$, the final Fe-core mass is monotonic
as the ZAMS mass increases, and agrees to within $\pm
0.1$\,\msun as the resolution increases. Whether the monotonic trend at
the highest mass resolutions remains monotonic with a significantly
finer grid of ZAMS masses is under consideration (I. Petermann et al,
in prep).

In comparing the mass loss models against those without, we find the iron core masses
are slightly smaller on average with mass loss. However the decrease in the mass of the cores
is much less than the decrease in the final mass. For instance the 30\,\msun models may lose
$\approx 10$\,\msun of material over their lifetime (Table \ref{table:variations_med}), yet the 
iron cores will only be $\approx 0.1\text{--}0.2$\,\msun smaller. Mass loss rates are uncertain for the mass
range considered here, but we have show that the size of the iron cores are 
only mildly sensitive to the total amount of mass lost.

Figure~\ref{fig:fe_core_cc} shows that the choice of reaction
network can directly affect the final value of $M_{\rm{Fe}}$. In turn,
this determines the location of the maximum infall velocity. This is
most notable in Figure~\ref{fig:ye_vel_ML} for the
$\mzams=15$\,\msun models, where the infall velocity occurs
$\approx$ 1.51 \,\msun and 1.55 \,\msun for the \netapprox and \net{79},
respectively. For the larger networks, the core infall velocity occurs
at arithmetic mean mass location of 1.4$^{+0.02}_{-0.04}$ \,\msun. 
The $20\text{--}30$\,\msun stars in
Figure~\ref{fig:ye_vel_ML} follow similar trends, with \netapprox and
\net{79} producing the largest deviations about the arithmetic mean Fe
core value of $\approx$ 0.07-0.13 \,\msun. The 15\,\msun models, with $\maxdm=0.005$,
mass loss, and either the \net{127} or \net{160} have anomalously 
low $M_{\rm{Fe}}<1.0$\,\msun, these
models achieve our definition of core collapse \replaced{($v>1000 \kmpers$)}{infall velocity} 
in a few spatial zones but they are not
undergoing a collapse and thus the evolution terminates early
before the core could grow to its full extent.

Figure~\ref{fig:ye_vel_ML} shows the \Ye and radial velocity profiles
at core-collapse for the $\mzams$=15, 20, 25, and 30\,\msun
models with mass loss and $\maxdm=0.01$\,\msun for different reaction
networks.  In most cases, we find that the models with the \netapprox
reaction network under-estimates the \Yec when compared to the
larger reaction networks. The converse is found for models
using the \net{79} network which tends to over-estimate \Yec.
For example, in the $\mzams=15$\,\msun case, the models
using the \net{127}, \net{160}, and \net{204} reaction networks
converge to \Yec$\approx$~0.43. However, models using the
smaller \netapprox and \net{79} reaction networks give values of
\Ye$\approx$~0.41 and \Ye$\approx$~0.46, respectively. A similar trend
is found for the $\mzams=20$\,\msun models. In the case of the
$\mzams=25$ and $30$\,\msun models, the \netapprox and \net{79}
reaction networks show less disagreement in \Yec 
with the values found for the larger networks. The final \Yec is set
by the final composition of the core, in the \netapprox this is a mixture of 
$^{56}$Fe and $^{60}$Cr, the \net{79}
are $\approx 95\%\, ^{56}$Fe (independent of mass), while in the larger networks 
the core becomes a mixture of iron groups elements. This is due to the \net{79} missing
$^{55}$Fe, $^{48-51}$V and $^{51-56}$Cr isotopes, which are present in the larger nuclear networks.
These additional isotopes provide alternate decay routes for the $^{56}$Fe in the core.

There is also considerable variation in the shape of the \Ye step at the
edge of the iron core. This is not a well defined jump in the \Ye value
but can show a number of sub steps. For instance, for the 15\,\msun
models in Figure~\ref{fig:ye_vel_ML} the \netapprox has a single large jump
at $1.62$\,\msun while with the larger nets the 
jump is at $1.3$\,\msun but shows substructure in the \Ye values.

\begin{figure*}[!htb]
\includegraphics[width=1.05\columnwidth,trim={0.5cm 1.25cm 0.5cm 1cm},clip]{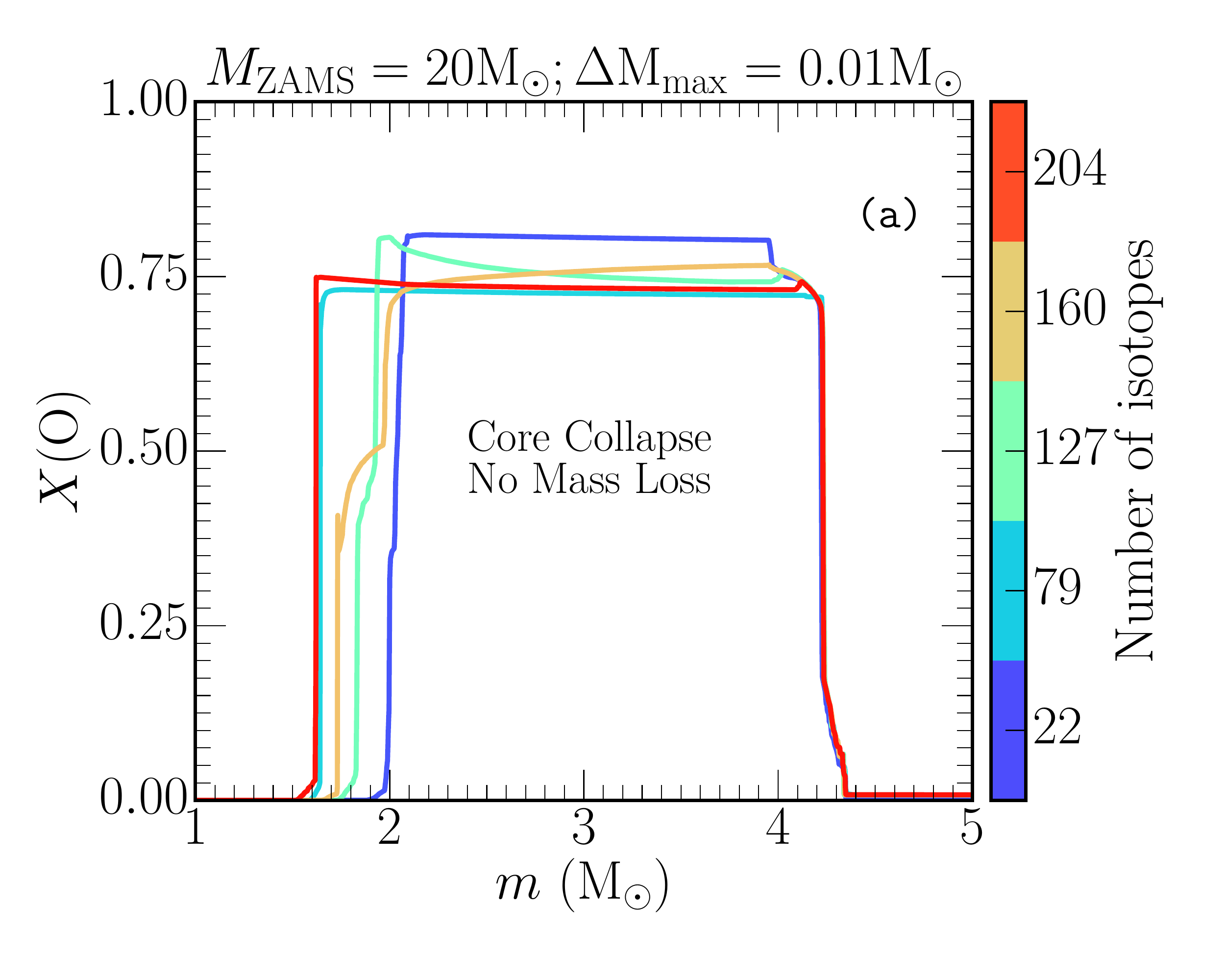}
\includegraphics[width=1.05\columnwidth,trim={0.5cm 1.25cm 0.5cm 1cm},clip]{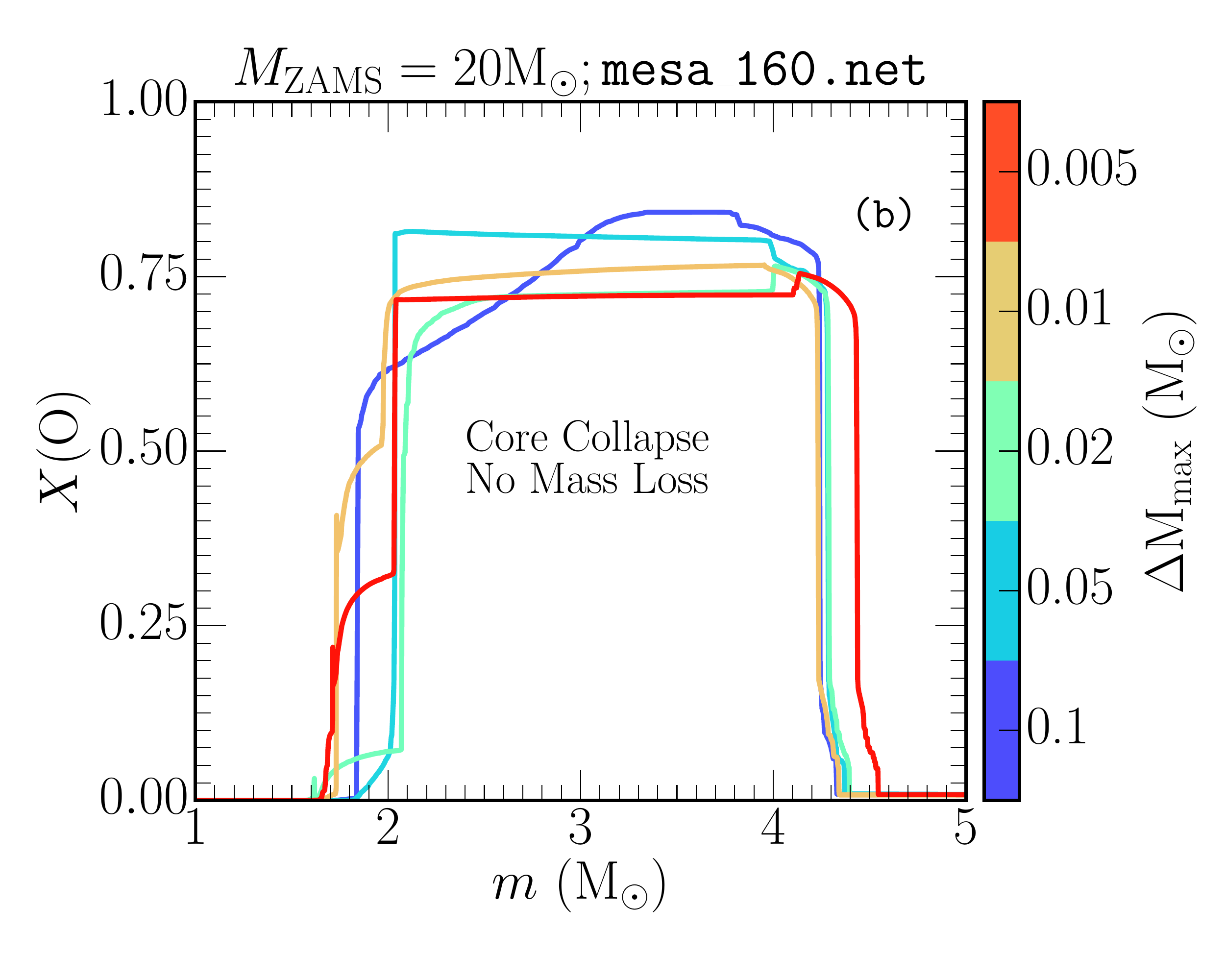}
\caption{
O mass fraction profile at core-collapse for the
$\mzams=20$\,\msun model without mass loss as a function of 
\texttt{(a)} the number of isotopes in the nuclear reaction network and 
\texttt{(b)} the mass resolution \maxdm.
}
\label{fig:m_o_at_cc}
\end{figure*}

\begin{figure*}[!htb]
\includegraphics[width=1.05\columnwidth,trim={0.5cm 1.25cm 0.5cm 1cm},clip]{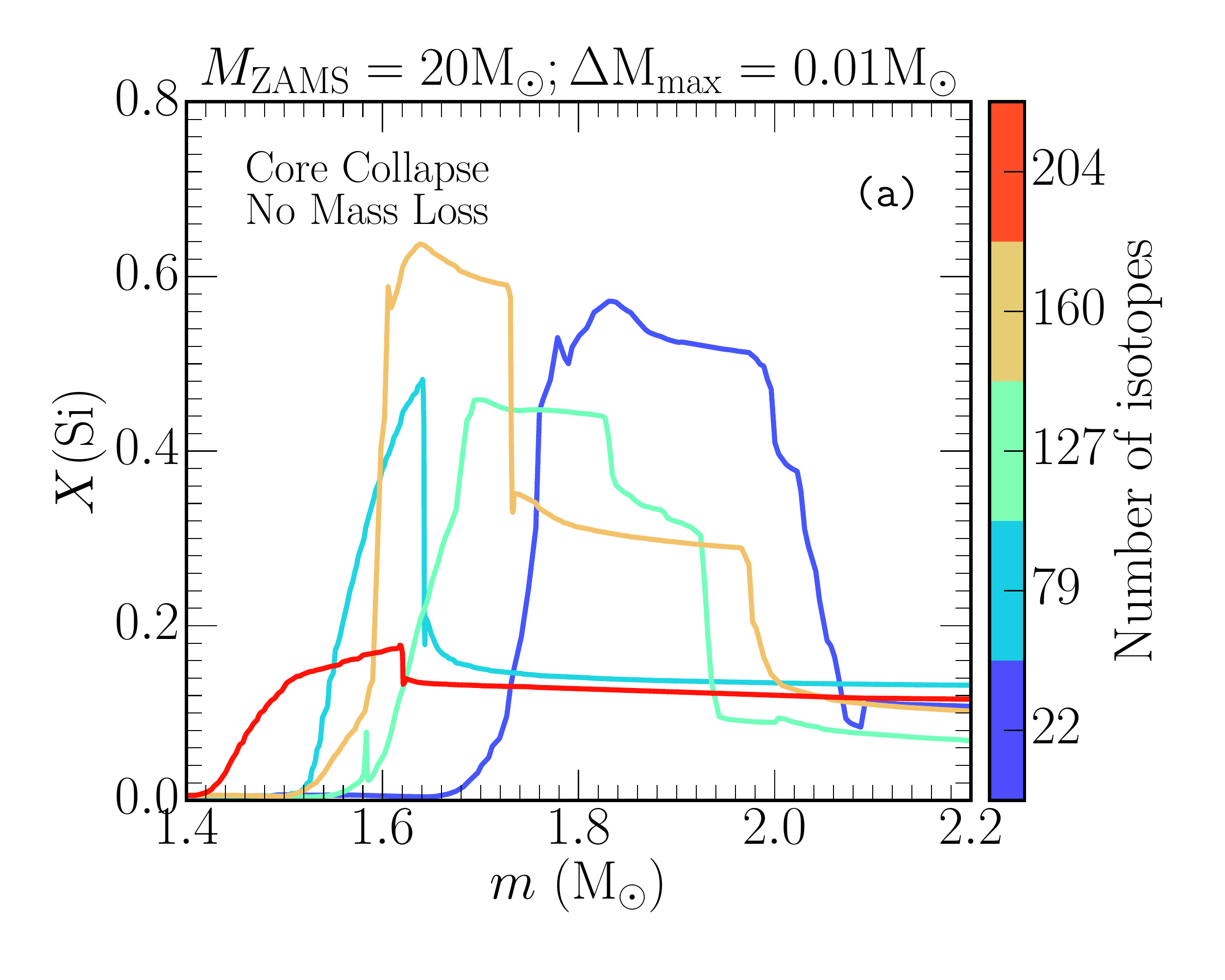}
\includegraphics[width=1.05\columnwidth,trim={0.5cm 1.25cm 0.5cm 1cm},clip]{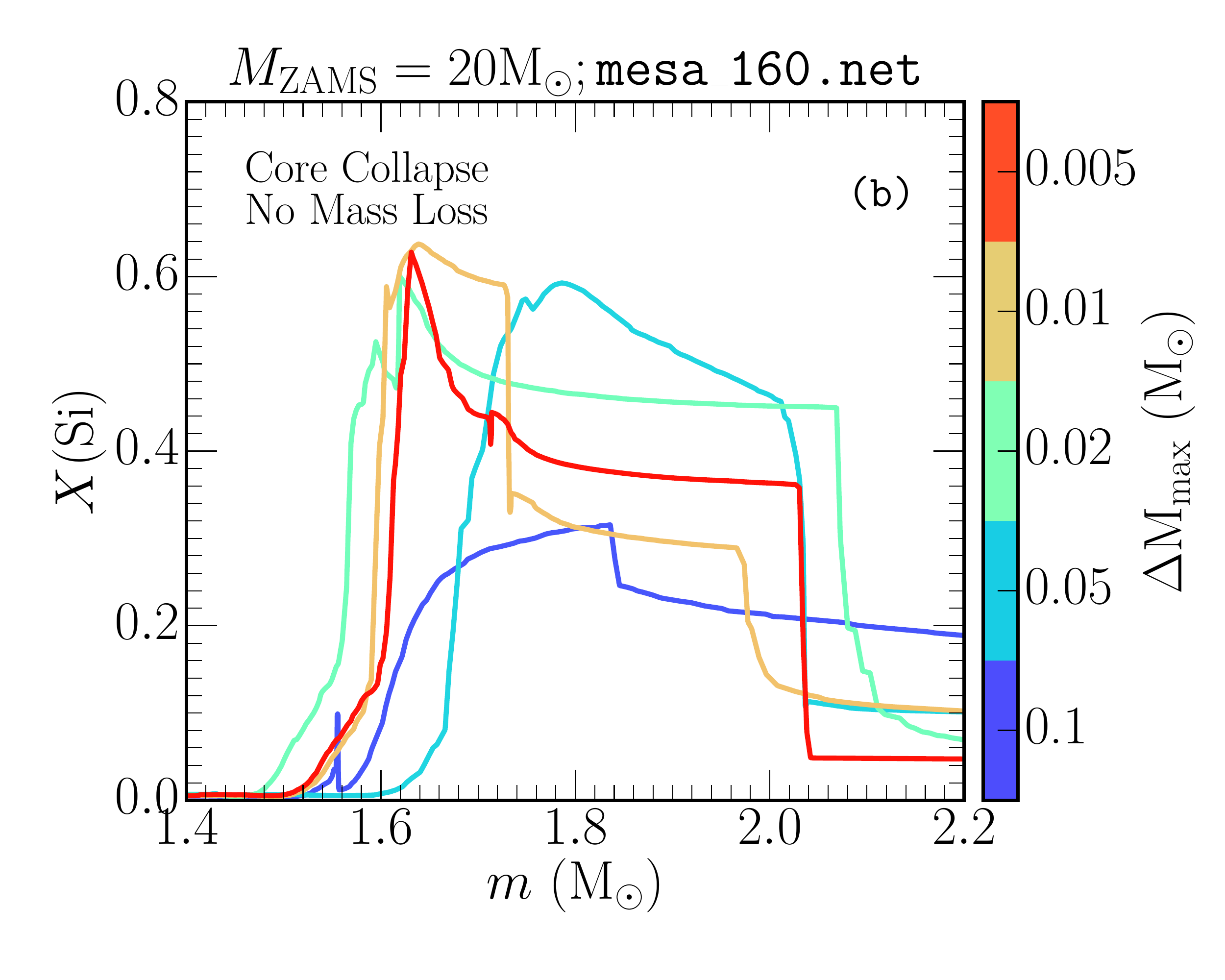}
\caption{
Si mass fraction profile at core collapse for the
$\mzams=20$\,\msun model without mass loss as a function of 
\texttt{(a)} the number of isotopes in the nuclear reaction network and 
\texttt{(b)} the mass resolution \maxdm.
}
\label{fig:m_si_at_cc}
\end{figure*}

\begin{figure*}[!htb]
\includegraphics[width=\textwidth]{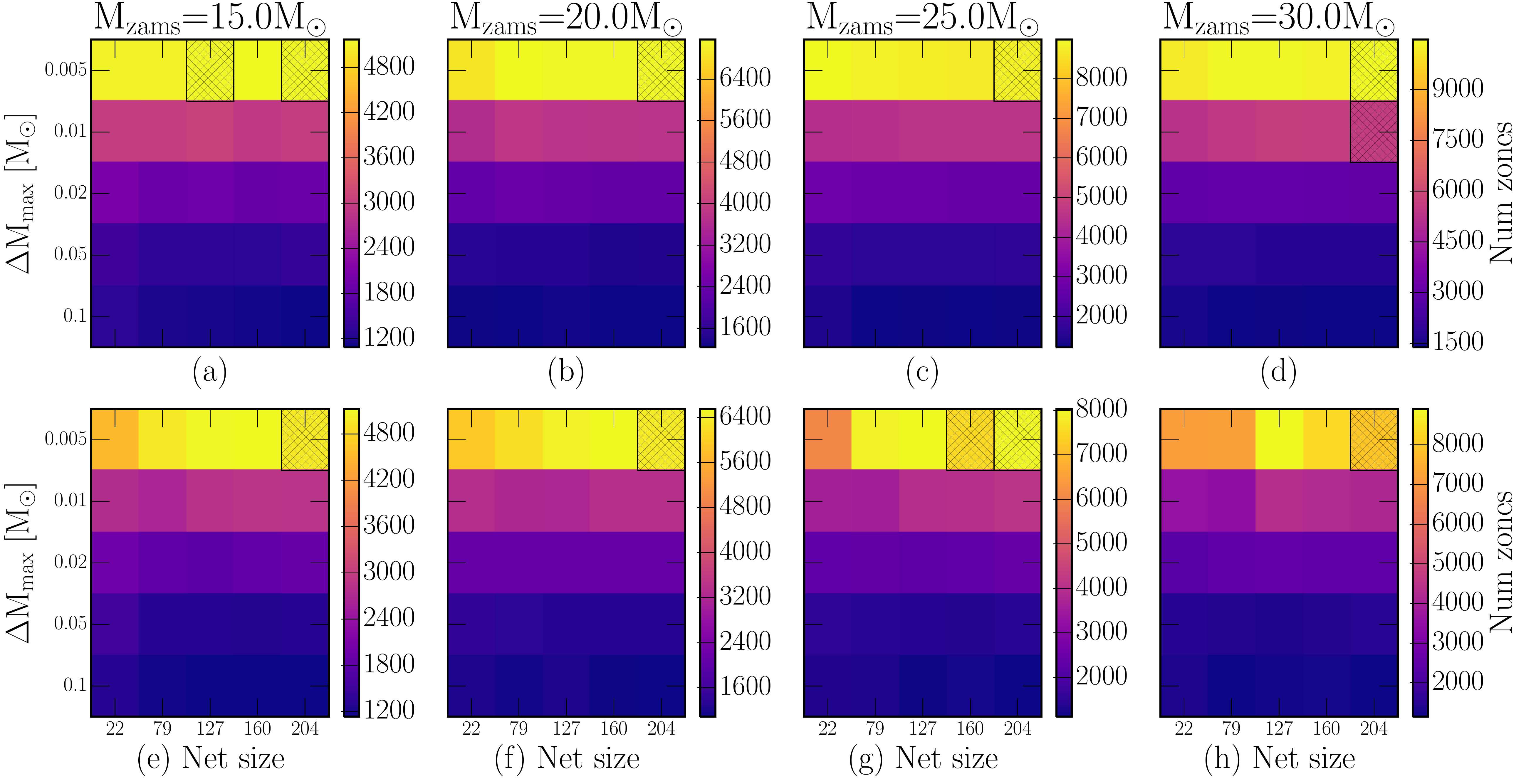}
\caption{
Total number of spatial zones at core-collapse as a function of the number of isotopes in the 
network and mass resolution.
Top: for \MESA models without mass loss. Bottom: with mass loss. Hatching indicates models
that did not reach core-collapse.
}
\label{fig:zones_cc}
\end{figure*}

Figure~\ref{fig:m_o_at_cc} shows the oxygen mass fraction profile at
core-collapse for a subset of the $\mzams=20$\,\msun
models. For each profile, all isotopes of oxygen in a particular network
are summed to give the total oxygen mass fraction as a function of mass
coordinate (see Figure~\ref{fig:nzplane} for the different isotopes of
oxygen included for each network). The left panel plots the $X$(O)
profile as a function of the number of isotopes in the network for the
$\mzams=20$\,\msun models without mass loss at a mass resolution of
$\maxdm=0.01$\,\msun. The right panel plots the $X$(O) profile as a
function of mass resolution for the $\mzams=20$\,\msun models 
without mass loss that use the \net{160} reaction network (this network follows $^{14-18}$O).

The upper mass boundary of the O shell is well constrained at
$4.4$\,\msun for the different networks at a fixed $\maxdm=0.01$\,\msun
models. While the lower mass boundary varies over $0.5$\,\msun, the total
mass of oxygen only varies between $1.9-2.0$\,\msun. For the models with
a fixed reaction network and variable \maxdm\ the upper mass boundary
varies over $\approx 0.2$\,\msun, similar to the spread in the lower
mass boundary. Most models show a top hat shape due to strong
convection occurring over the shell, while the $\maxdm=0.1$\,\msun, \net{160}
shows a different behavior due to an oxygen burning wave propagating from the base
of the O shell outwards. The almost step like feature seen in the lower
boundaries is due to a convective Si-burning shell, abutting into the
base of O-burning shell and the top of the Fe-core. This Si-burning
shell only partially burns the silicon and the oxygen before
core-collapse.

Similarly, in Figure~\ref{fig:m_si_at_cc} we show mass fraction
profiles for $X$(Si) for $\mzams=20$\,\msun models without
mass loss. This region shows considerably
more variation than the O shells in Figure \ref{fig:m_o_at_cc}. The
lower mass boundary of the silicon distribution denotes the edge of
the iron core, while the upper mass boundary denotes the edge between
the Si shell and the base of the O shell.

For fixed mass resolution of $\maxdm=0.01$\,\msun there are three
groupings in the left panel of Figure~\ref{fig:m_si_at_cc}; the \netapprox; 
the \net{79}, \net{127} and \net{160}; and
the \net{204}. The \netapprox model has the largest iron core due to
the most energetic shell Si-burning. This pushes the silicon shell outwards
and allows the shell to burn over a larger oxygen-rich region. The
larger networks are generally cooler in temperature so the Si-burning
shell does not extend as far outwards. The sudden drop in the silicon mass
fraction, at the upper boundary, seen in the \net{79}, \net{127} and \net{160} models is due
to the lower mass boundary of the O shell being convective which mixes
silicon outwards. The \net{204} shows a case when the iron core abuts
the O shell, leaving no distinct Si shell.

For a fixed \net{160} reaction network, Figure~\ref{fig:m_si_at_cc}
shows a trend for the Si shell to move inwards and concentrate at
lower mass coordinates as the resolution increases. This is due to 
the temperature and density of
the Si shell decreasing as the resolution increases.  This slows the
conversion of silicon into iron-group elements. For models with
\maxdm~$\leq$~0.02\,\msun the lower mass boundary of the silicon is sharply
defined at $\approx 1.6$\,\msun, with the core being composed primarily
of $^{54}$Fe up to the edge of the iron core.  For the
$\maxdm=0.1$\,\msun and $\maxdm=0.5$\,\msun models the lower mass
boundary is more inclined due to the increased presence of
$^{56}$Ni. In the $\maxdm=0.1$\,\msun case, this becomes a distinct
$^{56}$Ni shell between the iron core, which is mostly $^{54}$Fe, and
the silicon shell.  In the $\maxdm=0.05$\,\msun case $^{54}$Fe and
$^{56}$Ni are approximately equal components at $\approx$~25\% at the
interface of the $^{54}$Fe core and the Si shell.

Figure \ref{fig:zones_cc} shows the total number of mass zones at
core-collapse.  As \maxdm\ increases from $\maxdm=0.1$\,\msun to
0.005\,\msun there is a factor four increase in the number of
zones, which is much less than the factor 20 increase in \maxdm. This
is due to $\maxdm$ only providing an upper limit on the mass of a
cell. For instance, a $\mzams=15$\,\msun model with $\maxdm=0.1$\,\msun,
assuming all cells have the same \maxdm, needs only 150 spatial zones
-- much less than the 1200 zones present in our models.  On average we
find that the models with mass loss need slightly less zones at
core collapse. This is due to having a smaller mass star.  However, due to the
fact that the zones are unevenly distributed, being concentrated near the
core and where large changes in stellar properties occur,
the scaling is not linear with the star's mass. There are a number of
other \MESA controls that effect the spatial resolution, like \mesh,
which contribute to the total number of zones, and their relative
distribution. We recommend taking a critical view of the spatial
distribution of zones in stellar models and consider the effect of
increased resolution on their results.

Table \ref{table:variations_med} summarizes the pre-SN model properties
measured either during the evolution of the model or at core
collapse. We show the median value over all models with upper and lower bounds. 
The $\rm{He_{core}}, \rm{C_{core}}, \rm{O_{core}}, \rm{Si_{core}}$ summarize
the core masses at the point of the fuel ignition (See Figures
\ref{fig:he_core_startche_nml}, \ref{fig:c_core_start},
\ref{fig:o_core_start}, \ref{fig:si_core_start}, for a illustration of
their spread). Similarly the $\Yec$ shows the electron fraction at the
center of the star measured at the same time as the core
masses. The electron fraction does not drop much bellow 0.5 until after oxygen burning.
Once silicon burning commences the \Yec drops to $0.48\text{--}0.49$, with
larger initial mass stars having larger \Yec. At core collapse \Yec can drop to
$\Yec=0.43\text{--}0.44$, with larger initial masses having larger median
values, however the upper/lower bounds overlap between all the masses.

\explain{Moved from previous section}
\added{Table \ref{table:variations_med} lists the final masses ($\rm{M_{final}}$) 
at core-collapse,
but dominated by the mass at the TAMS, for models with
mass loss. Similarly to final ages, most of the variation in the final masses occurs 
as mass resolution increases (see Section \ref{sec:lifetimes}).  For
the $\mzams=15$\,\msun models the total variation in the final
mass is $\simeq$ 2\%.  As the ZAMS mass increases, this variation
increases to $\simeq$ 10\% for the $\mzams=30$\,\msun
models. In general, as the resolution increases the final mass
decreases, consistent with the findings from Section
\ref{sec:burn_h}, where the higher resolution models live longer due
to the change in behavior of the ICZ and SCZ.}

\begin{deluxetable*}{lllllllll}
\tablecolumns{9}
\tablewidth{0.8\linewidth}
\tablecaption{Median pre-SN model properties, with upper and lower bounds\label{table:variations_med}}
\tablehead{\colhead{Property} & \multicolumn{2}{c}{15\,\msun} & \multicolumn{2}{c}{20\,\msun} &\multicolumn{2}{c}{25\,\msun} &\multicolumn{2}{c}{30\,\msun} \\
 & $\Mdot=0$ & $\Mdot\neq 0$ & $\Mdot=0$ & $\Mdot\neq 0$ & $\Mdot=0$ & $\Mdot\neq 0$ & $\Mdot=0$ & $\Mdot\neq 0$}
\startdata
$\rm{He_{core}\ [M_{\odot}]}$\tablenotemark{a,b}            & $ 2.82^{2.82}_{2.79} $ & $ 2.77^{2.78}_{2.72} $  & $ 4.67^{4.70}_{4.59} $ & $ 4.57^{4.59}_{4.52} $  & $ 6.88^{7.28}_{6.80} $ & $ 6.56^{6.67}_{6.52} $  & $ 9.44^{9.89}_{9.15} $ & $ 8.67^{8.78}_{8.62} $  \\
$\rm{C_{core}\ [M_{\odot}]}$             & $ 2.51^{2.58}_{2.49} $ & $ 2.44^{2.53}_{2.43} $  & $ 4.19^{4.75}_{4.04} $ & $ 4.07^{4.08}_{3.69} $  & $ 6.02^{6.43}_{4.34} $ & $ 5.75^{5.92}_{5.53} $  & $ 8.28^{8.79}_{7.13} $ & $ 7.62^{7.82}_{7.22} $  \\
$\rm{O_{core}\ [M_{\odot}]}$             & $ 1.41^{1.43}_{1.35} $ & $ 1.40^{1.42}_{1.32} $  & $ 1.54^{2.47}_{1.43} $ & $ 1.57^{2.05}_{1.41} $  & $ 2.34^{3.04}_{1.74} $ & $ 1.81^{2.47}_{1.76} $  & $ 2.38^{3.18}_{2.14} $ & $ 2.39^{3.06}_{2.26} $  \\
$\rm{Si_{core}\ [M_{\odot}]}$            & $ 1.15^{1.38}_{1.02} $ & $ 1.15^{1.39}_{1.08} $  & $ 1.38^{1.65}_{1.30} $ & $ 1.40^{1.48}_{1.24} $  & $ 1.19^{1.61}_{0.91} $ & $ 1.40^{1.67}_{1.07} $  & $ 1.16^{1.64}_{1.08} $ & $ 1.15^{1.66}_{1.12} $  \\[0.05in]
\hline
$\rm{Ye_{c,He}}$\tablenotemark{b}                         & $ 0.505^{0.505}_{0.505} $ & $ 0.505^{0.505}_{0.505} $  & $ 0.505^{0.505}_{0.505} $ & $ 0.505^{0.505}_{0.505} $  & $ 0.505^{0.505}_{0.505} $ & $ 0.505^{0.505}_{0.505} $  & $ 0.505^{0.505}_{0.505} $ & $ 0.505^{0.505}_{0.505} $  \\
$\rm{Ye_{c,C}}$                          & $ 0.499^{0.500}_{0.499} $ & $ 0.499^{0.500}_{0.499} $  & $ 0.499^{0.500}_{0.499} $ & $ 0.499^{0.500}_{0.499} $  & $ 0.499^{0.500}_{0.499} $ & $ 0.499^{0.500}_{0.499} $  & $ 0.499^{0.500}_{0.499} $ & $ 0.499^{0.500}_{0.499} $  \\
$\rm{Ye_{c,O}}$                          & $ 0.499^{0.500}_{0.498} $ & $ 0.499^{0.500}_{0.498} $  & $ 0.499^{0.500}_{0.498} $ & $ 0.499^{0.500}_{0.498} $  & $ 0.498^{0.500}_{0.498} $ & $ 0.499^{0.500}_{0.484} $  & $ 0.498^{0.500}_{0.498} $ & $ 0.498^{0.500}_{0.490} $  \\
$\rm{Ye_{c,Si}}$                         & $ 0.486^{0.498}_{0.475} $ & $ 0.486^{0.498}_{0.475} $  & $ 0.488^{0.498}_{0.483} $ & $ 0.489^{0.498}_{0.483} $  & $ 0.491^{0.497}_{0.483} $ & $ 0.490^{0.499}_{0.482} $  & $ 0.491^{0.498}_{0.489} $ & $ 0.490^{0.497}_{0.488} $  \\
$\rm{Ye_{c,Fe}}$                         & $ 0.431^{0.461}_{0.419} $ & $ 0.432^{0.461}_{0.414} $  & $ 0.438^{0.462}_{0.425} $ & $ 0.438^{0.462}_{0.416} $  & $ 0.438^{0.462}_{0.414} $ & $ 0.438^{0.462}_{0.423} $  & $ 0.444^{0.462}_{0.437} $ & $ 0.442^{0.462}_{0.428} $  \\[0.05in]
\hline
$\tau_{\rm{H}}\ [\rm{Myr}]$\tablenotemark{c}              & $ 10.95^{10.96}_{10.93} $ & $ 10.99^{11.00}_{10.94} $  & $ 7.73^{7.74}_{7.72} $ & $ 7.78^{7.79}_{7.76} $  & $ 6.18^{6.51}_{6.16} $ & $ 6.24^{6.38}_{6.22} $  & $ 5.53^{5.62}_{5.26} $ & $ 5.34^{5.43}_{5.33} $  \\
$\tau_{\rm{He}}\ [\rm{Myr}]$             & $ 1.69^{1.71}_{1.48} $ & $ 1.74^{1.75}_{1.51} $  & $ 1.11^{1.72}_{1.05} $ & $ 1.10^{1.29}_{1.08} $  & $ 0.81^{1.19}_{0.77} $ & $ 0.82^{0.89}_{0.81} $  & $ 0.65^{0.73}_{0.63} $ & $ 0.69^{0.74}_{0.68} $  \\
$\tau_{\rm{C}}\ [\rm{yr}]$               & $ 78.37^{82.62}_{75.65} $ & $ 81.80^{87.04}_{76.15} $  & $ 111.36^{115.61}_{25.90} $ & $ 125.42^{131.19}_{28.52} $  & $ 27.06^{28.42}_{14.81} $ & $ 23.35^{26.98}_{22.53} $  & $ 16.56^{22.45}_{7.97} $ & $ 19.82^{22.62}_{15.11} $  \\
$\tau_{\rm{O}}\ [\rm{yr}]$               & $ 3.83^{5.05}_{3.28} $ & $ 4.08^{5.36}_{3.51} $  & $ 1.44^{2.76}_{0.16} $ & $ 1.28^{3.12}_{0.26} $  & $ 0.14^{0.29}_{0.08} $ & $ 0.31^{0.93}_{0.01} $  & $ 0.13^{0.19}_{0.02} $ & $ 0.14^{0.16}_{0.10} $  \\
$\tau_{\rm{Si}}\ [\rm{days}]$            & $ 3.43^{5.05}_{0.74} $ & $ 3.74^{5.60}_{0.75} $  & $ 0.81^{1.56}_{0.32} $ & $ 0.91^{2.04}_{0.27} $  & $ 0.61^{6.58}_{0.18} $ & $ 0.66^{1.59}_{0.16} $  & $ 0.35^{1.92}_{0.15} $ & $ 0.41^{1.97}_{0.21} $  \\
$\tau_{\rm{Fe}}\ [\rm{hr}]$              & $ 33.16^{57.11}_{11.21} $ & $ 36.38^{74.01}_{11.87} $  & $ 11.11^{35.22}_{3.66} $ & $ 11.63^{17.95}_{6.59} $  & $ 6.57^{25.25}_{4.04} $ & $ 9.55^{19.72}_{4.26} $  & $ 4.74^{24.33}_{3.23} $ & $ 4.77^{25.36}_{3.37} $  \\[0.05in]
\hline
$\rm{He_{shell}\ [M_{\odot}]}$\tablenotemark{d,e}           & $ 4.17^{4.22}_{4.16} $ & $ 4.09^{4.17}_{4.08} $  & $ 6.20^{6.88}_{6.09} $ & $ 6.05^{6.07}_{5.84} $  & $ 8.26^{8.71}_{8.12} $ & $ 7.95^{8.10}_{7.71} $  & $ 10.87^{11.30}_{9.66} $ & $ 9.99^{10.24}_{9.58} $  \\
$\rm{C_{shell}\ [M_{\odot}]}$            & $ 2.51^{2.58}_{2.49} $ & $ 2.44^{2.54}_{2.43} $  & $ 4.19^{4.75}_{4.04} $ & $ 4.07^{4.09}_{3.61} $  & $ 6.03^{6.44}_{5.84} $ & $ 5.76^{5.88}_{5.54} $  & $ 8.26^{8.73}_{7.09} $ & $ 7.57^{7.82}_{7.16} $  \\
$\rm{O_{shell}\ [M_{\odot}]}$            & $ 2.42^{2.53}_{2.18} $ & $ 2.33^{2.49}_{2.13} $  & $ 3.95^{4.14}_{3.27} $ & $ 3.82^{3.93}_{2.61} $  & $ 5.38^{5.69}_{4.08} $ & $ 5.25^{5.63}_{2.80} $  & $ 6.87^{8.20}_{5.64} $ & $ 6.47^{7.20}_{4.68} $  \\
$\rm{Si_{shell}\ [M_{\odot}]}$           & $ 1.59^{1.70}_{0.00} $ & $ 1.61^{1.65}_{1.24} $  & $ 1.80^{2.97}_{1.56} $ & $ 1.81^{2.11}_{0.00} $  & $ 1.90^{2.67}_{1.53} $ & $ 1.88^{2.22}_{1.62} $  & $ 2.27^{2.58}_{1.73} $ & $ 2.30^{3.14}_{1.50} $  \\
$\rm{Fe_{core}\ [M_{\odot}]}$           & $ 1.41^{1.55}_{0.75} $ & $ 1.42^{1.53}_{0.77} $  & $ 1.55^{1.86}_{1.35} $ & $ 1.57^{1.74}_{1.38} $  & $ 1.66^{1.88}_{1.35} $ & $ 1.59^{1.83}_{1.46} $  & $ 1.80^{1.90}_{1.58} $ & $ 1.79^{1.90}_{1.39} $  \\[0.05in]
\hline
$\rm{H_{env}\ [M_{\odot}]}$\tablenotemark{d,f}           & $ 6.96^{7.21}_{6.94} $ & $ 5.65^{6.14}_{5.59} $  & $ 8.41^{8.55}_{7.58} $ & $ 6.74^{7.18}_{6.50} $  & $ 9.76^{9.96}_{9.17} $ & $ 6.09^{7.22}_{4.69} $  & $ 10.47^{11.02}_{9.99} $ & $ 4.60^{6.10}_{0.00} $  \\
$\rm{M_{final}\ [M_{\odot}]}$\tablenotemark{d,g}                & \nodata & $ 13.02^{13.37}_{12.94} $  & \nodata & $ 17.26^{18.11}_{16.91} $  & \nodata & $ 18.95^{21.17}_{17.01} $  & \nodata & $ 20.27^{22.06}_{17.38} $  \\
$\rm{\log_{10}\, \left(R/R_{\odot}\right)}$\tablenotemark{d}  & $ 2.97^{2.99}_{2.96} $ & $ 2.98^{3.00}_{2.97} $  & $ 3.08^{3.08}_{3.06} $ & $ 3.07^{3.07}_{3.05} $  & $ 3.08^{3.09}_{3.05} $ & $ 3.07^{3.08}_{3.05} $  & $ 2.97^{3.02}_{2.92} $ & $ 2.99^{3.03}_{2.95} $  \\
$\rm{\log_{10}\, \left(T_{eff}/K\right)}$\tablenotemark{d}  & $ 3.50^{3.51}_{3.50} $ & $ 3.50^{3.52}_{3.49} $  & $ 3.52^{3.53}_{3.51} $ & $ 3.52^{3.53}_{3.51} $  & $ 3.56^{3.58}_{3.55} $ & $ 3.56^{3.73}_{3.54} $  & $ 3.65^{3.74}_{3.62} $ & $ 3.63^{3.67}_{3.60} $  \\
$\rm{\log_{10}\, \left(L/L_{\odot}\right)}$\tablenotemark{d}  & $ 4.91^{4.93}_{4.90} $ & $ 4.90^{5.01}_{4.88} $  & $ 5.18^{5.23}_{5.13} $ & $ 5.15^{5.20}_{5.09} $  & $ 5.36^{5.42}_{5.30} $ & $ 5.33^{6.03}_{5.25} $  & $ 5.47^{5.84}_{5.38} $ & $ 5.49^{5.56}_{5.40} $  \\
$\xi_{\rm{M}=2.5}$\tablenotemark{d,h}                      & $ 0.10^{0.15}_{0.04} $ & $ 0.08^{0.13}_{0.04} $  & $ 0.24^{0.63}_{0.10} $ & $ 0.24^{0.43}_{0.13} $  & $ 0.32^{0.66}_{0.16} $ & $ 0.27^{0.60}_{0.19} $  & $ 0.60^{0.69}_{0.31} $ & $ 0.58^{0.69}_{0.19} $  \\
\enddata
\tablenotetext{a}{Core mass values, see Figures \ref{fig:he_core_startche_nml}, \ref{fig:c_core_start}, \ref{fig:o_core_start}, \ref{fig:si_core_start}, \ref{fig:fe_core_cc} for definitions.}
\tablenotetext{b}{Measured at the corresponding ignition of each fuel, except for Fe which is measured at core collapse.}
\tablenotetext{c}{Approximate time to transition to the next major fuel source.}
\tablenotetext{d}{Measured at core collapse.}
\tablenotetext{e}{Outer mass coordinate where the element is the most abundant.}
\tablenotetext{f}{Mass of H-rich envelope.}
\tablenotetext{g}{Total mass of star.}
\tablenotetext{h}{Compactness parameter, with $\M=2.5$\,\msun.}
\end{deluxetable*}


The $\tau$ values in Table \ref{table:variations_med}  denote the
time between the depletion of the previous fuel to the core ignition of the
next. As we progress between each major fuel source the lifetime of the star
decreases rapidly, as expected. As the initial mass increases the lifetime for each 
fuel to burn decreases, and the fraction of the stars life spent burning each
fuel (other than hydrogen) decreases. For instance the MS lifetime of a 15\,\msun star 
is twice that of the 30\,\msun but eight times that during the final silicon shell burning. 
Also those 
stars that lose mass, live longer than their non mass losing companions due to the
slower evolution experienced by a lower mass star. 
$\rm{He_{shell}},\rm{C_{shell}},\rm{O_{shell}},\rm{Si_{shell}}$ denote
the locations of each fuel shell at core collapse, while the
$\rm{Fe_{core}}$ is the iron core mass at core collapse (see Figure
\ref{fig:fe_core_cc}). The shell locations increase
as the initial mass increases, with mass losing stars having smaller 
shell masses. The $\rm{Si_{shell}}$ lower bound of $0.0$ does not denote
that there is a distribution to $0.0$, but that a few models have no silicon shell
(see Figure~\ref{fig:m_si_at_cc}) while the other models sit near the median value.

Lastly we provide a summary of the surface of
the star at core collapse. $\rm{H_{env}}$ is the mass of H-rich
envelope, $\rm{M_{final}}$ is the final mass and
$\rm{R/R_{\odot}},\rm{T_{eff}},\rm{L/L_{\odot}}$ provide the surface
radius, temperature and luminosity. As the initial mass increases
the temperature and luminosity of the models increase, but the \added{surface} radius
\replaced{peaks at}{is maximum for} the $20\text{--}25$\,\msun models before decreasing for the
30\,\msun models. Mass loss also does not significantly effect the 
final radius, temperature or luminosity even though they can be 
significantly smaller and have much smaller $\rm{H_{env}}$ values.
The final compactness parameter, is given by,
\begin{equation}
\xi=\frac{\left(\M/\,\msun\right)}{R_{\M=2.5}/1000\rm{km}}
\end{equation}
where $\M=2.5$\,\msun and the radius is measured (in km) at the radius
where $\M=2.5$\,\msun \citep{oconnor_2011_aa}. In principle it should be measured at the
bounce, however \citet{sukhbold_2014_aa} showed that no substantial accuracy is lost if measured
in the pre-SN model. We see
a steady increase in the final value as the initial mass increases, with 
mass losing stars having slightly smaller values. All the parameters in
Table \ref{table:variations_med} 
show considerable spread in their values due to changes in the spatial resolution
and choice of nuclear network.

\subsection{Pre-SN Nucleosynthesis}\label{sec:nucleo}

\begin{figure*}[!htb]
\includegraphics[width=\columnwidth]{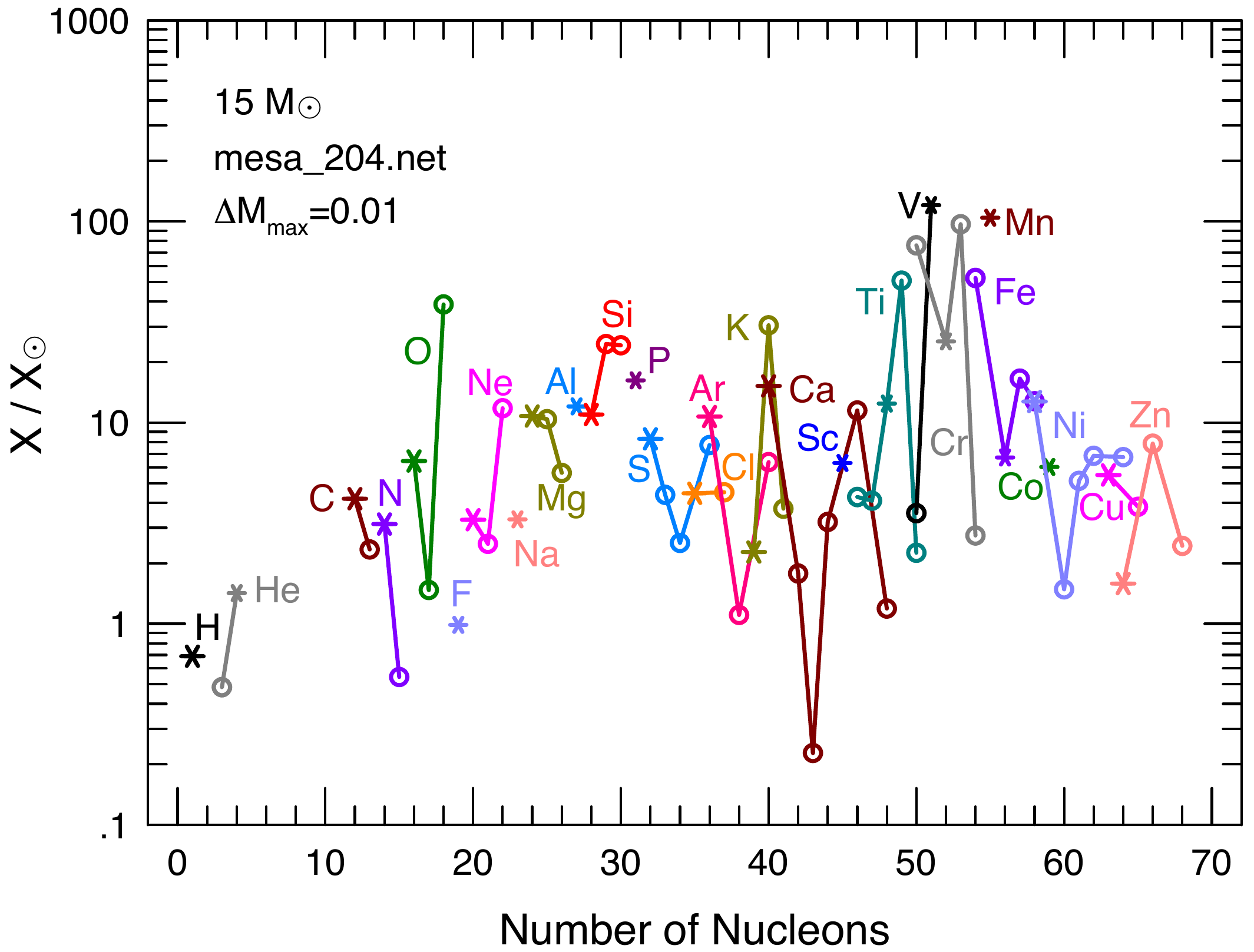}
\includegraphics[width=\columnwidth]{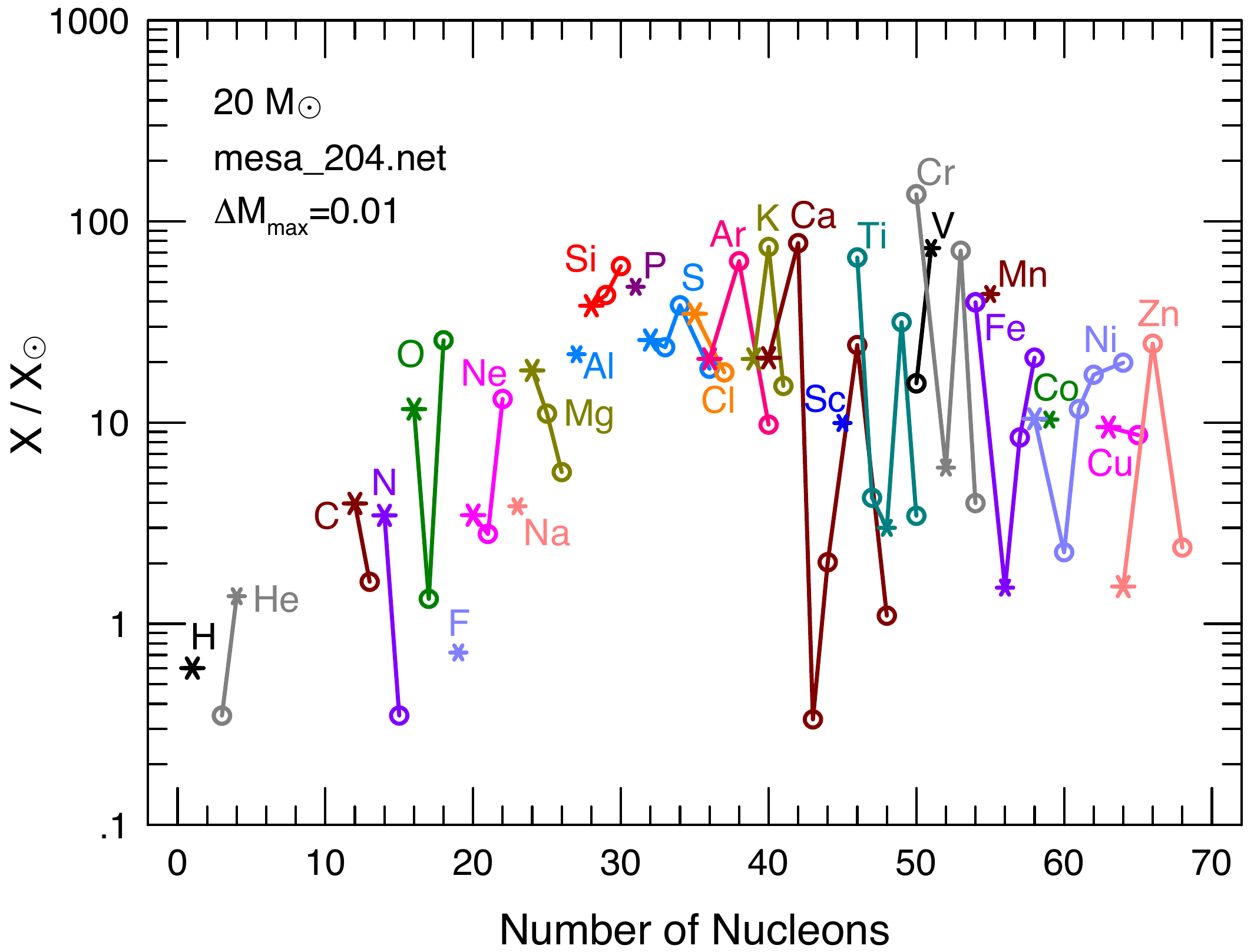}\\
\includegraphics[width=\columnwidth]{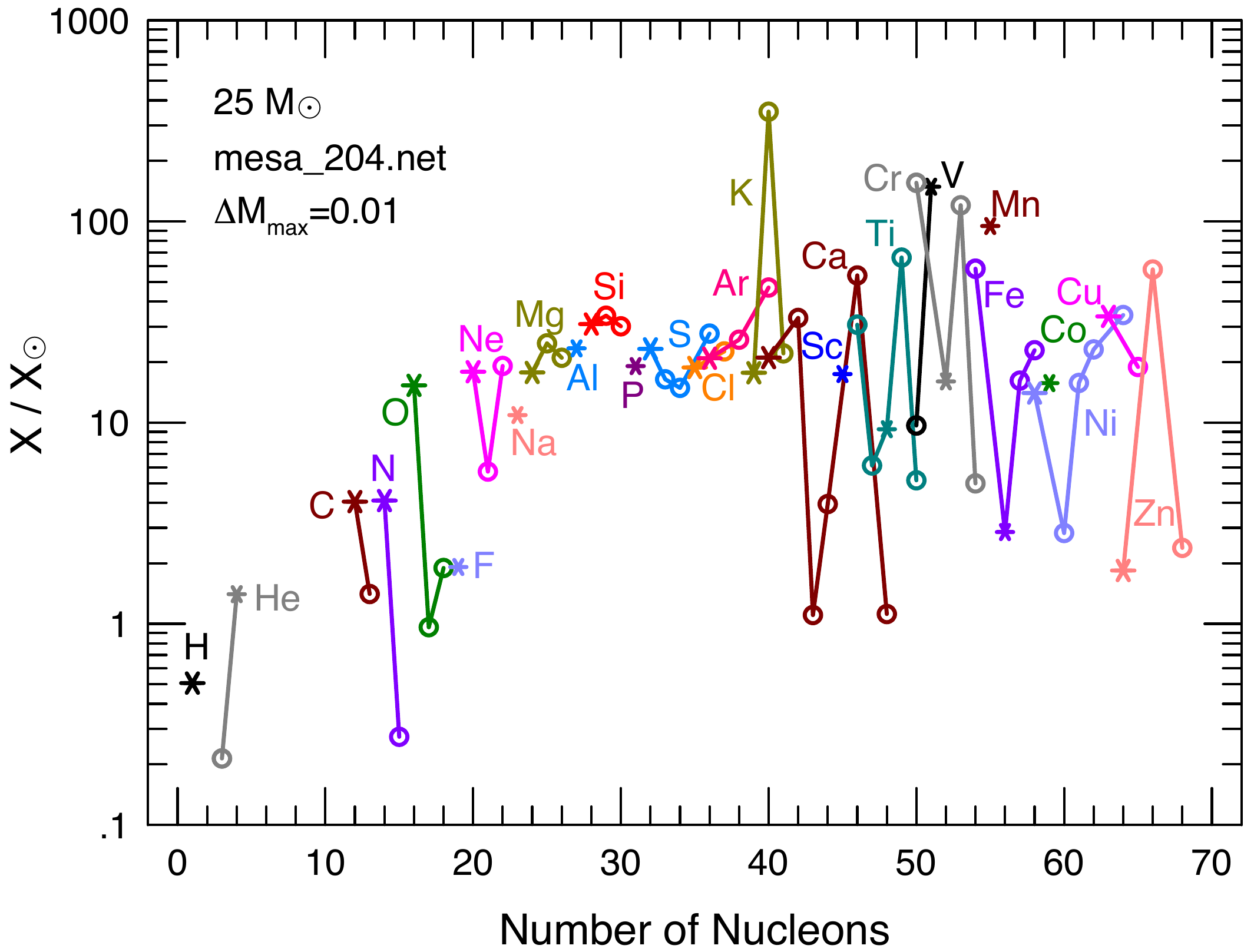}
\includegraphics[width=\columnwidth]{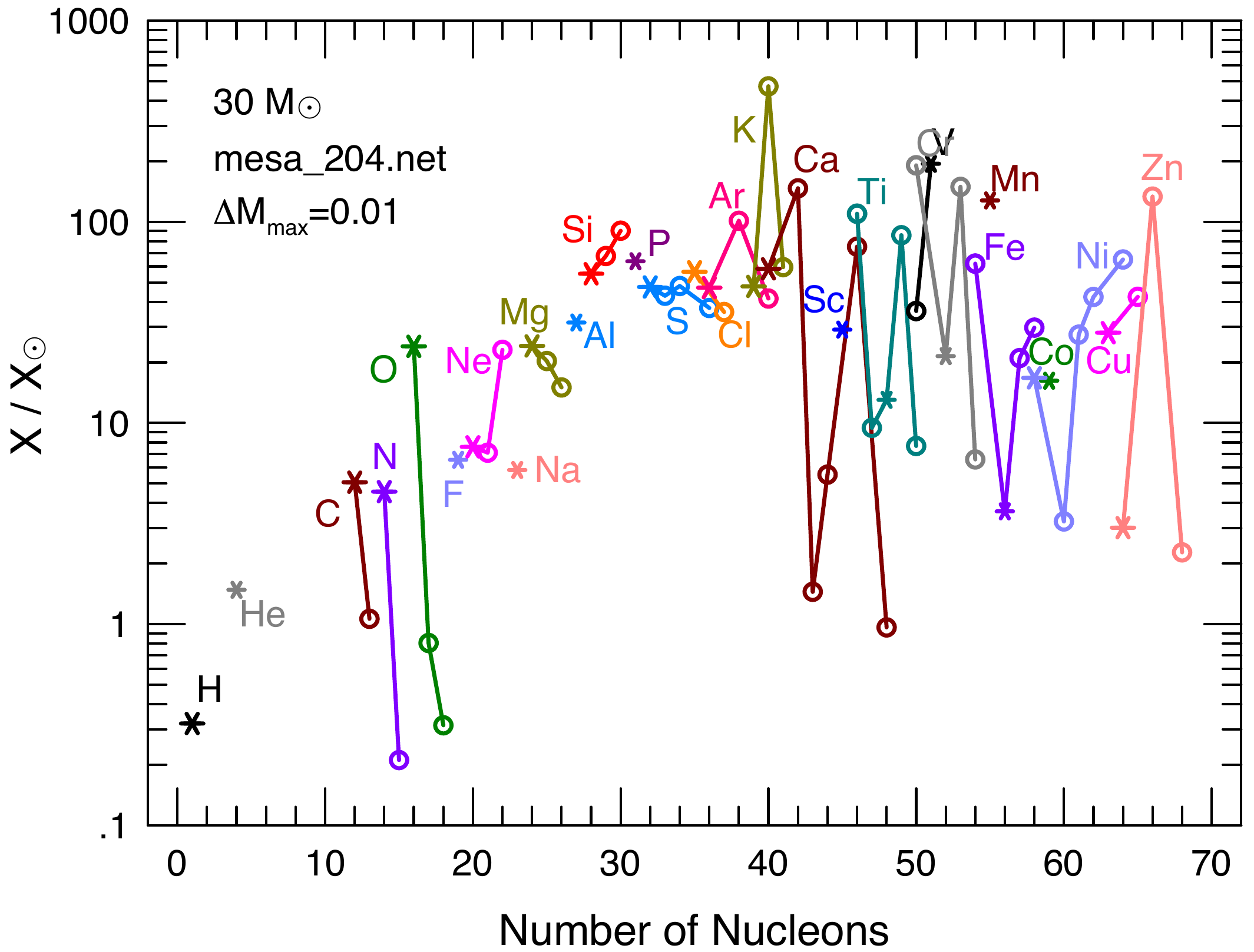}
\caption{
Stable isotopes from hydrogen to zinc for the $\mzams=15$, 20, 25, and
$30$\,\msun models with mass loss, $\maxdm=0.01$\,\msun, and the \net{204}
reaction network.  The x-axis is the atomic mass number.  The y-axis
is the logarithmic ratio of the model mass fraction to the
\citet{grevesse_1998_aa} solar mass fraction.  The most abundant
isotope of a given element is marked by an asterisk and isotopes of
the same element are connected by solid lines.
}
\label{fig:yields}
\end{figure*}

Figure \ref{fig:yields} shows the stable isotopes from hydrogen to
zinc for the pre-SN $\mzams=15$, 20, 25, and $30$\,\msun models with
mass loss, $\maxdm=0.01$\,\msun, and the \net{204} reaction
network. These pre-SN yields are for the stable isotopes outside the Fe-core.
The x-axis is the atomic mass number.  The y-axis is the logarithmic
ratio of the model mass fraction to the \citet{grevesse_1998_aa} solar
mass fraction.  The most abundant isotope of a given element is marked
by an asterisk and isotopes of the same element are connected by solid
lines.

Isotopes in Figure \ref{fig:yields} heavier that about calcium will be
strongly impacted by the explosion. Up to calcium, however, it
generally makes little difference whether the pre-SN or the exploded
yields are used \citep[e.g.,][]{timmes_1995_aa}. In a forthcoming
effort we anticipate exploding the pre-SN models examined in this
paper.  With this limitation of interpreting pre-SN yields in mind, the
overall average production factor of $\approx 20$ for the 
$\mzams=15$\,\msun model and rising to $\approx 100$ for the 
$\mzams=30$\,\msun model are commensurate with existing pre-SN yields
\citep[e.g.,][]{woosley_1995_aa,limongi_2003_aa,chieffi_2013_aa} and
the production factors needed by galactic chemical evolution models
\citep[e.g.,][]{gibson_2003_aa} to reproduce the solar composition.
Part of the spread below calcium is a consequence of uncertain physics 
in the \MESA models: the treatment of convective, semiconvective, 
and overshoot mixing, the parameterization of mass loss, nuclear reaction rates, and residual
uncertainty in the measured solar abundances. Post-SN the yields for isotopes with $\rm{A}>40$
will also depend sensitively on choice of explosion dynamics \citep{paxton_2015_aa}.

The light isotopes $^{6,7}$Li, $^{9}$Be, and $^{10,11}$B are not
plotted in Figure \ref{fig:yields} as the primary source of these
isotopes is commonly taken to be big-bang nucleosynthesis
\citep[e.g.,][]{coc_2012_aa,cyburt_2016_aa}, spallation of CNO nuclei in the
interstellar medium by cosmic rays 
\citep[e.g.,][]{reeves_1970_aa, olive_1992_aa,ramaty_2000_aa,fields_2000_aa,prantzos_2012_aa} 
or the $\nu$-process during the explosion
\citep[e.g.,][]{woosley_1990_aa,balasi_2015_aa}.

The isotopes $^{12,13}$C and $^{14}$N are under produced in Figure
\ref{fig:yields} by $\approx 3$ relative to the average overproduction
factor for the $\mzams=15$\,\msun model and by $\approx 20$ relative to
the average overproduction factor for the $\mzams=30$\,\msun model.
This is consistent with the bulk of these isotopes being produced
during CNO processing and dredged-up material from helium shell
flashes in intermediate and low mass stars
\citep[][]{iben_1978_aa,renzini_1981_aa,doherty_2014_aa,karakas_2016_ab}.

The solar $^{29}$Si abundance is $\approx 1.4$ times larger than the
solar $^{30}$Si abundance, in concordance with observations of some
individual stars \citep{peng_2013_aa} and pre-solar silicon-carbide
grains \citep[e.g.,][]{hoppe_2010_aa}, yet in tension with common models of 1D
core-collapse supernova, Type Ia supernovae, and AGB stars
\citep[e.g.,][]{timmes_1996_aa,lugaro_1999_aa,lewis_2013_aa,wasserburg_2015_aa}.
It is curious that the  $^{30}$Si abundance, normalized to solar in 
in Figure \ref{fig:yields}, is larger than the $^{29}$Si abundance
in the $\mzams=20$ and $30$\,\msun models but traditionally smaller in the 
$\mzams=15$ and $25$\,\msun models.

The isotope $^{40}$K has the largest production factor in the
$\mzams=25$ and $30$\,\msun models of Figure \ref{fig:yields}. This isotope
is also a special case as it is radioactive, but the half-life is long
enough \citep[1.248$\times$10$^{9}$ yr, ][]{malonda_2002_aa,
kossert_2004_aa} that it is included in compilations of the solar
composition. The overproduction factor may reflect an uncertain
solar abundance.

\section{Summary and Discussion}
\label{sec:discussion}

We have investigated the range of variation in properties of \MESA
pre-SN models with respect to changes in spatial resolution, number of
isotopes in the nuclear reaction network, and the inclusion of mass
loss.  To make this assessment we evolved 200 solar metallicity,
non-rotating, single-star models of initial mass $\mzams$=15, 20, 25,
and 30 \,\msun from the pre-MS to the onset of core-collapse.

We found that the choice of spatial resolution can have a larger impact on the final
states of these models, predominately by altering the state of the star during the MS. 
This small effect can then be compounded by the relative length of the MS, compared with later
evolution stages, to have large impacts on the stellar structure.  The choice of nuclear
network is also found not to be insignificant even during core helium burning where larger
nuclear networks can suppress CBPs. The combination of these effects compounds over the stellar
evolution leading to sometimes radically different behaviors between models with the same
mass, including differences; in the type of carbon and neon burning, behavior of C, Ne and O
shells during silicon burning; and the final mass of the iron core. In the remainder 
of this section, we discuss our findings and compare them with previous efforts.

\citet{Hirschi_2004_aa} considered the pre-SN evolution of 
non-rotating solar metallicity stellar models with 
$\mzams=12\textrm{--}60 $\,\msun from the ZAMS to core Si depletion. We compare
the median values of He$_{\rm{shell}}$ for our mass loss models
with $\mzams=$15, 20, and 25 \,\msun to their reported
He core mass values and find agreement of $\approx$ 
$\left|0.6\right| - \left|3.7\right| \%$. Additionally, we compare 
the median values of C$_{\rm{shell}}$ to their quantity, 
$M^{01}_{\rm{CO}}$ which denotes the mass coordinate where
the main fuel for CO burning ($^{4}$He) drops below 10$^{-2}$. 
Our mass loss models with $\mzams$=15, 20, and 25 \,\msun
agree to $\approx$ $\left|0.3\right| - \left|8.8\right| \%$ of their values.

The effect of nuclear and stellar input physics on pre-SN 
nucleosynthesis was considered by \citet{rauscher_2002_aa}. 
Their set of models were evolved using the 1D implicit
hydrodynamics code, \texttt{KEPLER} 
\citep{weaver_1978_aa,woosley_1988_aa,heger_2000_aa},
using a small nuclear network to generate the energy (similar to our
\netapprox) with a larger adaptive network for the nucleosynthesis following 
$\approx 700 \text{--} 2200$ nuclei from the MS to explosion.
Overall, our median final Fe core mass at core-collapse agrees
to within $\approx$ $\left|1.8\right| - \left|7.0\right| \%$ of their 
values. The largest difference is found when comparing
the $\mzams = 20$\,\msun model. This, however, is not surprising and
is stated in their investigation as possibly being a consequence of
the merging of the O-, Ne-, and C-shells $\approx 1$ day prior to 
collapse. We find considerable variation in the behavior of the O-, \hbox{Ne-}, and C-shells
with the choice of whether they merge dependent on the resolution and network chosen.
For instance our 25\,\msun and 30\,\msun models, the C-shell appears
to `merge' or dissolve prior to collapse. There is a large difference in the 
input physics of their models worthy of mention: the mass loss efficiency, convection
parameterization, convective overshoot and other mixing terms. 
On average, their $20$\,\msun and $25$\,\msun models lose $\approx 2\text{--}5$\,\msun 
more than our models.


\citet{limongi_2000_aa} studied $13, 15, 20, 25$\,\msun, solar metallicity models with the
\texttt{FRANEC} code \citep{chieffi_1998_aa} up to iron core collapse. Comparisons with
their final state models they find $\Yec= 0.432, 0.435, 0.436$ for their $\mzams=15,20,25$\,\msun
models, the 15\,\msun models match our results while their 20 and 25\,\msun models have slightly
lower \Yec values, though within our upper and lower bounds on \Yec. They do not
appear to have the ICZ we see in our models as they suppress overshoot and use
the Schwarzschild criterion. They also artificially suppress the formation of 
CBP during core helium
burning. Comparing the locations of the shell masses, our models in general predict shell locations
to be $\approx0.1$\,\msun greater than those of \citet{limongi_2000_aa}. This may be 
due to differences in the choice of mixing parameters as well as the differences in behavior
of the ICZ and SCZ between our models.

\citet{sukhbold_2014_aa} studied the compactness of pre-SN cores
in stars between $\mzams=15\text{--}65$\,\msun with \texttt{KEPLER}. Comparing the
compactness parameter, with their S-series models 
(non-rotating, solar metallicity, with mass loss), we have good agreement
for the 15, 20 and 25\,\msun models for the compactness parameter measured when
$v=1000$ \kmpers. However for the 30\,\msun models, our value of $\xi_{\M=2.5}\approx 0.6$
is much larger then their value of $\xi_{\M=2.5}\approx 0.2$. Though this value is 
consistent with the lower edge of the values we find. 
\texttt{KEPLER} models do not reach $\xi_{\M=2.5}\approx 0.6$ until
$\mzams \approx 35$\,\msun, though they point out that above $\mzams=30$\,\msun 
uncertainties in mass loss can affect the results.

\citet{eid_2004_aa} studied $\mzams=15,20,25$ and $30$\,\msun stars with mass loss
up to the end of central oxygen burning using the code described in \citet{the_2000_aa}.  
On average, comparing the core masses at the ignition of each fuel we find our models
to be $0.5$\,\msun heavier than those of \citet{eid_2004_aa}.  This can be explained
as they use the Schwarzschild criterion everywhere and only study using Ledoux for 
convective boundaries in a  
$25$\,\msun model. When they do use the Ledoux criteria (which we use everywhere)
they find their core masses grow by $0.8$\,\msun which places them 
within the values we find. Comparing the timescale to burn each fuel for the 15\,\msun and
the 25\,\msun models, we find our values to be commensurate with their values, except
for the length of carbon burning. For the 25\,\msun models without mass loss we find 
$\tau_{c}\approx 23$~yr, while they find $\tau_{c}\approx 1860$~yr. Differences
may be due to
the definition used. Comparing with their Figure 9, for the time over which carbon has 
vigorously ignited, they have carbon burning $\approx 100$~yr however their burning is
convective at the center, while we find carbon burning in the 25\,\msun to be 
radiative, likely due to the smaller core mass from using the Schwarzschild criteria.


One should bear in mind the limitations of our studies: our neglect of
rotation and magnetic fields; ambiguous fidelity to the underlying 3D
physics in the \MESA 1D models: the treatment of convective
\citep{trampedach_2014_aa}, semiconvective \citep{moore_2016_aa}, and
overshoot \citep{kitiashvili_2016_aa} mixing; the parameterization of
mass loss \citep[e.g.,][]{surlan_2012_aa,madura_2013_aa}; potentially
underestimated contributions from iron in the opacity
\citep{blancard_2012_aa,krief_2016_aa,krief_2016_ab,turck-chieze_2016_aa,colgan_2016_aa};
and unaccounted for uncertainties in the nuclear reaction rates
\citep[e.g.,][]{sallaska_2013_aa,fields_2016_aa}.

\acknowledgements 

This project was supported by NASA under the Theoretical and
Computational Astrophysics Networks (TCAN) grant NNX14AB53G, by NSF under the
Software Infrastructure for Sustained Innovation (SI$^2$) grant 1339600 and
grant PHY 08-022648 for the Physics Frontier Center ``Joint Institute
for Nuclear Astrophysics - Center for the Evolution of the Elements''
(JINA-CEE). 
L.D acknowledges financial support from 
``Agence Nationale de la Recherche'' grant ANR-2011-Blanc-SIMI-5-007-01.
C.E.F. acknowledges partial support from Michigan 
State University under the College of Natural Sciences Early Start
Fellowship Program. 
F.X.T. acknowledges sabbatical support from 
the Simons Foundation.

\software{
\texttt{MESA} \citep{paxton_2011_aa,paxton_2013_aa,paxton_2015_aa}, 
\texttt{Python} \href{https://www.python.org}{python.org},
\texttt{matplotlib} \citep{hunter_2007_aa}, 
\texttt{NumPy} \citep{der_walt_2011_aa}}

\facilities{Arizona State University Research Computing \texttt{Saguaro} and \texttt{Ocotillo} Systems.}


\bibliographystyle{aasjournal}

\bibliography{paper}

\end{document}